\documentclass[a4paper,11pt]{article}
\pdfoutput=1
\usepackage{jheppub}
\usepackage[T1]{fontenc}
\usepackage[utf8]{inputenc}
\usepackage{multirow}
\usepackage{hyperref}
\usepackage{graphicx,subcaption}
\usepackage{amsmath}
\usepackage{mathrsfs}
\usepackage{siunitx}

\newcommand\GeV{\ \mathrm{GeV}}

\newcommand\TeV{\ \mathrm{TeV}}
\providecommand*{\dd}{\mathop{}\!\mathrm{d}}

\allowdisplaybreaks

\begin{document}

\hfill{ACFI-T19-02}

\title{\boldmath Electroweak Baryogenesis with Vector-like Leptons and Scalar Singlets}

\author[a]{Nicole~F.~Bell,}
\author[a]{Matthew~J.~Dolan,}
\author[a,1]{Leon~S.~Friedrich,\note{Corresponding author}}
\author[b,c,d]{Michael~J.~Ramsey-Musolf,}
\author[a]{and Raymond~R.~Volkas,}
\affiliation[a]{ARC Centre of Excellence for Particle Physics at the Terascale,
  School of Physics, The University of Melbourne, Victoria 3010, Australia}
\affiliation[b]{Amherst Center for Fundamental Interactions, Department of
  Physics, University of Massachusetts Amherst, Amherst, MA 01003, USA and California Institute of Technology, Pasadena, CA 91135 USA}
 \affiliation[c]{Tsung-Dao Lee Institute and  School of Physics and Astronomy, Shanghai Jiao Tong University, 800 Dongchuan Road, Shanghai, 200240 China}
\affiliation[d]{Kellogg Radiation Laboratory, California Institute of Technology, Pasadena, CA 91125 USA}
\emailAdd{n.bell@unimelb.edu.au} \emailAdd{matthew.dolan@unimelb.edu.au}
\emailAdd{leon.friedrich@unimelb.edu.au}
\emailAdd{mjrm@physics.umass.edu}
\emailAdd{raymondv@unimelb.edu.au}

\abstract{We investigate the viability of electroweak baryogenesis in a model with a first order electroweak phase transition induced by the addition of two gauge singlet scalars. A vector-like lepton doublet is introduced in order to
  provide CP violating interactions with the singlets and Standard
  Model leptons, and the asymmetry generation dynamics are examined using the vacuum
  expectation value insertion approximation. We find that such a model
  is readily capable of generating sufficient baryon asymmetry while satisfying
  electron electric dipole moment and collider phenomenology constraints.}

\maketitle
\flushbottom
\section{Introduction}
\label{sec:intro}

Throughout the observable universe there is significantly more matter than
antimatter, and establishing the dynamical origin of this asymmetry remains an open problem in physics. The baryon asymmetry of the universe, as determined by measurements of the cosmic microwave background made by the Planck experiment~\cite{planck2018}, indicates a baryon-to-entropy ratio
\begin{align}
  Y_B = \frac{n_B}{s} = (8.66 \pm 0.04) \cdot 10^{-11}.
\end{align}
Any mechanism that might explain the origin of this asymmetry must satisfy the
three Sakharov conditions~\cite{Sakharov}: There must be baryon number violating
processes; violation of both charge conjugation (C) invariance and its
combination with parity invariance (CP); and either out-of-equilibrium dynamics
or violation of CPT conservation. A range of mechanisms have been proposed that
can satisfy these criteria and thus account for the observed asymmetry.

Electroweak baryogenesis (EWBG) is one such mechanism, in which the asymmetry is
generated during a first-order electroweak phase transition (EWPT). The
nucleation and expansion of the bubbles of broken electroweak symmetry during
the early universe satisfy the out-of-equilibrium criterion, while CP-violating
(CPV) interactions with the bubble wall catalyse asymmetry generation through
electroweak sphaleron processes. In order for EWBG to successfully explain the
observed asymmetry we require a strongly first-order (SFO) electroweak phase
transition, or SFOEWPT. However, the Standard Model (SM) instead features a
crossover transition~\cite{crossover} which cannot provide the needed out of
equilibrium dynamics. Additionally, even with an SFOEWPT, the amount of CPV
present in the SM is not enough to yield the required
asymmetry~\cite{notEnoughCPV}

Thus, the viability of electroweak baryogenesis requires the presence of Beyond
the Standard Model (BSM) physics to generate both a strongly first order EWPT
and to provide additional sources of CP-violation. Since electroweak symmetry
breaking occurs at a temperature $T\sim 100 \GeV$, the corresponding BSM mass
scale must not be too much higher than the electroweak scale. Such extensions to
the SM should thus be testable at current or future
colliders~\cite{SingletColliderPheno1,SingletColliderPheno2,CurtinColliderPheno},
or via high sensitivity fundamental symmetry tests such as electric dipole
moment (EDM) searches, which place stringent constraints on additional sources
of CPV~\cite{EDMSummary}.

While it is straightforward to obtain a strongly first-order electroweak phase
transition by extending the SM scalar sector (the simple addition of a gauge
singlet scalar often
suffices~\cite{SingletEWPT1,SingletEWPT2,SingletColliderPheno1,SingletEWPT3,SingletColliderPheno2,SingletEWPT4}),
introducing new CP-violating phases in Higgs interactions while avoiding EDM
constraints can be challenging. These constraints can be avoided in models where
the EDMs arise at two-loop order, or where the CPV interactions are flavor
non-diagonal, vector-like, or ``partially secluded'' from the SM\footnote{By
  partial seclusion we refer to scenarios where the CPV asymmetries are
  generated in species that carry no SM gauge quantum numbers (such that there
  are no EDM effects). These asymmetries get transferred to the SM via particle
  number changing reactions.}. In the latter instance, the introduction of
additional scalars charged under SU(2) can result in electroweak symmetry
breaking (EWSB) which occurs via multiple, successive phase
transitions~\cite{StepInto,MorriseyTwoStep,TwoStep,ColorTwoStep}. In these
models EWBG may occur during the first EWSB transition, with a subsequent
transition resulting to the usual SM Higgs-phase. This class of models have the
benefit that the new CPV interactions can be hidden in the new scalar sector and
hence avoid the EDM constraints~\cite{TwoStep,ColorTwoStep}.

One can also consider an EWBG scenario which combines elements of the above
possibilities. In what follows, we consider a model involving two scalar
singlets, $S_i$, and a vector-like lepton doublet, $\psi$, with the same quantum
numbers as the SM left handed lepton doublets. The vector-like nature of the new
leptons allows for CP-violating Yukawa-interactions with the gauge singlet
scalars and SM leptons, while also avoiding any contribution to anomalies. The
scalar singlets catalyse a single-step strongly first order EWPT, while their
interactions with the vector-like and SM leptons provide the necessary source of
CP-violation. The EDM constraints for this scenario are somewhat relaxed as:
\begin{itemize}
\item The electron EDM $d_e$ arises at two-loops due to restricting the CPV
  interactions to the third generation.
\item The scalar singlets couple to the first generation leptons only via mixing
  with the SM Higgs boson (a form of ``partial seclusion''), so the two-loop EDM
  involves a small mixing angle.
\item Finally, the new CPV terms in the weak currents which contribute to EDMs
  in similar models~\cite{VLQEDM} do not contribute due to the lack of light
  right handed neutrinos.
\end{itemize}

A model similar to the one outlined in this paper has recently been considered
in Ref.~\cite{ChaoSpontCP}. In that work, a complex scalar singlet interacts
with vector-like quark singlets, and CPV arises spontaneously from the vacuum
expectation value (VEV) of the complex scalar singlet, which changes during the
electroweak phase transition. This, and other similar models, often involve
vector-like fermions which have large Yukawa couplings to the SM
Higgs~\cite{VLFDiphotonEWPT,VLFDMEWPTEWBG,MJRM-CHAO-VL-EWBG,VLFEWPTEWBGEDM,VLFEWPT}.
This is desirable in the context of electroweak baryogenesis since large Yukawa
couplings can provide large CP violating interactions necessary for EWBG, and
may also play a role in acquiring a strongly first order
EWPT~\cite{VLFDiphotonEWPT,VLFDMEWPTEWBG}. However, such models lead to
significant corrections to Higgs boson properties, leading to important
constraints from on the diphoton branching ratio and production processes.

In contrast, the scenario considered here easily evades these constraints as the
SM Higgs coupling to the vector-like leptons is $\mathcal{O}(y_\tau)$ (or even
smaller) and plays no significant role in the asymmetry generation process. Both
classes of models have to contend with the non-observation of direct vector-like
lepton production at colliders, which implies a lower bound on their masses.
This lower bound cannot be much larger than the temperature at which the
electroweak phase transition takes place ($T\sim100\GeV$) in order for EWBG to
remain viable. An in-depth examination of the collider signatures of a minimal
vector-like lepton model has been undertaken in Ref.~\cite{VLLPheno}. Using
$8\TeV$ ATLAS data, Ref.~\cite{VLLPheno} places a lower bound of $\sim 270\GeV$
on the masses of vector-like lepton doublets that decay via mixing with the SM
leptons. However, the situation in our scenario is more complicated, as the
existence of the new scalars modifies the decay chain of the vector-like
leptons. We will also consider lepton universality, lepton flavor violation, and
electron EDM constraints. We find that that such a model can readily generate
the observed asymmetry while avoiding all current bounds, though future searches
for vector-like leptons at colliders will begin placing severe constraints.

The layout of the remainder of this paper is as follows. In
Section~\ref{sec:model} we define our model along with a discussion of the
interactions, mass mixing, and new CPV phases. Section~\ref{sec:pheno}
introduces some benchmark parameters and discusses various observables and
constraints. In Section~\ref{sec:ewbg} we discuss EWBG methodology, how we
derive and solve the transport equations, and our treatment of the EWPT. We
conclude and discuss future directions in Section~\ref{sec:summary}. Technical
details relating to the electron EDM and transport equations are given in
Appendices~\ref{app:EDM}~and~\ref{app:EWBG}, respectively.

\section{Model}
\label{sec:model}

We extend the SM by introducing two real gauge singlet scalars $S_1$ and $S_2$,
and a vector-like lepton doublet $\psi$. These fields transform under the SM
gauge group SU(3)$\times$SU(2)$\times$U(1) as
\begin{equation}
  S_i \sim (1,1,0), \quad \psi \sim
  \left(1,2,-\frac{1}{2}\right) \ .
\end{equation}
We use the notation
\begin{equation}
  \psi = \begin{bmatrix} N \\  E^-	\end{bmatrix}, \quad
  H = \begin{bmatrix} G^+ \\  \frac{1}{\sqrt{2}} (v_H + h + i G^0)	\end{bmatrix},\quad
  S_i =  v_{S_i} + s_i ,
\end{equation}
where the $s_i$ and $h$ are the dynamical fields, $G^0$ and $G^+$ are the
Goldstone bosons, and $v_X$ is the VEV of particle $X$. As the $\psi$ have the
same quantum numbers as the SM leptons, they can in principle mix and interact
with all of the SM lepton generations. Interactions with all of the lepton
generations would face strict constraints from lepton universality, flavor
violation, and electron EDMs. Furthermore, multi-generation interactions are not
necessary for successful electroweak baryogenesis. Hence, for simplicity we only
consider interactions between the $\psi$ and the third-generation of SM leptons.
The Lagrangian then includes the following Yukawa interactions terms,
\begin{align}
  - \mathscr{L}^Y_{\mathrm{\psi}} =
  & \sum_i \left[ \lambda_{L \psi i}  \;  \bar{L}_{3} \psi_R S_i  \ +\ \lambda_{\psi \psi i} \; \bar{\psi}_{L}  \psi_R S_i  \right]   \label{eq:LPsi}+ Y_{\psi}  \; \bar{\psi}_L  \left(i \sigma_2 H^*\right)  \tau_R  \ +\ \mathrm{h.c.},     
\end{align}
where $L_3 = \left(\nu_\tau , \tau_L^- \right)^T$ is the SM third generation
left-handed lepton doublet. These Yukawa interactions result in mixing between
the $\psi$ and SM leptons, and provide the CPV necessary for EWBG. There are
also two possible Dirac mass terms
\begin{equation}
 - \mathscr{L}^M_{\mathrm{\psi}} = M_{\psi} \bar{\psi}_L \psi_R\ +\ M_{L \psi} \bar{L}_3 \psi_R\ +\ \mathrm{h.c.}
 \label{eq:massTerms}
\end{equation}
The $M_{L \psi}$ mass mixing term can be removed via a redefinition of the
$\psi_L$ and $L_3$ fields,
\begin{equation}
  \label{eq:MLRemove}
  \begin{aligned}
    \begin{bmatrix}
    \psi_L
    \\
    L_3
    \end{bmatrix}
    &\rightarrow
    \begin{bmatrix}
    \psi_L'
    \\
    L_3'
    \end{bmatrix}
    =
    \mathcal{U} \cdot
    \begin{bmatrix}
    \psi_L
    \\
    L_3
  \end{bmatrix}, \qquad
  \mathcal{U} = 
    \begin{bmatrix}
     e^{ i \phi_1 }\cos{\theta} & \  e^{ i \phi_2}\sin{\theta}
     
    \\
    - e^{ i \phi_1 }\sin{\theta} & \ e^{ i \phi_2 }\cos{\theta}
    \end{bmatrix} 
    , \\
    \sin{\theta} &= \frac{\left\lvert M_{L \psi}\right \rvert }{\sqrt{ \left \lvert {M_{\psi}}  \right \rvert^2 + \left \lvert {M_{L \psi}}  \right \rvert^2  }}, \qquad \phi_1= \mathrm{Arg}\left(M_\psi\right), \qquad \phi_2= \mathrm{Arg}\left(M_{L \psi}\right),
    \end{aligned}
  \end{equation}
  such that eq.~\ref{eq:massTerms} becomes,
\begin{equation}
 - \mathscr{L}^M_{\mathrm{\psi}} = \sqrt{ \left \lvert {M_{\psi}}  \right \rvert^2 + \left \lvert {M_{L \psi}}  \right \rvert^2  }\, \bar{\psi}'_L \psi_R' +\ \mathrm{h.c.}
\, = \, M_\psi' \bar{\psi}'_L \psi_R' +\ \mathrm{h.c.}
 \label{eq:massTerms}
\end{equation}
We can then perform a redefinition of the Yukawa couplings,
\begin{equation}
  \label{eq:yukawatransform}
  \begin{bmatrix}
    Y_\psi & Y_\tau
  \end{bmatrix}
  \rightarrow 
  \begin{bmatrix}
    Y_\psi' & Y_\tau'
  \end{bmatrix}
  =
  \begin{bmatrix}
    Y_\psi & Y_\tau
  \end{bmatrix}
  \cdot
  \mathcal{U}^\dagger \, , 
\end{equation}
and similarly for $\lambda_{L \psi i}$ and $\lambda_{\psi \psi i}$. Aside from
removing $M_{L \psi }$ this transformation leaves the form of the
Lagrangian unchanged and we drop the primes.

\subsection{Scalar Potential}

We now examine the scalar potential and outline how its parameters are related
to the zero temperature VEVs and scalar masses. The scalar potential is
\begin{equation} 
  \begin{aligned}
    \label{eq:potential}
    V_0 \ =\
    -\ &\mu_H^2 H^\dagger H \ +\ \lambda_H (H^\dagger H)^2 
    \\
    +\ &\sum_i a_i S_i \ -\ \frac{1}{2} \sum_{i,j} a_{ij}^2 S_i S_j  \ + \  \frac{1}{3} \sum_{i,j,k} a_{i j k} S_{i} S_{j} S_{k} \ + \  \frac{1}{4} \sum_{i,j,k,l} a_{i j k l} S_i S_j S_k S_l  
    \\
    +\ &\frac{1}{2} \sum_{i,j}  b_{ij} S_i S_j H^\dagger H \ +\  \sum_i b_{i} S_i H^\dagger H,
  \end{aligned}
\end{equation}
where the sums go over the two scalar singlets, and the couplings are symmetric
in all indices. As the $S_i$ are gauge singlets they can be shifted by a
constant ($S_i \rightarrow S_i + c$), and if the couplings are suitably
redefined, this will leave the form of the Lagrangian unchanged. This shift
freedom is frequently used to eliminate the linear terms $a_i S_i$. However, in
order to simplify the resulting mass matrices that need to be diagonalised, we
instead use this freedom to relocate the global minimum of the effective
potential at zero temperature to a point where $v_{S_i} = 0$. This choice
implies
\begin{equation}
\left. \frac{\partial V}{\partial v_{S_i}} \right. = a_i + \frac{1}{2} v_H^2 b_{i} = 0,
\end{equation}
which we use to fix $a_i$. We then require that the global minimum resembles
that of the SM, resulting in a SM-like Higgs with mass and VEV,
  \begin{equation}
    (v_H, v_{S_1}, v_{S_2}) = (246, 0 , 0 )\GeV, \qquad m_h = 125\GeV.
  \end{equation}
This fixes $\lambda_H = \frac{\mu_H^2}{2 v_H^2}$ and $\mu_H^2$ such that we obtain the correct Higgs mass after diagonalisation of the mass matrix. Similar to the shift, we can also always perform a rotation of the singlets. We choose to rotate such that the scalar singlet mass mixing terms disappear,  which fixes $a_{12}$ such that
\begin{equation}
  a_{12} = - \frac{1}{2} b_{12} v_H^2.
\end{equation}

The remainder of the scalar potential couplings are then free parameters,
subject to the requirement that the SM Higgs-phase is indeed the global minimum
of the effective potential at zero temperature. We are interested in the
scenario where this potential enables baryogenesis by having a strongly
first-order phase transition during which both of the scalar singlets have
changing VEVs. In this paper we focus on the transport and asymmetry generation
dynamics and do not perform a thorough scan of the scalar potential parameter
space. We utilise a single EWPT benchmark point which fixes these scalar
potential parameters and satisfies the SM-like global minimum requirement (see
Section~\ref{sec:ewpt}). A more detailed study of the phase transition is left
to future study.

\subsection{New CPV phases}

We now briefly discuss the new CPV phases that appear in our model and how we
parameterise them. Some of the phases in the Yukawa couplings can be removed by rephasing  fields via the transformation
\begin{equation}
  \label{eq:rephasings1}
  \begin{aligned}
    & \psi_R \rightarrow \psi_R e^{i \alpha},  && \psi_L \rightarrow \psi_L e^{i \beta},
    \\
    & L_{3} \rightarrow L_{3} e^{i \gamma},    && \tau_{R} \rightarrow \tau_R e^{i \delta}.
  \end{aligned}
\end{equation} With these rephasings we find that the Lagrangian is unchanged if we take
\begin{equation}
  \label{eq:rephasings2}
  \begin{aligned}
    \lambda_{L \psi i } &\rightarrow \lambda_{L \psi i } e^{i (\gamma-\alpha)},  & Y_\tau &\rightarrow Y_\tau e^{i (\gamma - \delta)}, 
    \\
    \lambda_{\psi \psi i} &\rightarrow \lambda_{\psi \psi i} e^{i (\beta-\alpha)} ,
    & Y_\psi &\rightarrow Y_\psi e^{i (\beta -\delta)},\\
    M_{\psi} &\rightarrow M_{\psi} e^{i (\beta-\alpha)}, &&
  \end{aligned} 
\end{equation} such that we have the following invariant phases,
\begin{equation}
  \label{eq:invariants}
  \begin{aligned}
    \delta_1 &= \mathrm{Arg}(\lambda_{L \psi 1} \lambda_{L \psi 2}^*),& 
    \delta_2 &= \mathrm{Arg}(\lambda_{\psi \psi 1} M^*_{\psi}), \\
    \delta_3  &= \mathrm{Arg}(\lambda_{\psi \psi 2} M^*_{\psi}),& 
    \delta_4 &= \mathrm{Arg}(Y_\tau^* Y_{\psi} \lambda_{L \psi 2} M^*_{\psi}).
  \end{aligned}
\end{equation}
We will find that $\delta_{1}$--$\delta_{3}$ will appear within the EWBG
calculation as the sources of the CP-violation necessary to generate the
asymmetry. $\delta_4$ can contribute to electron EDM constraints but will not
directly contribute to the asymmetry generation.
From here on we choose to re-phase our
couplings such that the following conditions are satisfied
\begin{equation}
  \label{eq:phaseFix1}
  \mathrm{Arg}\left( M_\psi \right)\; =\;  
  \mathrm{Arg}\left( Y_\tau \right) \; = \; 
  \mathrm{Arg}\left( \lambda_{L \psi 2} \right) \; = \; 0 
\end{equation}
which is equivalent to setting
\begin{equation}
  \label{eq:phaseFix2}
  \begin{aligned}
    \delta_1 &= \mathrm{Arg}(\lambda_{L \psi 1}),  
    & \delta_2 &=\mathrm{Arg}(\lambda_{\psi \psi 1}),  \\
    \delta_3 &=\mathrm{Arg}(\lambda_{\psi \psi 2}) , 
    & \delta_4 &=\mathrm{Arg}(Y_{\psi}) . \\
  \end{aligned}
\end{equation}
The choice of this specific phase will simplify some of the expressions in the
following sections.

\subsection{Mass and Mixing Matrices}
\label{subsec:MassMix}

We now consider the diagonalisation of the mass matrices. The neutral scalar and
lepton mass terms are given by,
\begin{subequations} 
  \label{eq:massMatrices}
  \begin{equation}
    \frac{1}{2} \phi^T \mathcal{M}^2_\phi \phi + \left(  \bar{{\mathcal{E}}}_L \mathcal{M}_{\mathcal{E}} {\mathcal{E}}_R + \bar{\mathcal{N}}_L \mathcal{M}_{\mathcal{N}} \mathcal{N}_R  + \mathrm{h.c.}\right),
  \end{equation}
  where,
  \begin{align}
    \phi
    &=
      \begin{bmatrix}
        h \\ s_1 \\ s_2
      \end{bmatrix}
    , &
        \mathcal{E} &=
                      \begin{bmatrix}
                        E \\ \tau
                      \end{bmatrix}
    , &
        \mathcal{N}_L &=
                        \begin{bmatrix}
                          N_L \\ \nu_\tau
                        \end{bmatrix}
    , &
        \mathcal{N}_R &=
                        \begin{bmatrix}
                          N_R \\ 0
                        \end{bmatrix}
    , 
  \end{align}
  \begin{equation}
    \mathcal{M}^2_{\phi}
     =
      \begin{bmatrix}
        - \mu_H^2 + 3 \lambda v_H^2  \ &\ v_H b_1  \ &\   v_H b_2 
        \\
        v_H b_1  \ &\ - a_{11}^2 + \frac{1}{2} b_{11} v_H^2  \ &\  0
        \\
        v_H b_2  \ &\ 0       \ &\  - a_{22}^2 + \frac{1}{2} b_{22} v_H^2 
      \end{bmatrix},
    \end{equation}
    \begin{align}
                                   \mathcal{M}_{{\mathcal{E}}}
                   &=
                     \begin{bmatrix}
                       M_{\psi}
                       \ &\ 
                       \frac{v_H Y_\psi}{\sqrt{2}}
                       \\
                       0
                       \ &\ 
                       \frac{v_H Y_\tau}{\sqrt{2}}
                     \end{bmatrix},
                   &   
                     \mathcal{M}_{{\mathcal{N}}}
                                 &=
                                   \begin{bmatrix}
                                     M_{\psi}
                                     \ &\ 
                                     0
                                     \\
                                     0
                                     \ &\ 
                                     0
                                   \end{bmatrix}.&&
  \end{align}
\end{subequations}
These mass matrices can be diagonalised by a redefinition of the fields
\begin{align}
  \phi' = \mathcal{P} \phi
  , \qquad
  {\mathcal{E}}'_{L/R} = \mathcal{U}^\dagger_{L/R} {\mathcal{E}}_{L/R}
  ,
\end{align}
such that
\begin{subequations}
  \label{eqs:massDiag}
  \begin{align}
    \mathcal{P} \mathcal{M}^2_\phi \mathcal{P}^T &=  \mathrm{Diag}(m^2_{h},m^2_{s_1},m^2_{s_2}),
    \\ 
    \mathcal{U}^\dagger_{L} \mathcal{M}_{\mathcal{E}} \mathcal{U}_{R} &=  \mathrm{Diag}(m_E, m_\tau) \label{eq:leptonMassDiag}.
  \end{align}
\end{subequations}
The $\mathcal{M}_{\mathcal{N}}$ matrix is already diagonal and real, such that
no redefinitions are necessary. However, the form of the weak currents after
diagonalisation can be simplified by choosing to re-phase the neutral
vector-likes $N'_{L/R} = N_{L/R} e^{i \delta_4}$. In the limit where the scalar
mass mixing terms $v_H b_i$ are small relative to the scalar mass differences,
the elements of $\mathcal{P}$ are well-approximated by,
\begin{subequations}
\begin{align}
    \mathcal{P}_{i i} & \approx 1, \\ 
    \mathcal{P}_{1 i} & \approx - \mathcal{P}_{i 1} \approx \frac{(M_{\phi}^2)_{1 i}}{(M_{\phi}^2)_{11} - (M_{\phi}^2)_{ii}} , \\
    \mathcal{P}_{2 3} &\approx \frac{(M_{\phi}^2)_{1 3} (M_{\phi}^2)_{1 2}}{ \left( (M_{\phi}^2)_{22} - (M_{\phi}^2)_{33} \right)\left((M_{\phi}^2)_{11} - (M_{\phi}^2)_{22}\right)} , \\
    \mathcal{P}_{3 2} &\approx \mathcal{P}_{2 3} \frac{(M_{\phi}^2)_{11} - (M_{\phi}^2)_{22}}{ (M_{\phi}^2)_{11} - (M_{\phi}^2)_{33}} .
\end{align}
\end{subequations}

For the fermions, in the case where $M_\psi$ in $\mathcal{M}_{\mathcal{E}}$ is
much larger than any other term, we can approximate the mixing matrices via
\begin{equation}
  \mathcal{U}_{L,R}
  =
  \begin{bmatrix}
    e^{i \delta_4}\cos\theta_{L,R} \ &\  -e^{i \delta_4} \sin \theta_{L,R} \\
      \sin\theta_{L,R}\ &\  \cos\theta_{L,R} 
  \end{bmatrix},
\end{equation}
where
\begin{equation}
\label{eq:thetaRL}
  \theta_R \approx \frac{v_H \lvert Y_\psi \rvert}{\sqrt{2} M_\psi}, \qquad 
  \theta_L \approx \frac{ v_H^2 Y_{\tau} \lvert Y_\psi \rvert}{2 M_\psi^2}, 
\end{equation}
and $\delta_4$ is given in eq.~\eqref{eq:phaseFix2}. The $\tau$ mass defined
in eq.~\eqref{eq:leptonMassDiag} will be shifted slightly from its SM
relation with Higgs $Y_\tau$ Yukawa and is given by,
\begin{align}
  m_\tau \approx \frac{Y_\tau v_H}{\sqrt{2}} \left(  1 - \frac{v_H^2 \lvert Y_\psi\rvert^2  }{4 M_\psi^2}   \right).
\end{align}
As we want the physically measured $m_\tau$, we fix $Y_\tau$ as a function of
$M_\psi$ and $Y_\psi$ to give the correct value. If $\lvert Y_\psi \rvert \sim
Y_\tau$ and $M_\psi>200\GeV$ the correction to $Y_\tau$ will be orders of
magnitude smaller than the current uncertainty in $Y_\tau$ inferred from Higgs
decay measurements~\cite{higgstautau}. The mass of the charged component of
$\psi$ will also be shifted slightly,
\begin{align}
  m_E \approx M_\psi \left(  1 + 
  \frac{   v_H^2 \lvert Y_\psi \rvert^2 }{4 M^2_\psi} \right).
\end{align}
This will lead to a small mass splitting between the neutral and charged
vector-like leptons which will be further enhanced by radiative corrections;
however, we will still generally refer to the masses of the $\psi$ components as
simply $M_\psi$.

Using these mixing matrices one is able to write out the new forms of the weak
current and Yukawa interactions after diagonalisation. These are given in
Appendix~\ref{app:EDM}, where we have employed the notation of
Ref.~\cite{mjrmEDMmssm}, whose electron EDM expressions will be used in the next
section. As the mixing angles are small, the diagonalisation matrices were all
selected such that the mass basis states are labelled by their primary
components, {\it e.g}., $\tau'$ will be the physical lepton that consists
primarily of $\tau_L$ and $\tau_R$. Throughout the phenomenology section we will
refer to the mass basis and simply drop the primes.

\section{Phenomenology}
\label{sec:pheno}

In order to quantitatively investigate the phenomenology of this model we
consider the four benchmark scenarios defined in Tables~\ref{tab:Benchmarks1}
and \ref{tab:Benchmarks2}. The resulting mixing parameters are provided in
Table~\ref{tab:Benchmarks3}. The scalar potential parameters of benchmark points
$A$ and $C$ were obtained from a scan for a suitable electroweak phase
transition, as discussed in Section~\ref{sec:ewbg}. Benchmark $B$ is a variant
of $A$ with slightly different Yukawa couplings which modify the decays of the
new particles leading to significant effects on the collider phenomenology.
Benchmark $D$ is a variant of $C$ with no mixing between the physical Higgs and
the new scalars. We will study the constraints from Higgs physics, direct
collider searches sensitive to the vector-like leptons, and precision lepton
phenomenology.

\begin{table}
  \centering
  \begin{tabular}{|c||c|c|c|c|} 
    \hline
    Benchmark  & $A$ & $B$ &  $C$ & $D$\\
    \hline
    \hline
    $m_{s_1}$ ($\mathrm{GeV}$) & \multicolumn{2}{c|}{$157.5$}    & \multicolumn{2}{c|}{$127.6$}       \\\hline
    $m_{s_2}$ ($\mathrm{GeV}$) &  \multicolumn{2}{c|}{$136.4$}    & \multicolumn{2}{c|}{$205.1$}       \\\hline
    $a_{1111}$ & \multicolumn{2}{c|}{$2.80$}  & \multicolumn{2}{c|}{$2.44$}\\\hline
    $a_{2222}$ & \multicolumn{2}{c|}{$2.87$}  &  \multicolumn{2}{c|}{$1.30$}\\\hline
    $a_{1122}$ & \multicolumn{2}{c|}{ $-4.32$} & \multicolumn{2}{c|}{$-1.36$}\\\hline
    $b_{11}$ & \multicolumn{2}{c|}{$0.511$}  & \multicolumn{2}{c|}{$0.962$}\\\hline    
    $b_{22}$ & \multicolumn{2}{c|}{$1.36$} & \multicolumn{2}{c|}{$1.67$}\\\hline
    $b_{1}$ ($\mathrm{GeV}$)   &  \multicolumn{2}{c|}{$0.0164$}  & $0.01$   & $0$  \\\hline
    $b_{2}$ ($\mathrm{GeV}$)   & \multicolumn{2}{c|}{$0.00615$} &
                                                                  $0.5$  & $0$ \\\hline
    $\left\lvert  \lambda_{L \psi 1}\right\rvert$ & $0.0089$ & $0.05$ & \multicolumn{2}{c|}{$0.036$}\\\hline
    $\left\lvert \lambda_{L \psi 2} \right\rvert$ &\multicolumn{2}{c|}{$0.036$} & \multicolumn{2}{c|}{$0.05$}   \\\hline
    $\left\lvert \lambda_{\psi \psi i} \right\rvert$ &\multicolumn{2}{c|}{$3$} & \multicolumn{2}{c|}{$3.5$}   \\\hline
    $\delta_1$ &\multicolumn{2}{c|}{$-\frac{1}{3} \pi$} & \multicolumn{2}{c|}{$\frac{1}{2} \pi$}   \\\hline
    $\delta_2$ &\multicolumn{2}{c|}{$\frac{3}{4} \pi$} & \multicolumn{2}{c|}{$-\frac{4}{5} \pi$}   \\\hline
    $\delta_3$ &\multicolumn{2}{c|}{$-\frac{3}{4} \pi$} & \multicolumn{2}{c|}{$\frac{4}{5} \pi$}   \\\hline
  \end{tabular}
  \caption{Benchmark point parameters. Parameters not listed here are
    the same for all the benchmark points, and are given in
    Table~\ref{tab:Benchmarks2}. Phases were selected to maximise the baryon
    asymmetry. }
  \label{tab:Benchmarks1}
\end{table}

\begin{table}
  \centering
  \begin{tabular}{|c|c|c|c|c|c|c|c|c|} 
    \hline
        $M_\psi$    & $Y_\psi$           & $\delta_4$             & $b_{12}$ & $a_{i j k}$ & $a_{i j j j}$
    \\\hline
         $500\GeV$ & $0.1 \cdot Y_\tau$ & $\frac{\pi}{2}$ &  $0$     &    $0$       & $0$             
    \\\hline
  \end{tabular}
  \caption{Additional scalar and Yukawa coupling parameters that are the same
    across all benchmark points.}
  \label{tab:Benchmarks2}
\end{table}

\subsection{Vector-Like Lepton Collider Phenomenology}
\label{sec:colliderPheno}
The charged and neutral components of the vector-like lepton doublet can be
produced at colliders via Drell-Yan processes. In minimal vector-like lepton
models they decay into a SM lepton (a $\tau$ or $\nu_{\tau}$) plus either a SM
Higgs or weak gauge boson via the mixing induced by the SM Higgs Yukawa
$Y_\psi$. This results in a lower bound on minimal vector-like lepton doublet
masses $M_\psi \gtrsim 270 \GeV$~\cite{VLLPheno} (using 8 TeV data). However, in
our scenario if the gauge singlet scalars are lighter than the vector-like
leptons, the leptons may instead decay via the singlet scalars. This is the case
for all of our benchmark points. This is motivated by the requirement that both
of the scalars have changing VEVs during the electroweak phase transition, which
generally requires them to have non-negligible negative mass terms. When
combined with perturbativity bounds on the $b_{ii}$ couplings, this leads to an
upper bound on the zero temperature scalar masses $m_{s_i}^2 \approx - a_{ii} +
\frac{1}{2} b_{ii} v_H^2 \lesssim - a_{ii} + 2 \pi v_H^2$. We thus expect that
if we have the desired phase transition the scalars will generally be lighter
than the vector-like leptons.

\begin{table}
  \centering
  \begin{tabular}{|c||c|c|c|} 
    \hline
    Benchmark  & $A,B$ &  $C$ & $D$\\ \hline
    $\mathcal{P}_{12}$ & $ -4.39\cdot 10^{-4} $ & $ -3.71\cdot 10^{-3} $ & $ 0 $ \\\hline
    $\mathcal{P}_{13}$ & $ -5.09\cdot 10^{-4} $ & $ -4.65\cdot 10^{-3} $ & $ 0 $ \\\hline
    $\mathcal{P}_{23}$ & $ 1.07\cdot 10^{-7} $ & $ -1.77\cdot 10^{-5} $ & $ 0 $ \\\hline
    $\mathcal{P}_{32}$ & $ 3.30\cdot 10^{-7} $ & $ -4.40\cdot 10^{-7} $ & $ 0 $ \\\hline
    $\theta_L$ &\multicolumn{3}{c|}{$ 1.27\cdot 10^{-6} $} \\\hline
    $\theta_R$ &  \multicolumn{3}{c|}{$ 3.56\cdot 10^{-4} $ } \\\hline
   \end{tabular}
  \caption{Mixing parameters for the benchmark points as defined in Section~\ref{subsec:MassMix}. Benchmark points $A$ and $B$ have
    the same mixing angles. $\theta_{L,R}$ are the same across all the benchmark
  points as they only depend on $M_\psi$ and $Y_\psi$.}
  \label{tab:Benchmarks3}
\end{table}

The dominant decay modes of the vector-like leptons are depend on the relative
sizes of the Yukuwa couplings with the scalar singlets $\lambda_{L\psi i}$, and
the Yukawa coupling with the SM Higgs $Y_\psi$ ,and the mixing angles $\theta_L$
and $\theta_R$. If $Y_\psi$ dominates, then the main decay modes of the new
leptons will be as for the minimal vector-like lepton doublet model. On the
other hand, if $Y_\psi$ and the associated mixings are small relative to
$\lambda_{L\psi i}$ then $\psi$ will mainly decay into a SM lepton and a singlet
$s_i$. In the benchmark points we consider this is the case, as can be seen in
Table~\ref{tab:decays}, which shows the main branching ratios of the vector-like
leptons and scalar singlets for benchmarks $A$ and $D$.

The decays of the scalar singlets depend on whether the singlets mix with the SM
Higgs boson. If they do, then they inherit their decays from the Higgs decay
channels, with the Higgs mass set to that of the appropriate singlet. This is
the case for benchmark points $A$, $B$ and $C$ . Since the scalar singlets for
these points have masses between $127$ and $200\GeV$ their decays are mainly
into $b\bar{b}$, $W W^{(*)}$ and $Z Z^{(*)}$. For benchmark point $D$ the
singlet mixing with the SM Higgs is turned off and the singlets decay into
$\tau^+ \tau^-$ pairs through the $\lambda_{L\psi i}$ coupling and $E$-$\tau$
mixing. We have checked that singlet lifetimes in all cases do not lead to
displaced vertices at the LHC or constraints from BBN.

Therefore, in $E^+ E^-$ pair production the final state will always contain a $\tau^+ \tau^-$ pair and the decay products of two singlets. As the singlets frequently decay into $WW^{(*)}/ZZ^{(*)}$, which may in-turn decay leptonically, we expect multilepton searches featuring signal regions with multiple $\tau$ leptons to be the most sensitive to our scenario.

\begin{table}
  \centering
  \begin{tabular}{|c|c|c|c|} 
    \hline
    \multicolumn{4}{|c|}{Benchmark point $A$} \\\hline
    Particle  & Total Width & Decay Products &  Branching Fraction \\ \hline
    \multirow{2}{*}{$E^{\pm}$} & \multirow{2}{*}{$5.74\cdot 10^{-3}\GeV$}
                            & $\tau^{\pm} s_2$ & $94.1\%$ \\ & 
                            & $\tau^{\pm} s_1$ & $5.57\%$ \\\hline
    \multirow{2}{*}{$N$} & \multirow{2}{*}{$5.75\cdot 10^{-3}\GeV$}
                            & $\nu_{\tau} s_2$ & $93.9\%$ \\ &
                            & $\nu_{\tau} s_1$ & $5.56\%$ \\\hline
    \multirow{4}{*}{$s_{1}$} & \multirow{4}{*}{$1.27\cdot 10^{-9}\GeV$}
                            & $W W^{*}$ & $78.8\%$ \\ &
                            & $b \bar{b}$ & $14.7\%$ \\ &
                            & $Z Z^{*}$ & $5.4\%$ \\ &
                            & $\tau^+ \tau^-$ & $1.12\%$ \\\hline
    \multirow{4}{*}{$s_{2}$} & \multirow{4}{*}{$3.41\cdot 10^{-9}\GeV$}
                            & $b \bar{b}$ & $66.0\%$ \\ &
                            & $W W^{*}$ & $28.8\%$ \\ &
                            & $Z Z^{*}$ & $3.64\%$ \\ &
                            & $\tau^+ \tau^-$ & $1.65\%$ \\\hline
    \multicolumn{4}{c}{}\\\hline
    \multicolumn{4}{|c|}{Benchmark point $D$} \\\hline
    Particle  & Total Width & decay products &  Branching fraction \\ \hline
    \multirow{2}{*}{$E^{\pm}$} & \multirow{2}{*}{$1.43\cdot 10^{-2}\GeV$}
                            & $\tau^{\pm} s_2$ & $60.3\%$ \\ &
                            & $\tau^{\pm} s_1$ & $39.5\%$ \\\hline
    \multirow{2}{*}{$N$} & \multirow{2}{*}{$1.43\cdot 10^{-2}\GeV$}
                            & $\nu_{\tau} s_2$ & $60.2\%$ \\ &
                            & $\nu_{\tau} s_1$ & $39.4\%$ \\\hline
    \multirow{1}{*}{$s_{1}$} & \multirow{1}{*}{$7.96\cdot 10^{-12}\GeV$} 
                            & $\tau^+ \tau^-$ & $100\%$ \\\hline
    \multirow{3}{*}{$s_{2}$} & \multirow{3}{*}{$2.47\cdot 10^{-11}\GeV$} 
                            & $\tau^+ \tau^-$ & $99.9\%$ \\ &
                            & $\tau^+ \tau^- s_1$ & $0.05\%$ \\ &
                            & $\nu_{\tau} \bar{\nu}_{\tau} s_1$ & $0.05\%$ \\\hline
   \end{tabular}
   \caption{List of particle widths, decay modes, and branching fractions as
     calculated by \texttt{MadGraph5} (via \texttt{MadWidth}) for benchmark points $A$ and $D$.
     Benchmark points $A$ and $D$ were selected in order to illustrate the
     effect of $s_i$-$h$ mixing, which is not present in benchmark point $D$.}
  \label{tab:decays}
\end{table}

To examine in detail the constraints arising from collider analyses we
have implemented our model in \texttt{FeynRules}~$2.3.29$~\cite{asperge,feynrules}. Using
the UFO format~\cite{UFO} we import our model into
\texttt{MadGraph5\_aMC@NLO}~$2.6.1$~\cite{madgraph,madWidth} in order to
generate events at parton level, before showering them using
\texttt{Pythia}~$8.230$~\cite{pythia1,pythia2}. These are then fed into
\texttt{CheckMATE}~$2.0.26$~\cite{checkmate,CLS, fastjet2,cmjet,fastjet1}, which
includes an interface with \texttt{Delphes}~$3.4.1$~\cite{delphes}, in order to
determine the most relevant ATLAS or CMS analyses and constraints on our model.

QCD corrections to the pair production of SU(2) triplet leptons in type-III seesaw models at $13\TeV$ leads to $K$-factors ranging from $1.17$--$1.25$ for lepton masses ranging from $100$--$1000\GeV$~\cite{LeptonKFactor}. Based on this result we scale the $\psi$ production cross sections passed to \texttt{CheckMATE} using a $K$-factor of $1.2$. 

\begin{figure}[t]
  \centering
  \includegraphics[width=0.49\textwidth]{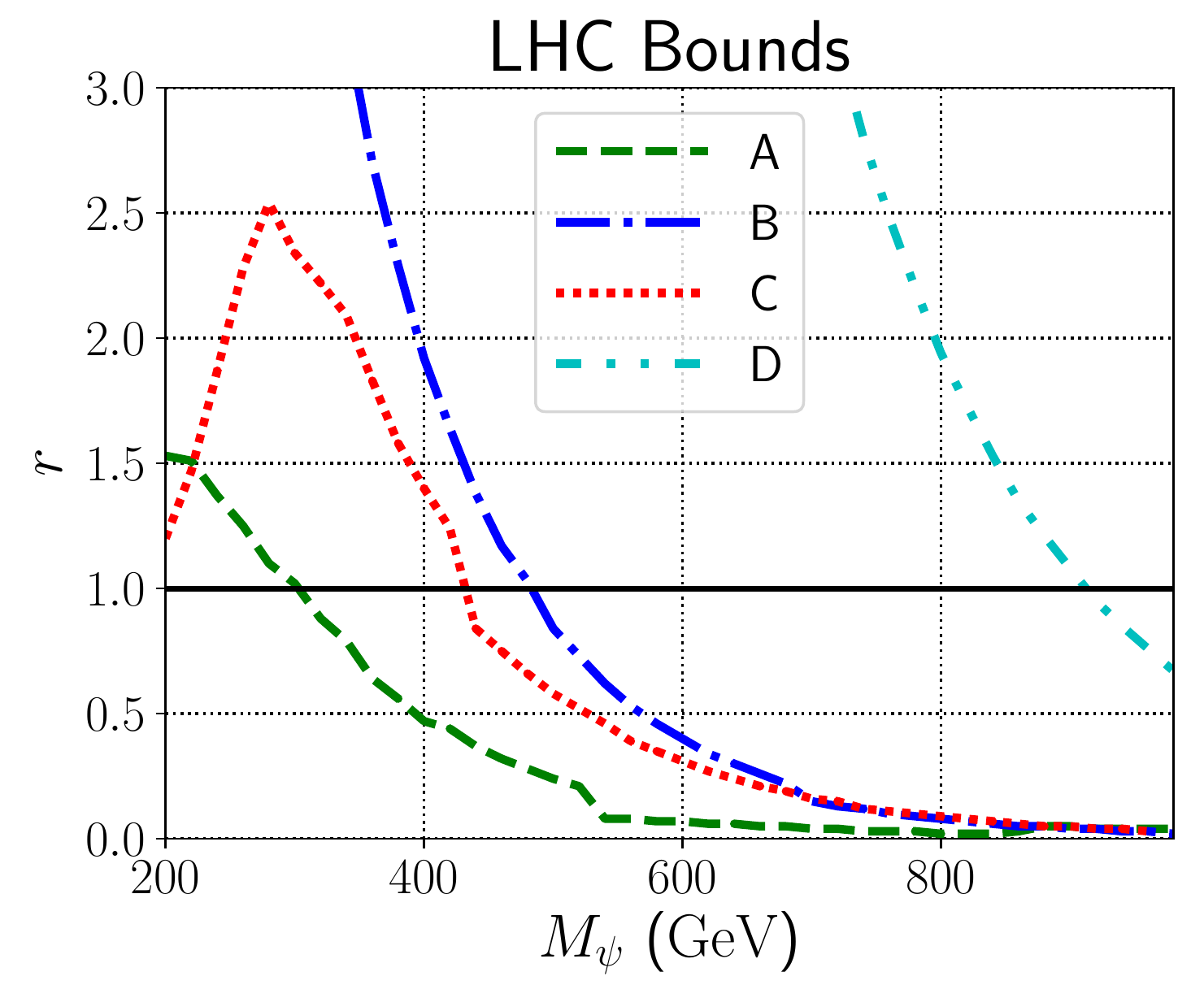}
  \caption{Exclusion limits for benchmark points $A-D$ with varying $M_\psi$
    generated using \texttt{CheckMATE}. The ordinate $r$ is the ratio of
    predicted to observed signal events for the most constraining analysis.
    Points above the $r=1$ line are excluded. For benchmark point $C$,
    the initial increase in $r$ for low $M_\psi$ is due to relaxation of the kinematic suppression
    of $\psi$ decays to the heavier scalar singlet. }
  \label{fig:PhenoResults}
\end{figure}

For the benchmark points listed in Table~\ref{tab:Benchmarks1} we varied
$M_\psi$ from $200$ to $980\GeV$ in steps of $20\GeV$ and generated two million
events of $pp\to \mathcal{EE},\mathcal{EN},\mathcal{NN}$ for each parameter
point. \texttt{CheckMATE} evaluates the expected number of signal events $S$ and
associated error $\Delta S$ per signal region for each of the implemented
analyses and then evaluates the ratio of the $90\%$ confidence lower limit on
the number of expected signal events $S - 1.64 \Delta S$ and the $95\%$
confidence upper limit on experimentally observed number of events $S_{95}$.
This ratio,
\begin{equation}
  r = \frac{S - 1.64 \Delta S}{S_{95}},
\end{equation}
provides a conservative quantification of the constraints on a given model, with
any point having $r>1$ considered excluded~\cite{checkmate}. The resulting
$r$-values are shown in Fig.~\ref{fig:PhenoResults}. As expected, the most
important analyses are generally ATLAS or CMS searches for charginos and
neutralinos with multilepton final states, specifically the searches in
Refs.~\cite{atlas170603731,atlas170807875,CMSSUS16039} which involve
$36\rm{fb}^{-1}$ of data taken at 13~TeV.

The exclusion limits on benchmark $D$, where the singlets have no mixing with
the SM Higgs boson, are noticeably stronger than the rest. In this case the
singlets always decay to $\tau^+\tau^-$, so that each signal event involves
$4-6$ $\tau$ leptons. This is strongly constrained by signal regions of the
above CMS and ATLAS analyses involving hadronic $\tau$ leptons, missing energy,
and either 0 or 1 pairs of opposite sign-same flavour leptons.

On the other hand, when mixing with the SM Higgs is present, the $\psi$ mass can
be brought as low as $350$--$500\GeV$. The lowest mass bound of the considered
benchmark points is given by benchmark point $A$. The relative sizes of the
Yukawa couplings in benchmark $A$ cause the $\psi$ to decay primarily to just
the lightest scalar singlet $s_2$. This decays mostly to $b\bar{b}$, leading to
more jets than leptons and thus weaker constraints on the vector-like lepton
masses.

We have also calculated the production cross-sections for scalar singlets, which
may in principle be constrained by Higgs boson searches and measurements. We
find that Higgs searches have no impact for our benchmark points, due to the
very small mixing between the scalars and the SM Higgs, as given in
Table~\ref{tab:Benchmarks3}. The singlet production cross-sections in standard
Higgs production channels are $\sim 10^{-3}\,\rm{fb}$ for benchmarks $A$ and
$B$, and hence unobservable even at the HL-LHC. The cross section for benchmark
$C$ is larger at $\sim 0.1 \rm{fb}$, but this leads to no constraints from
currently available searches. In principle the scalar mixing could be larger,
which could lead to significant constraints that require a more in-depth
analysis. See, {\it e.g.}, Refs.~\cite{No:2013wsa,xSM} and references therein.
However, a larger mixing does not reduce the lower bound on the $\psi$ masses
and is thus uninteresting in the context of the viability for this model to
generate a baryon asymmetry. We have used small mixing terms for simplicity.

Singlet pair production is also possible. In the SM, Higgs pair production is
due to a triangle and box diagram, which interfere destructively. Singlet pair
production via an intermediate off-shell SM Higgs (via the triangle diagram) is
not suppressed by the singlet-Higgs mixing angle. However, the box diagram is
suppressed by two powers of the mixing angle and is thus negligible in our
model. We therefore expect that this leads to an $\mathcal{O}(1)$ correction to
the dihiggs production cross-section at the LHC. Given that this process is not
expected to be observed until the end of the high-luminosity LHC run, this is
unconstrained by current LHC searches~\cite{ATLASHiggsCubic}. We refer the
reader to
Refs.~\cite{xSM,No:2013wsa,Kotwal:2016tex,SingletPairAtHLLHC,SingletPairAtHELC}
for discussions on singlet pair production at current and future colliders.

Finally, the vector-like leptons also lead to a correction to the Higgs diphoton
decay rate. However, as we consider the scenario where the $\psi$ Higgs Yukawa
is of the order of the $\tau$ Yukawa, the correction is negligible compared to
the SM contributions and current experimental accuracy.

We conclude  that scalar mixing is critical for a low mass
bound and that $M_\psi \gtrsim 500\GeV$ is likely a strong enough requirement to
satisfy current collider bounds, though the specific lower bound will depend on
$\frac{\lambda_{L \psi 1}}{\lambda_{L \psi 2}}$ and may in some cases be larger.

\subsection{Electron Electric Dipole Moment}

The  electron EDM $d_e$ provides a strong constraint on new CP violation. For large phases
this will lead to constraints of the sizes of the Yukawa couplings, scalar
mixing, and $\psi$-$L$ mixing. Current experimental measurements place the upper
bound on the electron EDM~\cite{ACMEEDM} at,
\begin{equation}
  \label{eq:ACME}
  \lvert d_e \rvert < d_e^{\mathrm{expt}} = 1.1 \cdot 10^{-29} \
  \mathrm{e}\cdot\mathrm{cm} \ (90\%\ \mathrm{CL}).
\end{equation}
We consider contributions to the
electron EDM via the two two-loop Barr-Zee style diagrams shown in Fig.~\ref{fig:EDM}. The relevant EDM formulae used to obtain numerical results
are provided in Appendix~\ref{app:EDM}.

\begin{figure}[t]
    \centering
    \begin{subfigure}[b]{0.3\textwidth} 
      \includegraphics{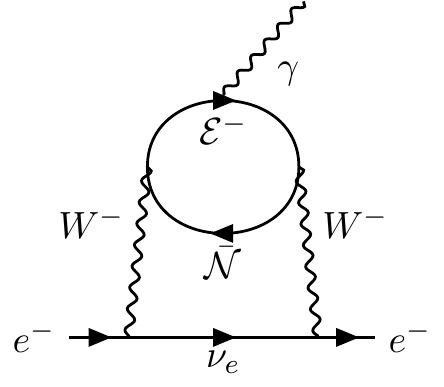}
    \caption{$d_{e}^{W^- W^- }$}
    \label{fig:EDM1}
    \end{subfigure}
    \begin{subfigure}[b]{0.3\textwidth} 
      \includegraphics{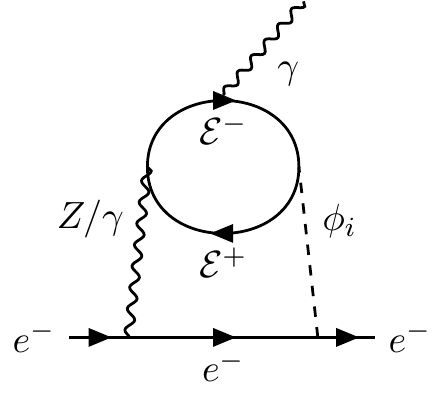}
    \caption{$d_{e}^{Z \phi_i}$, $d_{e}^{\gamma \phi_i}$}
    \label{fig:EDM2}
    \end{subfigure}
  \caption{Electron EDM contribution arising from new interactions. CPV comes from either
    phases in the weak currents or phases in the new scalar-fermion interactions
    which arise through scalar mixing. Mirrored versions of the second diagram ({\it i.e.}, $Z
    \leftrightarrow \phi_i$, $\mathcal{E}^- \leftrightarrow \mathcal{E}^+$) also
    contribute.}
    \label{fig:EDM}
\end{figure}

The diagram in Fig.~\ref{fig:EDM1} could produce EDM contributions due to new
CPV interactions in the charged weak currents, as is the case in vector-like
quark models~\cite{VLQEDM}. This diagram relies on a relative phase between the
left- and right-handed charged weak current interactions with $d_{e}^{W^- W^- }
\propto \mathrm{Im}\left[M^L \left(M^R\right)^\dagger\right]$, where $M^{L,R}$
are the left- or right-handed weak current mixing matrices that are given in
Appendix~\ref{app:mixing}. However, in our scenario, these matrices are real,
such that there is no EDM contribution from this diagram. This can be attributed
to the lack of a right handed $\nu_{\tau}$ for the $N_R$ to mix with. We do,
however, obtain nonzero EDM contributions driven by scalar mixing, as in diagram
in Fig.~\ref{fig:EDM2}. These contributions are similar to those that appear in
supersymmetric models with chargino and neutralino loops and thus we make use of
the formulae provided in Ref.~\cite{mjrmEDMmssm} to evaluate these
contributions.

Substituting the scalar and $\psi$ masses into the EDM loop functions yields expressions for the EDM
contributions as a function of phases, Yukawa interactions, and mixing angles,
though the latter are not independent of the masses. Doing so for benchmark $A$
or $B$ and retaining only terms first order in mixing angles we obtain
\begin{subequations}
\begin{align} 
  \label{eq:analyticEDM}
\sum_i d_{e}^{\gamma \phi_i} &\approx
                             -\left( 56.2 \, \lvert  \lambda_{\psi \psi 1 }\rvert \,\mathcal{P}_{1 2} \,\sin \delta_2
                           + 21.2 \,\lvert  \lambda_{\psi \psi 2}\rvert \,\mathcal{P}_{1 3} \,\sin \delta_3 \right) \cdot 10^{-29} \
  \mathrm{e}\cdot\mathrm{cm} ,
                           \\
 \sum_i d_{e}^{Z \phi_i} &\approx
                            \left(7.90  \,\lvert  \lambda_{\psi \psi 1 }\rvert \,\mathcal{P}_{1 2} \,\sin\delta_2
                           + 7.60  \,\lvert  \lambda_{\psi \psi 2}\rvert \,\mathcal{P}_{1 3} \,\sin\delta_3\right) \cdot 10^{-29} \
  \mathrm{e}\cdot\mathrm{cm},
\end{align}
\end{subequations}
where $\mathcal{P}_{ij}$ are the components of the scalar mixing matrix defined
in Section~\ref{subsec:MassMix}, with numerical values given in
Table~\ref{tab:Benchmarks3}.

The contribution of the CPV phases $\delta_1$ and $\delta_4$ to the EDM are
suppressed relative to the $\delta_{2}$ and $\delta_{3}$ contributions by an
additional factor of the $L$--$\psi$ mixing angles $\theta_{L/R}$. This is
because the diagram shown in Fig.~\ref{fig:EDM2} that involves the Yukawa
couplings $\lambda_{L \psi i}$ and $Y_{\psi}$ associated with these phases
requires non-zero $L\leftrightarrow\psi$ flavor changing neutral currents. These
are proportional to the mixing angles $\theta_{L/R}$. This is not the case for
the $\lambda_{\psi \psi i}$ couplings associated with the $\delta_2$ and
$\delta_3$ phases.

  Full numerical electron EDM results for benchmarks $A$--$C$ are given in
  Table~\ref{tab:EDMResults}. The predicted values for benchmark $A$ and $B$ are
  two orders of magnitude less than the current bound, whereas benchmark $C$ is
  on the edge of being excluded at $90\%$ confidence level. The EDM is much
  larger in benchmark $C$ as it has significantly larger $b_i$ couplings which
  lead to scalar mixing. No results are given for benchmark $D$ as it does not
  feature scalar mixing, such that there is no two-loop electron EDM
  contribution and a higher order calculation is required. An increase in the
  sensitivity of EDM experiments by two orders of magnitude would place severe
  constraints on the scalar mixing matrix and the $\lambda_{\psi \psi i}$ couplings. 

\begin{table}[t]
  \centering
  \begin{tabular}{|c||c|c|c|} 
\hline
    Benchmark  &   $A$, $B$  & $C$   \\
    \hline
    $d_e$ ($\mathrm{e}\cdot\mathrm{cm}$)
               &  $3.02\cdot 10^{-31}$    & $1.07\cdot 10^{-29}$ \\\hline
    $\frac{d_e}{d_e^{\mathrm{expt}}}$
               &  $ 0.0275$   & $0.978$ \\\hline 
  \end{tabular}
  \caption{Numerical electron EDM results for benchmark points, including the
    ratio with respect to the current experimental bound given in
    eq.~\eqref{eq:ACME}. The results of benchmark points $A$ and $B$ are
    approximately equal as they only differ in the $\lambda_{1 L \psi}$ Yukawa
    coupling which plays a subdominant role here.}
  \label{tab:EDMResults}
\end{table}


\subsection{Lepton Phenomenology}

We will briefly discuss corrections to lepton universality and $Z$/$W$
couplings. Because the new particles couple only with third generation leptons,
we do not obtain a contribution to lepton flavour violating decays. However, the
new particles modify the couplings of the weak gauge bosons to $\tau$ and
$\nu_\tau$. The weak currents and Yukawa interactions are provided in
Appendix~\ref{app:mixing}. Due to the mixing between $\tau$ and $E$ the $Z
\tau_R \tau_R$ and $W^{\pm} \tau_L \nu_\tau$ interactions are modified,
\begin{align}
  j^\mu_{Z} \supset  \frac{s_w^2}{c_w}   \bar{\tau} \gamma^\mu \mathbb{P}_R \tau \ &\rightarrow \ \frac{1}{c_w} \left( s_w^2 + \frac{1}{2} \sin^2 \theta_R \right) \bar{\tau} \gamma^\mu \mathbb{P}_R \tau ,                                                                                         \label{eq:zmix}
  \\
  j^\mu_{W^-} \supset  \frac{1}{\sqrt{2}} \bar{\tau} \gamma^\mu \mathbb{P}_L \nu_{\tau} \ &\rightarrow \ \frac{1}{\sqrt{2}} \cos \theta_L  \bar{\tau} \gamma^\mu \mathbb{P}_L \nu_{\tau}  ,
\end{align}
where $c_w$ and $s_w$ are the sine and cosine of the weak mixing angle, and
$\mathbb{P}_{L,R}$ are the standard left- and right-projection operators. There
is no correction to the $Z \tau_L \tau_L$ vertex. The corrections to the
weak-current couplings lead to small deviations from lepton universality. As an
example, the $\tau$ decays will be modified due to the non-unitarity of the
standard $3 \times 3$ Pontecorvo-Maki-Nakagawa-Sakata (PMNS) matrix
$U_{\mathrm{PMNS}}$. The $\tau\rightarrow \nu \nu \ell$ decay rate $\Gamma_\tau$
is proportional to
\begin{equation}
 \Gamma_\tau  \propto \sum_{i=1}^3 \lvert (U_{\mathrm{PMNS}})_{3 i} \rvert^2  \approx 1 - \theta_{L}^2 \ ,
\end{equation}
where the unitarity of the PMNS matrix is recovered in the $\theta_L\to 0$ limit. 
Note that these universality-violating effects are second order in the mixing angles $\theta_L$ and
$\theta_R$. Using the mixing angles in the benchmark points listed in
Table~\ref{tab:Benchmarks3}, the corresponding universality-breaking corrections will be of
$\mathcal{O}\left( 10^{-7} \right)$.

\begin{figure}[t]
    \centering
      \includegraphics{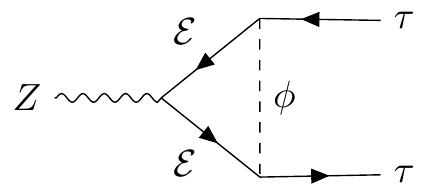}
  \caption{Vertex corrections to $Z$ couplings with $\tau$ and $\nu_\tau$ as a
    result of the new particles. Cancellation of divergences necessitates the inclusion of diagrams with Goldstone bosons.}
    \label{fig:LeptonLoop}
\end{figure}

There will also be a contribution to the 1-loop $Z\tau\tau$ coupling from vertex
corrections of the types shown in Fig.~\ref{fig:LeptonLoop}. While superficially
divergent, the divergences cancel after including diagrams with Goldstone bosons
and accounting for the $\tau$ wave-function renormalisation. Evaluating the
relevant loop diagrams using {\it Package-{\bf X}}~\cite{packageX} and inserting
the benchmark couplings and mixing angles, we find that this vertex correction
is $\sim 0.01 \frac{\lvert Y_\psi \rvert^2}{16 \pi^2} \sim
\mathcal{O}\left(10^{-9}\right)$. This correction is many orders of magnitude
smaller than those found in vector-like quark models~\cite{VLQZVert}, as the
latter feature factors of ${m_t}/{M_\psi}$ rather than ${m_\tau}/{M_\psi}$, and
generally consider much larger Yukawa couplings than those considered here.
Corrections to the weak gauge couplings of this order of magnitude are easily
small enough to avoid existing constraints on lepton universality and
$Z$-coupling measurements~\cite{PDG2016,LeptonPheno1}.

\section{Electroweak Phase Transition and Baryogenesis}
\label{sec:ewbg}

This section is separated into three parts: an overview of EWBG, a
discussion about the EWPT and bubble nucleation, and a discussion of the
transport dynamics around the moving bubble wall. Our methodology for EWBG
closely follows that of Refs.~\cite{ResRelax,TwoStep}. 

\subsection{Overview of EWBG}

Electroweak sphalerons are anomalous processes that violate $B+L$ while
preserving $B-L$. Prior to electroweak symmetry breaking, the rate for these
processes is relatively rapid, being proportional to the temperature $T$. The
presence of a non-zero chemical potential, $\mu_L$, for any fermions carrying
SU(2$)_L$ charge can act to bias the electroweak sphalerons to produce a net
$B+L$ charge density. If the EWSB transition is sufficiently out-of-equilibrium,
the net $B+L$ charge will be preserved in regions of broken electroweak symmetry
before minimisation of Gibbs free energy drives $B+L\rightarrow 0$.

A first order electroweak phase transition can provide the necessary conditions
for this $B+L$ production and preservation if the effects of CP-violation
generate a sufficiently large left-handed fermion chemical potential and if the
degree of sphaleron \lq\lq quenching\rq\rq\, is sufficiently strong in regions
of broken electroweak symmetry. The transition proceeds via the nucleation and
expansion of the bubbles of broken electroweak symmetry. In the present
scenario, the VEVs of the SM Higgs doublet and the scalar singlets evolve,
varying across the bubble walls. The corresponding CP-violating Yukawa
interactions generate a non-zero $\mu_L$, as described in detail below. This
left-handed charge diffuses ahead of the expanding bubbles, thereby biasing the
electroweak sphalerons into net $B+L$ generation. The expanding bubbles capture
the resulting asymmetry, preserving it if the sphaleron transitions are
sufficiently quenched inside the bubbles.

The degree of preservation is governed by the broken phase 
EW sphaleron rate $\Gamma_{\mathrm{WS}}$
that is exponentially suppressed as
\begin{equation}
 \Gamma_{\mathrm{WS}}(T) = A(T) e^{-  E_\mathrm{WS}/T}
\end{equation}
where $E_\mathrm{WS}$ is the weak sphaleron energy and where the prefactor $A$
is a function of $T$. The value of $E_\mathrm{WS}$ can be related to a
$T$-dependent energy scale ${\bar v}(T)$ associated with electroweak symmetry
breaking. For a discussion of the relationship between ${\bar v}(T)$ and the
scalar field VEV and the associated issue of gauge invariance, see
Ref.~\cite{MJRMgaugeDep}. It is conventional to write
\begin{equation}
E_\mathrm{WS} =   \frac{4\pi B}{g}\, {\bar v(T)}\ \ \ ,
\end{equation}
where $g$ is the SU(2$)_L$ gauge coupling and $B$ is a computable function of
$g$ and the other couplings in the gauge-Higgs sector. Note that both $A$ and
$B$ will vary depending on the representation of the EW symmetry breaking
scalar~\cite{Origsphaleron,singletSphaleron,sphaleronEnergyForRepresentations}.
Preservation of the $B+L$ asymmetry requires a sufficiently large
$E_\mathrm{WS}/T$ at the the bubble nucleation temperature $T_N$, which
typically lies just below the critical temperature of the transition $T_C$. A
first-order phase transition that satisfies this requirement is dubbed
``strong'', a characterisation that is usually translated into the requirement
that
\begin{equation}
\frac{{\bar v}(T_C)}{T_C} \gtrsim \mathcal{O}(1) \ \ \ .
\end{equation}
For recent discussions of sphaleron rate computations in perturbation theory and
the associated theoretical uncertainties see, {\it e.g}.,
Refs.~\cite{MJRMgaugeDep,singletSphaleron,sphaleronEnergyForRepresentations,MJRMgaugeDep2}.

The dynamics of the expanding bubbles, together with the CPV transport dynamics,
constitute the crucial elements for generating the left-handed number density
that catalyses $B+L$ generation. In what follows, we concentrate on the
transport dynamics but note here the importance of the bubble expansion rate,
characterised by the wall velocity $w$. On the one hand, one must have $w > 0$
in order to generate any asymmetry (see below). On the other hand, the expansion
must be sufficiently slow to allow the left-handed number density to diffuse
ahead of the advancing wall and ``seed'' the EW sphalerons before they are
quenched in the bubble interior. For a detailed discussion of the bubble
dynamics and wall velocity, we refer the reader to
Refs.~\cite{wallVelocityKonstandin,wallVelocity,newKonstandinWall}.

For purposes of analysing the CPV transport dynamics, it is conventional to
treat the bubble wall as a flat plane moving with a constant $w$. Under this
assumption, the primary inputs needed are the specific CPV interactions, $T$,
$w$, and the bubble wall profile as a function of the direction normal to the
wall. As a result, the procedure of evaluating the asymmetry generated by a
given model during the EWPT can generally be split into two steps: verifying the
presence of a strongly first-order electroweak phase transition and obtaining
the moving bubble wall information, and then solving a system of
out-of-equilibrium transport equations to compute the final baryon asymmetry.
For a more thorough review of EWBG
see~\cite{TroddenReview,MorrisseyMJRMReview,GrahamReview}.

\subsection{Electroweak Phase Transition}
\label{sec:ewpt}

In this section we outline our treatment of the electroweak and the selection of
the benchmark points listed in Section~\ref{sec:pheno}. Our goal is not to
perform a comprehensive study of the model parameter space that yields a
strongly first-order EWPT but rather to identify illustrative parameter choices
for use in the treatment of the transport dynamics.

To that end, we will employ the high-temperature effective potential which is
acquired by adding thermal mass terms to the tree level potential,
\begin{equation}
    V^{\mathrm{high-}T} = V_0 + \delta m^2_{H}  H^\dagger H T^2 + \frac{1}{2} \delta m^2_{S_1} S_1^2 T^2  + \frac{1}{2} \delta m^2_{S_2} S_2^2 T^2,
\end{equation}
where $V_0$ is the tree level potential given in eq.~\eqref{eq:potential}, and
the thermal masses $\delta m^2_{X}$ are computable in terms of the benchmark
model parameters and are listed in Table~\ref{tab:ThermalMasses} within
Appendix~\ref{app:FTFT}. This treatment has the advantage of being directly
gauge-independent, numerically fast to evaluate, and still yields results
comparable to a more thorough gauge-independent treatment, as seen in
Ref.~\cite{MJRMgaugeDep2}. Note that, in this context, the scalar field VEVs are
manifestly gauge invariant, and we will identify these VEVs with the
corresponding scale ${\bar v}$ introduced above.

In order to find benchmark points that yield the desired phase transition, we
performed a random scan over the scalar potential parameter space and utilised a
modified version of the \texttt{CosmoTransitions}
package~\cite{CosmoTransitions} to determine $T_C$, the nucleation bubble wall
profiles, and the bubble nucleation probability density per unit
time~\cite{LindeFTFTVacuumDecay},
\begin{equation}
  \Gamma_{\mathrm{nuc}} = T^4 \left( \frac{S}{2 \pi T} \right)^{3/2} e^{- S/T} \, ,
\end{equation}
where $S$ is the three-dimensional Euclidean bubble action.
The expected number of bubbles in a Hubble volume is then given by~\cite{grahamGravWaves},
\begin{equation}
  \left\langle N(T)  \right\rangle = \int_{T}^{T_c} \frac{\dd T}{T} \frac{\Gamma_{\mathrm{nuc}}}{H(T)^4}.
\end{equation}
The corresponding nucleation temperature is defined by the criterion
\begin{align}
  & \left\langle N(T_N)  \right\rangle \gtrsim 1
  \\ \implies \ &
  \Gamma_{\mathrm{nuc}}(T_N) \gtrsim H(T_N)^4 = \left( \frac{4 \pi^3 g_\star T^4}{45 m^2_{\mathrm{pl}}} \right)^2 \\
  \implies \ &
      \frac{S}{T_N} - \frac{3}{2} \mathrm{ln}  \frac{S}{T_N} \lesssim  173.8 - 2 \mathrm{ln} g_\star - 4 \mathrm{ln} \frac{T_N}{\mathrm{GeV}}, 
\end{align}
where $g_\star$ is the effective number of relativistic degrees of freedom and
$m_{\mathrm{pl}}$ is the Planck mass. This $T_N$ is then used throughout the
remainder of the EWBG calculation.

\begin{figure}[t]
    \centering
    \begin{subfigure}[b]{0.49\textwidth} 
  \includegraphics[width=1\textwidth]{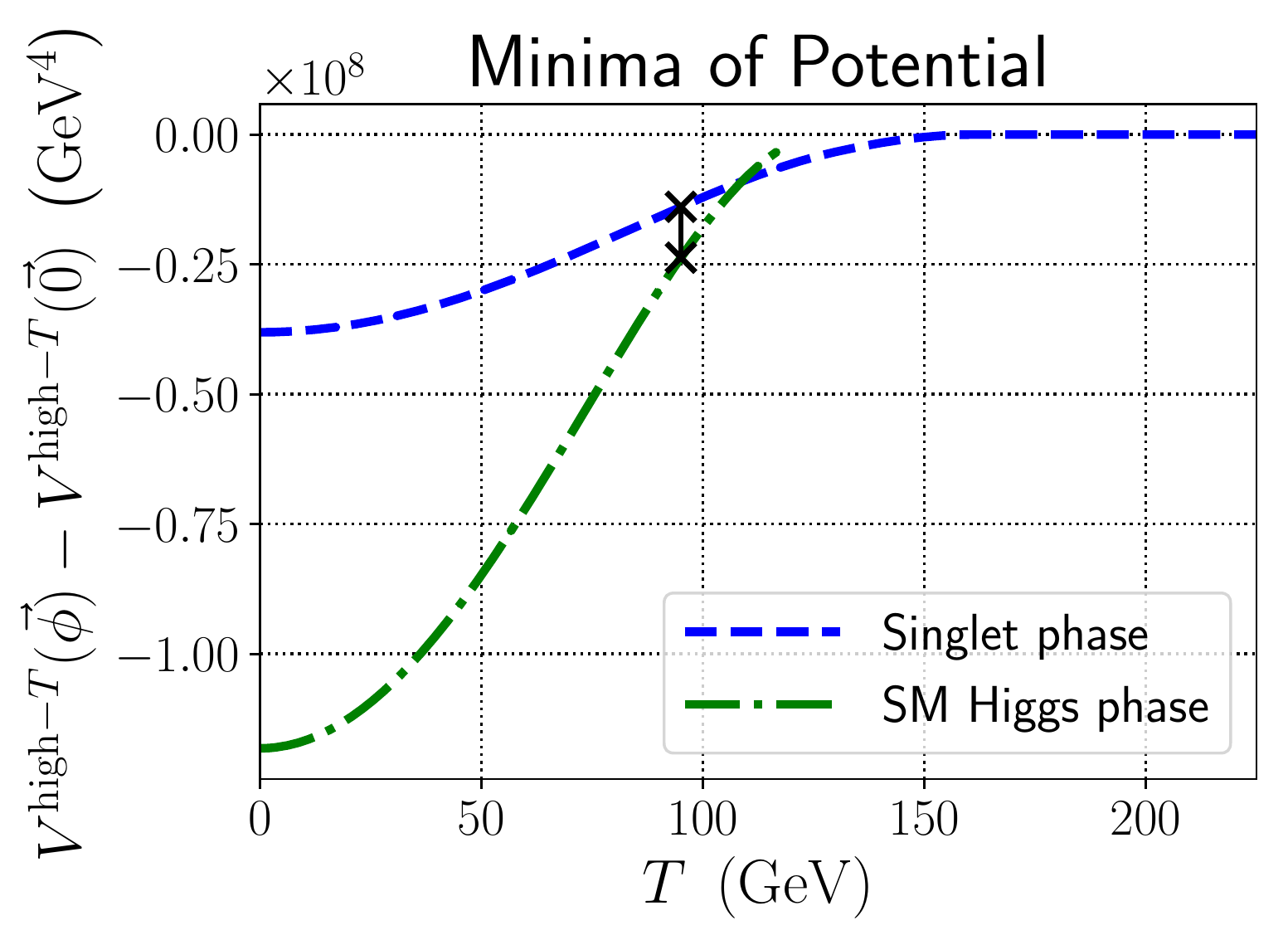}
    \caption{}
    \label{fig:EWPTResultsMinima}
    \end{subfigure}
    \begin{subfigure}[b]{0.49\textwidth} 
  \includegraphics[width=1\textwidth]{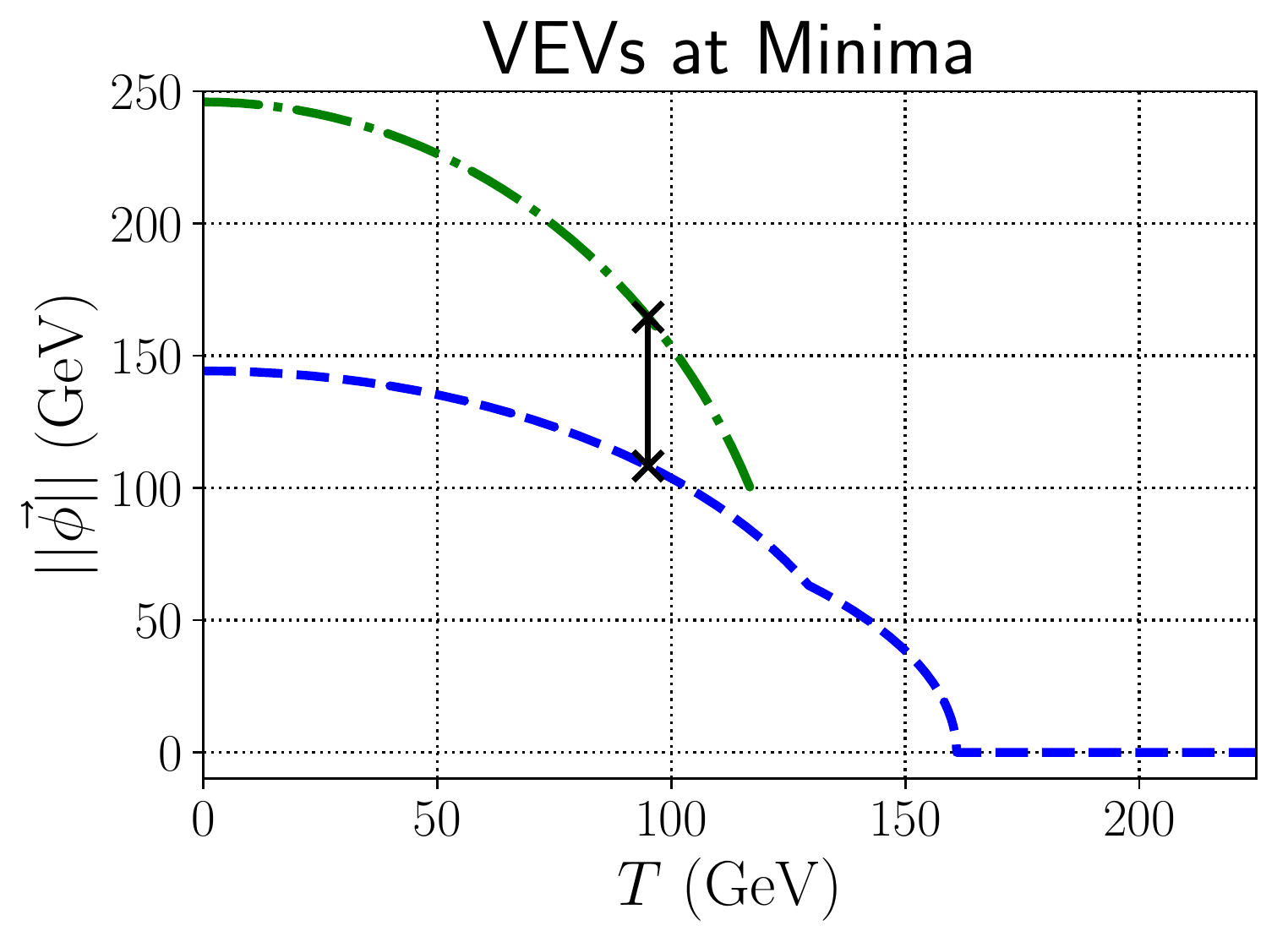}
    \caption{}
    \label{fig:EWPTResultsVEVS}
    \end{subfigure}
    \caption{Phase transition results for benchmark point $A$ showing, as a
      function of temperature: (\subref{fig:EWPTResultsMinima}) relative depth
      of minima of potential, and (\subref{fig:EWPTResultsVEVS}) distance of
      minima from the origin. The crosses and black lines indicate the nucleation temperature.}
  \label{fig:EWPTResults}
\end{figure}

The results of using \texttt{CosmoTransitions} to trace the minima of the
effective potential for benchmark point $A$ are shown in
Fig.~\ref{fig:EWPTResults}. As the temperature decreases benchmark point $A$
exhibits two second-order phase transitions where the scalar singlets gain their
VEVs. Eventually there is a strongly first-order electroweak where the Higgs
gains a VEV and the scalar singlets lose theirs. Such a phase transition will
result in a potentially observable gravitational wave
signal~\cite{GW1,GW2,GW3,GW4,GW5}. It should be noted that if the preceding
phase transitions where the scalar singlets gained their VEVs were also first
order, these transitions would also produce gravitational waves. When combined
with the gravitational waves from the Higgs phase transition this would lead to
a multi-peaked gravitational wave power spectrum that is characteristic of
multi-step phase transitions~\cite{grahamGravWaves,multistepGravWave}.

Determining the wall velocity and moving bubble wall profiles is a complicated task requiring
solving a set of integro-differential equations~\cite{wallVelocityKonstandin,wallVelocity}.
However both nucleation and moving wall profiles are generally well approximated
by hyperbolic tangent  profiles with a characteristic wall width $L$,
\begin{equation}
  v_i(z) = \frac{\Delta v_i}{2}\left( 1 \pm \tanh\left( \frac{z\ - \ {\delta z}_{i}}{L_{i}} \right) \right), \qquad i \in \{H,S_1,S_2\}, \label{eq:vevProfile}
\end{equation}
where $z$ is the distance from the planar bubble wall and $z>0$ corresponds to
the inside of the bubble where the SM Higgs has a non-zero VEV.

The finite width of the wall can be neglected in the limit where the width
is small relative to the distances that particles diffuse ahead of the wall. In
the thin wall limit the reaction rates and CPV source terms can be well approximated by
step and delta functions, respectively. The typical reaction-diffusion distances
for particle species are given by $\sqrt{D/\Gamma}$, where $D$ is a
diffusion constant arising due to scattering interactions with the plasma, and
$\Gamma$ is a dominant relaxation rate that drives a species back towards
equilibrium. In our model the particles responsible for biasing sphalerons to
generate the asymmetry are leptons, which diffuse much further ahead of the
wall than heavy quarks do. This is in part due to larger
diffusion constants as a result of leptons not being charged under SU(3), and in part
due to smaller relaxation rates as a result of smaller Yukawa couplings. Hence,
we find $L \ll \sqrt{D/\Gamma}$, and will neglect the finite width of the
wall.

\begin{figure}[t]
    \centering
  \includegraphics[width=0.49\textwidth]{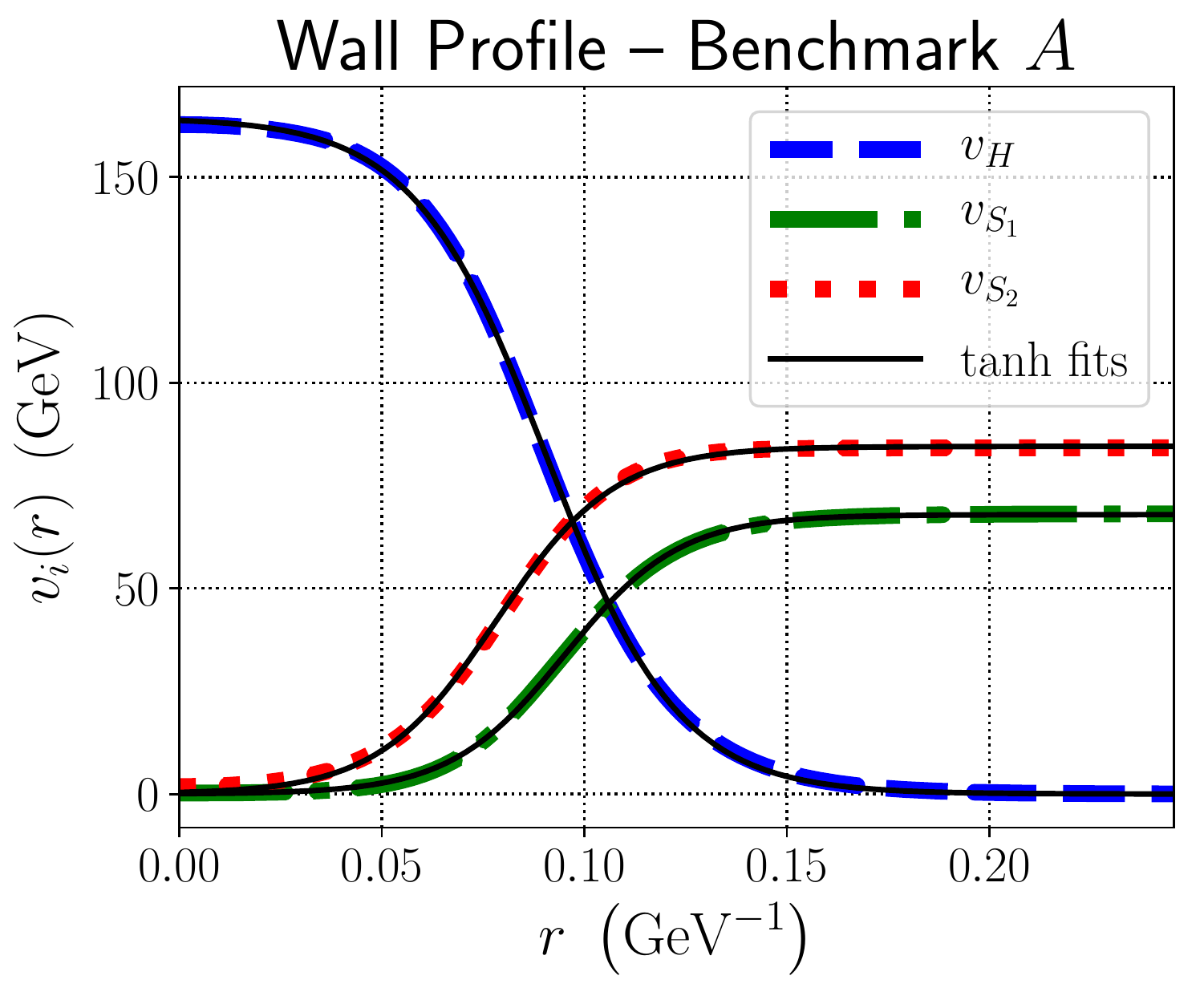}
  \includegraphics[width=0.49\textwidth]{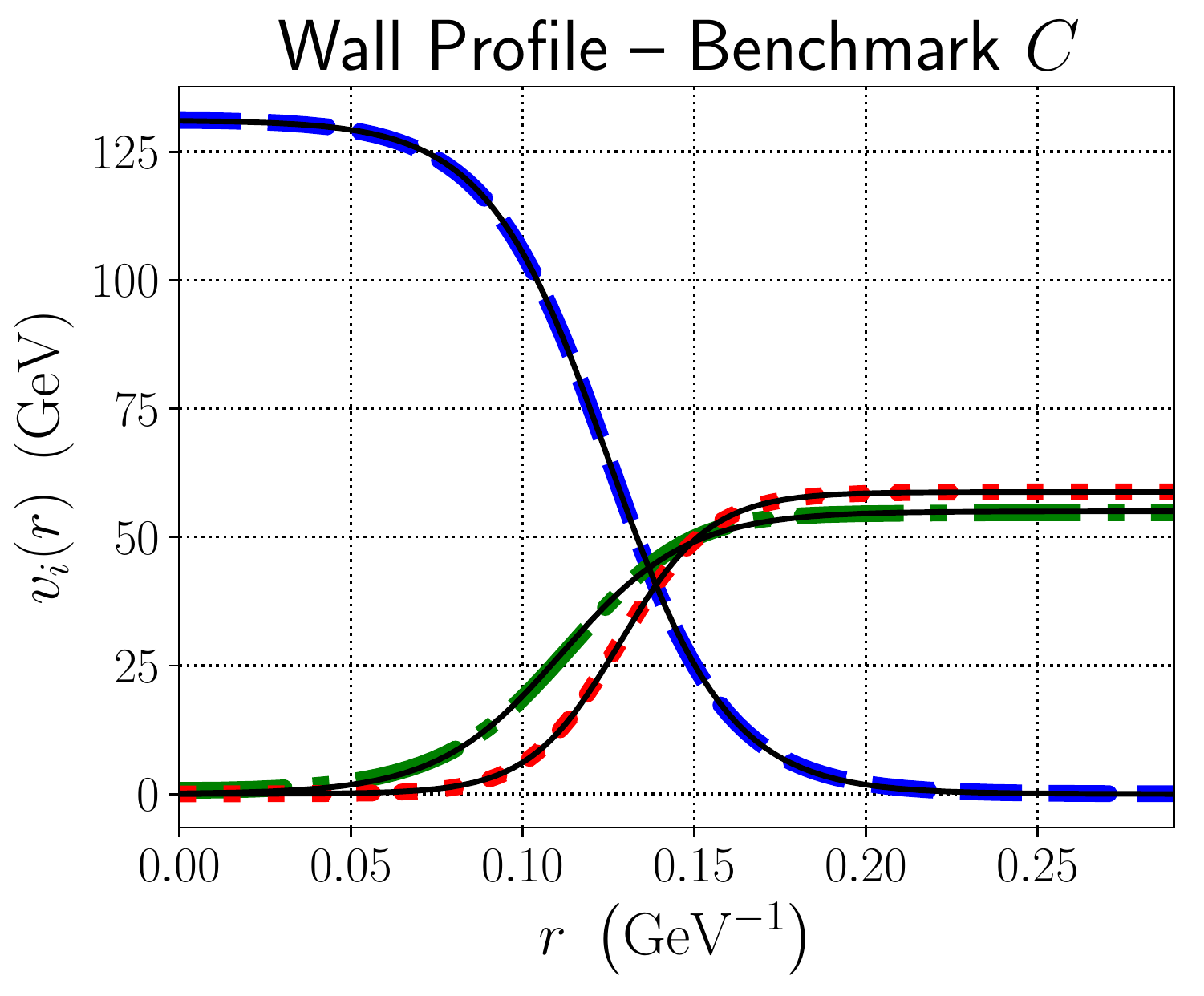}
  \caption{The radial nucleation bubble wall VEV profiles of the electroweak phase transition  for
    benchmark points $A$ and $C$. $r=0$ is the centre of the bubble. The thin
    black lines are fitted hyperbolic tangent functions, eq.~\eqref{eq:vevProfile}. The profiles for
    benchmarks $B$ and $D$ are approximately equivalent to those of $A$ and $C$,
    respectively.}
  \label{fig:EWPTResultsVEVProfile}
\end{figure}

However, as will be shown in Section~\ref{sec:QTE}, the magnitude of the CPV
source term is sensitive to the relative magnitudes of the VEVs within the
bubble wall, and thus an accurate expression for the moving wall profiles is
still necessary to evaluate the source term. For this purpose we utilise the
nucleation bubble wall profiles as obtained by \texttt{CosmoTransitions}. The
profiles for benchmarks $A$ and $C$ are shown in
Fig.~\ref{fig:EWPTResultsVEVProfile} along with fitted $\tanh$-functions.
Parameters for the fitted profile functions, as defined in
eq.~\eqref{eq:vevProfile}, are given in Table~\ref{tab:EWPTBenchmark} along with
the nucleation temperatures. The differences in the phase transitions between
benchmarks $A$ and $B$ are negligible, and similarly for $C$ and $D$. Studies of moving wall dynamics typically find wall velocities $w
\sim$ $0.1$--$0.3$~\cite{wallVelocityKonstandin,wallVelocity,newKonstandinWall},
so for our benchmarks we will take $w = 0.1$.

\begin{table}
  \centering
  \begin{tabular}{|c||c|c|c|c|c|c|c|c|c|c|} 
    \hline
    & $T\ (\mathrm{GeV})$ &  $L_{H} T$ & $L_{S_1} T$ & $L_{S_2} T$ & ${\delta z}_{H} T$ & ${\delta z}_{S_1}T $ & ${\delta z}_{S_2} T$ & $\frac{\Delta v_{H}}{T}$ & $\frac{\Delta v_{S_1}}{T}$ & $\frac{\Delta v_{S_2}}{T}$ 
    \\\hline \hline
    $A$ & 95 & 3.11 & 2.68 & 2.76 & 0 & 0.441 & -1.17 & 1.73 & 0.715 & 0.889 \\\hline
    $C$ & 105 & 3.68 & 3.8 & 2.75 & 0 & -1.38 & 0.361 & 1.25 & 0.525 & 0.561 \\\hline
  \end{tabular}
  \caption{EWPT Temperature and rescaled VEV-profile parameters for
    eq.~\eqref{eq:vevProfile}. The parameters for benchmarks $B$ and $D$ are
    approximately equivalent to those of $A$ and $C$, respectively.}
  \label{tab:EWPTBenchmark}
\end{table}

Note that in general the nucleation VEV profiles differ from the moving wall
profiles. In particular, the wall widths of moving walls tend to be smaller than
the nucleation widths by a factor of $2$--$4$~\cite{wallVelocityKonstandin}.
However as mentioned earlier, we ignore the finite width of the wall and will
ignore this scaling. It should be noted that other studies of EWBG have found
that the asymmetry scales as $1/L$. This is a result of the formalism
used to derive the CPV source terms. As will be discussed in the next section,
our CPV source term is proportional to the first derivatives of the VEVs. Studies using
the semi-classical force approximations instead find source terms proportional to the
second derivative~\cite{semiclassicalForce, semiclassicalForce2}. This would
introduce an additional factor of $1/L$ and explains the discrepancy.

The presence of non-zero singlet VEVs in the unbroken phase will introduce a
mass term that leads to mixing between the $\psi$ and SM leptons.
The singlet VEVs modify the Dirac mass terms,
\begin{equation}
  M_{\psi} \rightarrow M_{\psi} + \lambda_{\psi \psi 1} v_{S_1} +
  \lambda_{\psi \psi 2} v_{S_2}, 
\qquad
  M_{L \psi} \rightarrow M_{L \psi} + \lambda_{L \psi 1} v_{S_1} +
  \lambda_{L \psi 2} v_{S_2} , 
\end{equation}
where $M_{\psi L}$ is the mass term that we removed via a field redefinition in
Section~\ref{sec:model}. This new mixing term can once again be removed in the
singlet phase via the same transformation, eq.~\eqref{eq:MLRemove}, with
$M_{\psi}$ and $M_{\psi L}$ appropriately modified. We perform such a rotation
and work in the mass basis within the singlet phase. As a result, the $\psi$ and $L_3$
Yukawa couplings that will appear in the following sections will be primed to
denote the fact that they correspond to the transformed Yukawas, as given by
eq.~\eqref{eq:yukawatransform}, not the zero temperature Yukawas. A benefit of
the singlet VEVs is that if the $\lambda_{\psi \psi i}$ couplings have the right
phases, they can drive the $\psi$ to be lighter in the early universe than they
are at zero temperature, which results in an enhanced asymmetry. The benchmark
$\lambda_{\psi \psi i}$ and their phases were selected such that this is the
case.


\subsection{Quantum Transport Equations}
\label{sec:QTE}

The baryon asymmetry is generated by electroweak sphalerons acting on the net
density of fermions charged under SU(2) that are diffusing ahead of the bubble wall. Computing
this number density requires solving a set of transport equations of the form
\begin{equation}
  \partial_{\lambda} J^\lambda_i (x) = - \sum_\Gamma \frac{T^2}{6} \Gamma_{i j \ldots}(x) (\mu_i - \mu_j \pm \ldots ) + S_i^{\mathrm{CPV}}(x) \, , \label{eq:generalRates}  
\end{equation}
where $J^\lambda_i$ and $\mu_i$ are the number density current and chemical
potential of particle species $i$. The $\Gamma_{i j
  \ldots}$ are  equilibration rates that arise due to various interactions
in the plasma, while $S_i^{\mathrm{CPV}}$ is a CPV source term arising from
interactions with the bubble wall. The number density and chemical potential for species $i$ are related as
\begin{equation}
\begin{aligned}
  n_i &= N_i - \bar{N}_i   = g_i \int \frac{\dd^3 k}{(2 \pi)^3} \left[ n_f(\omega_i(k), \mu_i) - n_{f}(\omega_i(k), -\mu_i) \right]
        , \\
      &\approx \frac{k_i(m_i/T) T^2}{6} \mu_i \ + \ \mathcal{O}\left(\mu_i^3/T^3\right) \label{eq:muTransform}
        , \\
  k_i\left( m_i/T\right) &= g_i \frac{6}{\pi^2} \int_{m_i/T}^\infty  \dd x \frac{x e^x}{(e^x + 1)^2} \sqrt{x^2 - \frac{m_i^2}{T^2}},
\end{aligned}
\end{equation}
where $n_f(\omega,\mu)$ is the fermion distribution function and $g_i$ is the
number of degrees of freedom associated with the particle species. Treating the
bubble as a flat plane moving with speed $w$, employing the diffusion
approximation ${\vec J} = - D{\vec\nabla} n$, and using
eq.~\eqref{eq:muTransform}, we can re-write eq.~\eqref{eq:generalRates} as
\begin{equation}
  w \frac{\partial}{\partial z} n_i(z) - D_i  \frac{\partial^2}{\partial^2 z}n_i(z) = - \sum_{\Gamma} \Gamma_{i j \ldots}(z) \left(\frac{n_i}{k_i} - \frac{n_j}{k_j} \pm \ldots \right) + S_i^{\mathrm{CPV}}(z) \label{eq:generalrRates2} \, .
\end{equation}
We use the diffusion constants $D_i$ derived by Ref.~\cite{DiffusionTerms},
which are provided in Appendix~\ref{app:FTFT} along with the thermal masses and
widths. The fermion diffusion constants were derived for massless particles and
arise from their gauge interactions. Even though the $\psi$ are massive, for now
we simply assume $D_\psi \approx D_{L_i}$ and will briefly investigate the
dependence of the final asymmetry on $D_\psi$.

\begin{figure}[t]
    \centering
    \begin{subfigure}[b]{0.3\textwidth} 
    \centering
      \includegraphics{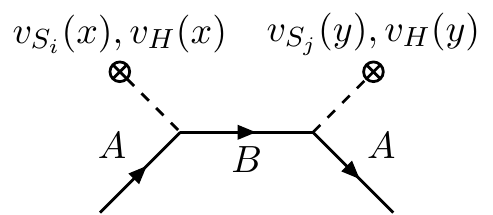}
    \caption{VEV-Insertion}
    \label{fig:VIA1}
    \end{subfigure}
    \qquad
    \begin{subfigure}[b]{0.3\textwidth} 
    \centering
      \includegraphics{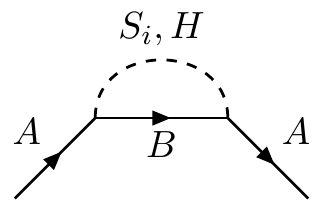}
    \caption{Yukawa-loop}
    \label{fig:VIA2}
    \end{subfigure}
    \caption{Self-energy diagrams that contribute terms to the transport equations, including;
      (\subref{fig:VIA1}) VEV-Insertions with spatially varying VEVs that will
      provide a CPV source term $S^{CPV}_{A B}$ and an equilibration rate $\Gamma^{\pm}_{A B\,S_i,H}$ acting on the fermions, and (\subref{fig:VIA2})
      Yukawa-loops which just provide an equilibration rate $\Gamma_{A B\, S_i,H}$ acting on all of the particles in the diagram.}
  \label{fig:VIA}
\end{figure}

Calculating the rates $\Gamma_{ijk}$ and CPV source terms $S^{\mathrm{CPV}}$
requires the use of out-of-equilibrium finite-temperature field theory. We will
use rates and source terms calculated using the VEV-insertion approximation
(VIA)~\cite{ResEWBG,ResRelax,ResRelaxFollowup}, which itself relies upon the
Schwinger-Keldysh closed time path framework. The VIA formalism utilises
self-energy diagrams, such has the one shown in Fig.~\ref{fig:VIA}, to compute
the reaction rates and source terms for the species in the thermal bath through
the use of VEV-insertions. A more thorough treatment necessitates a
VEV-resummation, resulting in a spatially varying mass matrix and flavour
oscillations.
The VIA is an estimate of the contribution to the source due to flavor mixing that arises at first order in a gradient expansion across the bubble wall. The VIA formalism may overestimate the generated asymmetry, and
the results ought to be considered illustrative of what may be expected from
such a model. For a more thorough review of the theoretical issues see
Ref.~\cite{MorrisseyMJRMReview}. For a discussion of terms arising at second order in gradients, we refer the reader to Ref.~\cite{KPSW}. We also refer the reader to Refs.~\cite{FlavQTE} and \cite{ResFlavQTE} for discussion of these issues in the context of flavor mixing involving scalar fields. 
We do not give detailed derivations of the
rates and source terms in this paper. Instead we provide an overview of rates
involved along with the full system of transport equations, and then give
analytic formulae and numerical results for the rates and source terms in
Appendix~\ref{app:QTE}.

When performing the VIA derivations, the type of diagram shown in
Fig.~\ref{fig:VIA1} gives rise to two equilibration rates and a CPV source term.
These equilibration rates, which we denote as $\Gamma^{ \pm}_{A B\,S_i,H} \,
(\mu_{A} \pm \mu_{B} )$, are proportional to the product of the VEVs of $S_i$ or
$H$, and hence have a spatial dependence across the bubble wall. The CPV source
term arising from the VEV insertion, denoted $S_{A B}^{CPV}$, is non-zero only
within the bubble wall when the VEVs are changing. The absorptive parts of
Yukawa-loops, shown in Fig.~\ref{fig:VIA2}, give rise to equilibration rates,
$\Gamma_{A B\,S_i,H} \, (\mu_{A} - \mu_{B} \pm \mu_{S_i,H} )$. These rates also
acquire some spatial dependence as the masses of the particles in the diagram
will vary across the bubble wall. Additionally, strong and EW sphalerons will
also act on the number densities. The effect of the sphalerons is to introduce
terms proportional to
\begin{equation}
  \frac{T^2}{6} \Gamma_{\mathrm{WS}} \sum_i \left( 3 \mu_{Q_i} + \mu_{L_i} \right)  \qquad \mathrm{and} \qquad
  \frac{T^2}{6}  \Gamma_{\mathrm{SS}}  \sum_i \left( 2 \mu_{Q_i} - \mu_{u_i} - \mu_{d_i} \right), \label{eq:sphalerons1}
\end{equation}
where $\Gamma_{\mathrm{WS}}$ is the EW sphaleron rate, $\Gamma_{ss}$ is the
strong sphaleron rate, $u_i$ and $d_i$ are the three generations of right-handed
SU(2) singlet quarks, $Q_i$ are the three generations of left-handed quark
doublets, and $L_i$ are the three generations of left-handed SM lepton doublets.
We approximate the sphaleron rates as~\cite{GammaWS2,GammaWS1,GammaSS},
\begin{equation}\label{eq:sphalerons2}
  \Gamma_{\mathrm{WS}}  = 6 \kappa_{\mathrm{WS}} \, T \alpha_w^5 , \qquad  \Gamma_{\mathrm{SS}} =  6 \kappa_{SS} \, T \alpha_s^4, 
\end{equation}
with $\kappa_{\mathrm{WS}}\approx 20$, $\kappa_{\mathrm{SS}}\approx 14$, and
$\Gamma_{\mathrm{WS}}$ exponentially suppressed within the bubble such that it
is effectively zero as far as the transport dynamics are concerned. In summary
we have have a combination of VEV-insertion, Yukawa-loop, and sphaleron
equilibration rates driving the number densities back to zero while a CPV term,
which is non-zero only within bubble wall, provides the source for the number
densities that diffuse into and ahead of the bubble. EW sphalerons acting
outside of the bubble may generate a non-zero net baryon density.

We can simplify our system of transport equations by reducing the number of
species to be considered via some equilibrium considerations. Due to rapid weak interactions we take the
components of the SU(2) doublets to be in equilibrium, such that for the third generation
leptons we have
\begin{equation}
  \mu_{\tau_L} = \mu_{\nu_\tau} = \mu_{L_3}, \quad n_{L_3} = n_{\tau_L} + n_{\nu_\tau} , \quad k_{L_3} = k_{\tau_L} + k_{\nu_\tau},
\end{equation}
with similar relations holding for other doublets.

At this order our treatment will only provide a CPV source term for the SM third
generation leptons and $\psi$ via their scalar singlet Yukawas. Non-zero
densities for the other particles are introduced only via EW sphalerons, or via
the $\tau$ or $\psi$ Yukawa couplings to the Higgs. As the $\tau$ Yukawa is
relatively small, and as the benchmarks under consideration have an even smaller
$\psi$ Yukawa, we can simplify our system of equations by neglecting these
Yukawa couplings as their resulting reaction rates are 2--3 orders of magnitudes
smaller than the relaxation rates arising from the $\lambda'_{L \psi}$ couplings.
Then as the strong sphaleron rate equilibrates left and right handed quarks
faster than the weak sphalerons produce them, and as the sphalerons act equally
on all generations, we can use
\begin{subequations}
\begin{align}
\label{eq:QTEsimplification}
\mu_{\ell} = \mu_{L_1}= \mu_{L_2}, && k_\ell = k_{L_1} + k_{L_2}, && n_{\ell} = n_{L_1} + n_{L_2},
    \\
    \mu_{q} = \mu_{u_i} = \mu_{d_i} = \mu_{Q_i}, && k_q = \sum_i k_{Q_i} +  k_{u_i} + k_{d_i}, && n_q = \sum_i n_{Q_i} + n_{u_i} + n_{d_i}.
\end{align}
\end{subequations}
It should be noted that while the top quark Yukawa interactions are fast, they drive $\mu_{Q_3} - \mu_{d_3}  - \mu_H \rightarrow 0$. Hence they will not change these relations if there is no source for $\mu_H$. The first part of eq.~\eqref{eq:sphalerons1} then reduces to
\begin{equation}
    \Gamma_{\mathrm{WS}} \left( 9 \frac{n_q}{k_q}  + 2 \frac{n_\ell}{k_\ell} + \frac{n_{L_3}}{k_{L_3}} \right).
\end{equation}
When $\Gamma_{\mathrm{WS}}$ is much smaller than all other relevant reaction
rates, it is reasonable to decouple the weak sphaleron rate from the system of
transport equations and compute the resulting baryon asymmetry in two steps:
first, compute the left-handed chiral charge; second, use the latter to compute
the $B+L$ asymmetry from the sphaleron rate
equation~\cite{ResRelax,TwoStep,beautygenesis}. In the present case, however, we
will find that the weak sphaleron rate ($\sim 5 \cdot 10^{-4} \GeV$) may be
comparable to the $L_3$ and $\psi$ relaxation rates ($\sim 10^{-3}$--$10^{-2} \GeV$ in
benchmark A). Thus we will instead include the weak sphaleron effects into our
system of transport equations.

Using the previous approximations, we obtain the following set of coupled transport equations:
\begin{subequations}
\label{eq:QTEs}
\begin{align}
  w \frac{\partial}{\partial z}n_\psi(z) - D_\psi \frac{\partial^2}{\partial^2 z}n_\psi 
  =& 
  - \left( \Gamma_{{L_3} \psi S_i }^{-} + \Gamma_{ {L_3} \psi  S_1}  + \Gamma_{ {L_3} \psi  S_2}\right)
    \left( \frac{n_\psi}{k_\psi}  - \frac{n_{L_3}}{k_{L_3}} \right)
  \nonumber\\ & \
- \Gamma_{L_3 \psi S_i }^{+} \left( \frac{n_\psi}{k_\psi} + \frac{n_{L_3}}{k_{L_3}} \right)
       + S_{L_3 \psi}^{CPV},\\
  w \frac{\partial}{\partial z}n_{L_3} - D_{L_3} \frac{\partial^2}{\partial^2 z}n_{L_3}  =&
   \left( \Gamma_{{L_3} \psi S_i }^{-} + \Gamma_{{L_3} \psi  S_1}  + \Gamma_{{L_3} \psi  S_2}\right)
    \left( \frac{n_\psi}{k_\psi}  - \frac{n_{L_3}}{k_{L_3}} \right)
  \nonumber\\ & \
+ \Gamma_{L_3 \psi S_i }^{+} \left( \frac{n_\psi}{k_\psi} + \frac{n_{L_3}}{k_{L_3}} \right)
   - \Gamma_{\mathrm{WS}} \left( 9 \frac{n_q}{k_q} + 2 \frac{n_\ell}{k_\ell} + \frac{n_{L_3}}{k_{L_3}} \right)
       - S^{CPV}_{L_3 \psi} ,\\         
  w \frac{\partial}{\partial z}n_{\ell} - D_{\ell} \frac{\partial^2}{\partial^2 z}n_{\ell}  =&  
   - 2 \Gamma_{\mathrm{WS}} \left( 9 \frac{n_q}{k_q} + 2 \frac{n_\ell}{k_\ell} + \frac{n_{L_3}}{k_{L_3}} \right)
     ,\\
     w \frac{\partial}{\partial z} n_{q} - D_{q} \frac{\partial^2}{\partial^2 z}n_{q}  =&  
   - 9 \Gamma_{\mathrm{WS}} \left( 9 \frac{n_q}{k_q} + 2 \frac{n_\ell}{k_\ell} + \frac{n_{L_3}}{k_{L_3}} \right)
.\end{align}
\end{subequations}
The relevant equilibration rates and CPV source terms are given by,
\begin{subequations}
\label{eq:rates}
\begin{align}
  \label{eq:sourceFunc} S_{L_3 \psi}^{CPV}(z) &
 = - 2 \mathrm{Im}\left[\lambda'_{L \psi 1} {\lambda'}^*_{L \psi 2}\right]  \left( v_{S_1}(z)  \frac{\dd}{\dd t}v_{S_2}(z) -  v_{S_2}(z) \frac{\dd}{\dd t} v_{S_1}(z) \right) 
            \  \Lambda^0_{L_3 \psi},
  \\
  \Gamma_{L_3 \psi S_i }^{ \pm} &= \frac{12}{T^2} \left\lvert \lambda'_{L \psi 1 } v_{S_1}(z) + \lambda'_{L \psi 2} v_{S_2} (z) \right\rvert^2   \Lambda^{\pm}_{L_3 \psi},
  \\
  \Gamma_{L_3 \psi S_i} & =  \frac{12}{T^2} \left\lvert \lambda'_{L \psi i} \right\rvert^2  I_F(m_\psi, m_L, m_{S_i}),
\end{align}
\end{subequations}
The $\Lambda^{0,\pm}_{L_3 \psi}$ are numerically evaluated integrals that appear
when evaluating the rates arising from the VEV-insertion diagrams. Similarly the
$I_F$ function is a numerically evaluated integral arising from the Yukawa-loop
processes that are dependant on the thermal masses of the particles in the loop.
The thermal masses are given in Appendix~\ref{app:FTFT}, and the formulae for
$\Lambda^{\pm,0}$ and $I_F$ are provided in Appendix~\ref{app:QTE} along with
numerical values for the reaction rates in benchmark $A$. We also consider a
more complex system of transport equations, outlined in
Appendix~\ref{app:FullQTE}, which is suitable when the $\psi$ Yukawa is large
and we can no longer use some of the earlier simplifying assumptions.

  Note that due to the mixing induced by the singlet VEVs the couplings leading
  to a nonzero CPV source term $S^{CPV} \propto \mathrm{Im}[\lambda'_{L \psi 1}
  {\lambda'}^*_{L \psi 2} ]$, are a non-trivial function of the $\lambda_{L \psi
    i }$, $\lambda_{\psi \psi i }$, and their associated phases
  $\delta_1$--$\delta_3$.
The $\delta_4$ phase does not contribute since introducing its associated Yukawa
coupling $Y_\psi$, needs a Higgs-VEV-insertion. The only way to construct
self-energy diagrams of the type seen in Fig.~\ref{fig:VIA1} is to have two such
insertions, leading to CPV sources proportional to terms like
$\mathrm{Im}\left[Y'_\psi {Y'}^*_\psi\right] =
0$. 

This system of four linearly independent ODEs are solved numerically using a
technique similar to the one outlined in appendix C of Ref.~\cite{TwoStep}. The
approach outlined there treats the VEV and mass dependent rates as a step
function across the bubble wall. We go one step further and also treat the
source term as a delta function, which introduces negligible error for the
bubble wall lengths that we consider. A more thorough treatment only becomes
important once the scale of the wall width $L$ approaches that of the typical
diffusion-reaction distance, when we no longer have $L \ll \sqrt{D/\Gamma}$.
Once the transport equations are solved, the asymmetry is given by dividing the
baryon-density deep inside of the bubble by the entropy density $s$,
\begin{equation} 
  Y_B =  \frac{n_q(+\infty)}{3 s} \, , 
  \qquad
  s  = \frac{2 \pi^2 g_{\star}}{45} T^3 \, , \qquad g_\star=108.75.
\end{equation}

\subsection{EWBG Results}

\begin{figure}[t]
  \centering
  \includegraphics[width=0.49\textwidth]{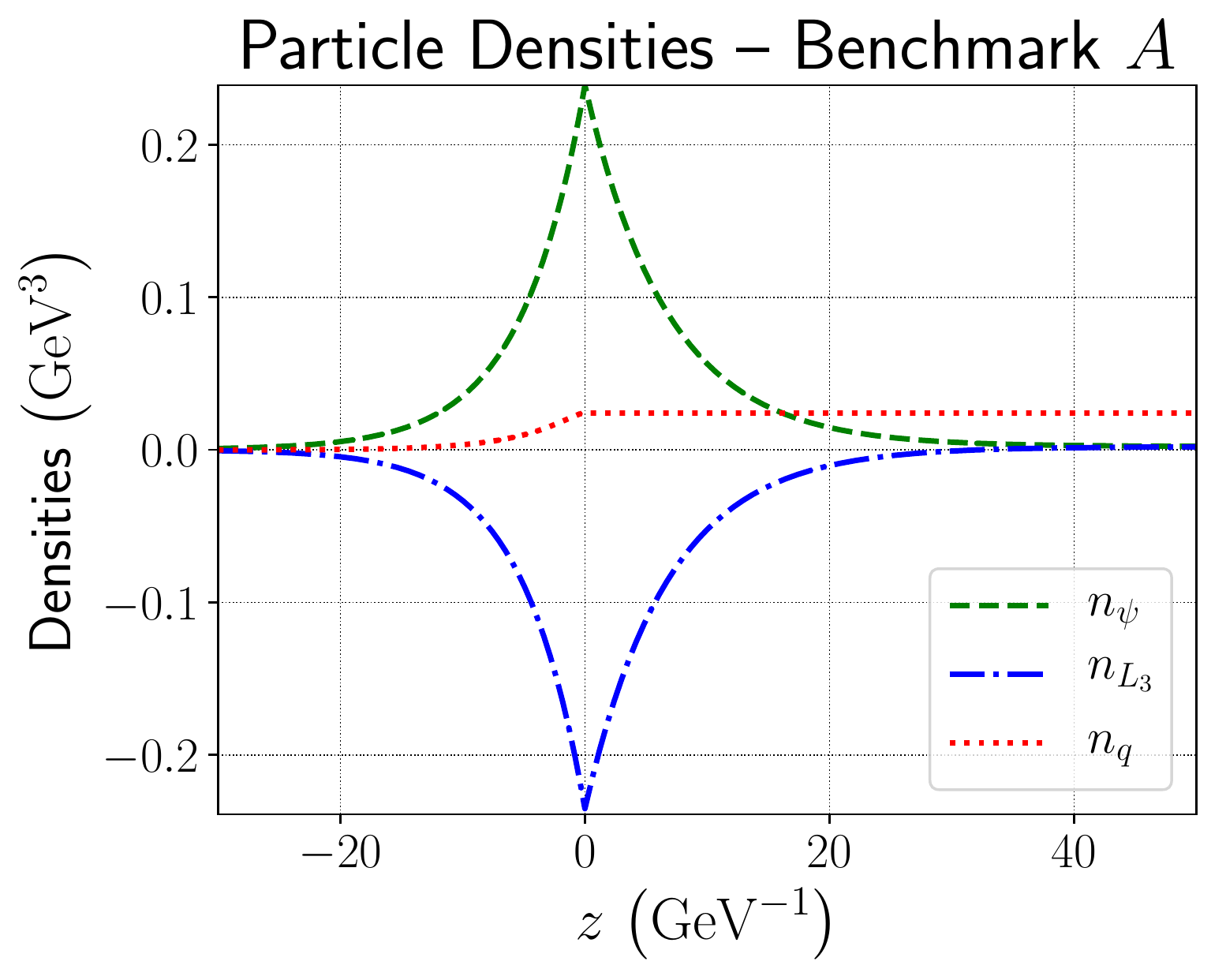}
  \caption{Number densities around the bubble wall for benchmark $A$ using the
    approximate system of transport equations. Using the full system of
    equations leads to no visual difference. The green dashed, blue dash-dotted
    and red dotted lines indicate the $\psi$, $L_3$, and total quark number
    densities, respectively. The region $z>0$ corresponds to the inside of the
    bubble, where SU(2) is broken.}
  \label{fig:EWBGDensities}
\end{figure}

\begin{figure}[t]
  \centering
  \begin{subfigure}[b]{0.49\textwidth}
    \includegraphics[width=\textwidth]{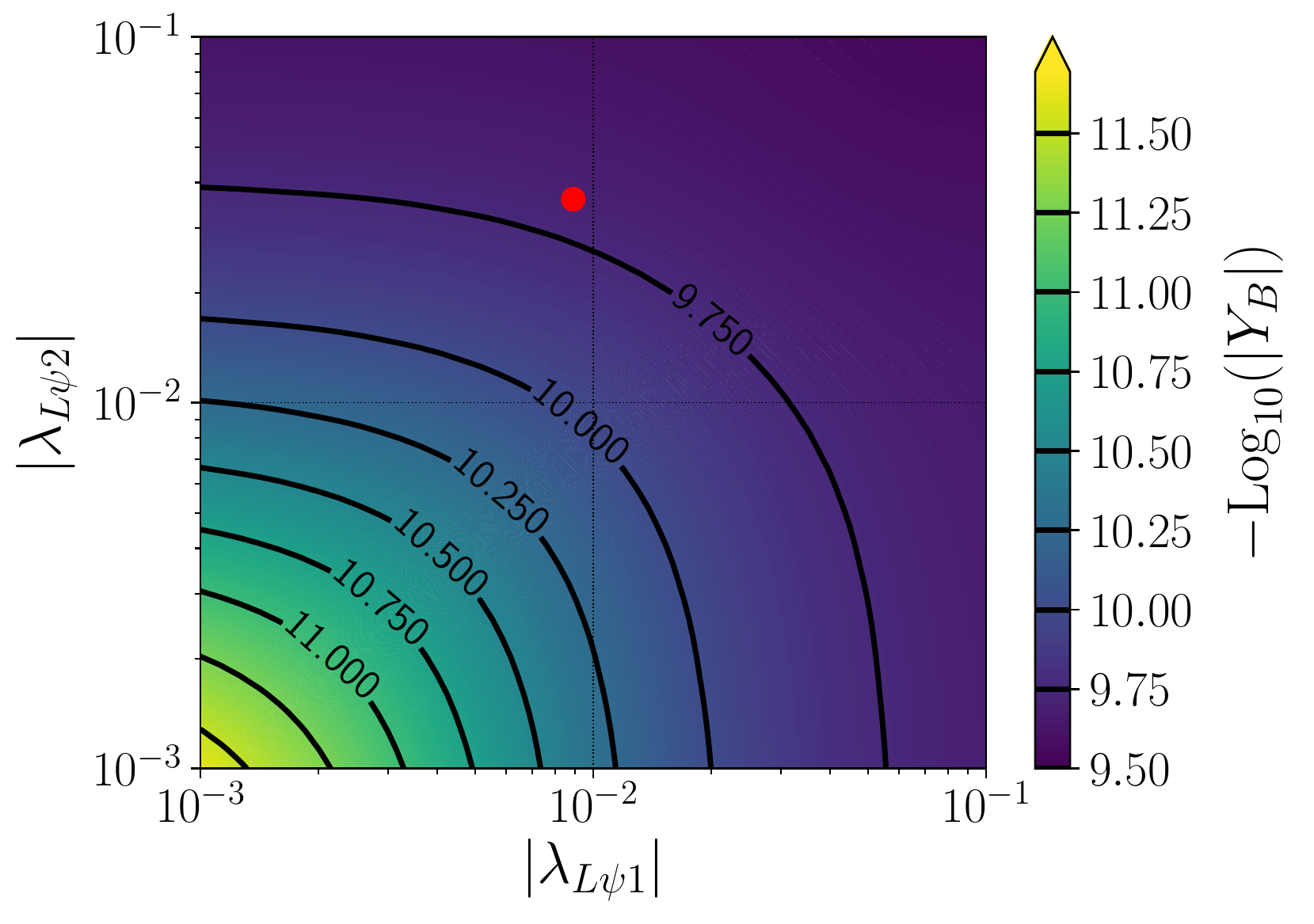}
    \caption{}
    \label{fig:YukawaContour}
  \end{subfigure}
  \begin{subfigure}[b]{0.49\textwidth}
    \includegraphics[width=\textwidth]{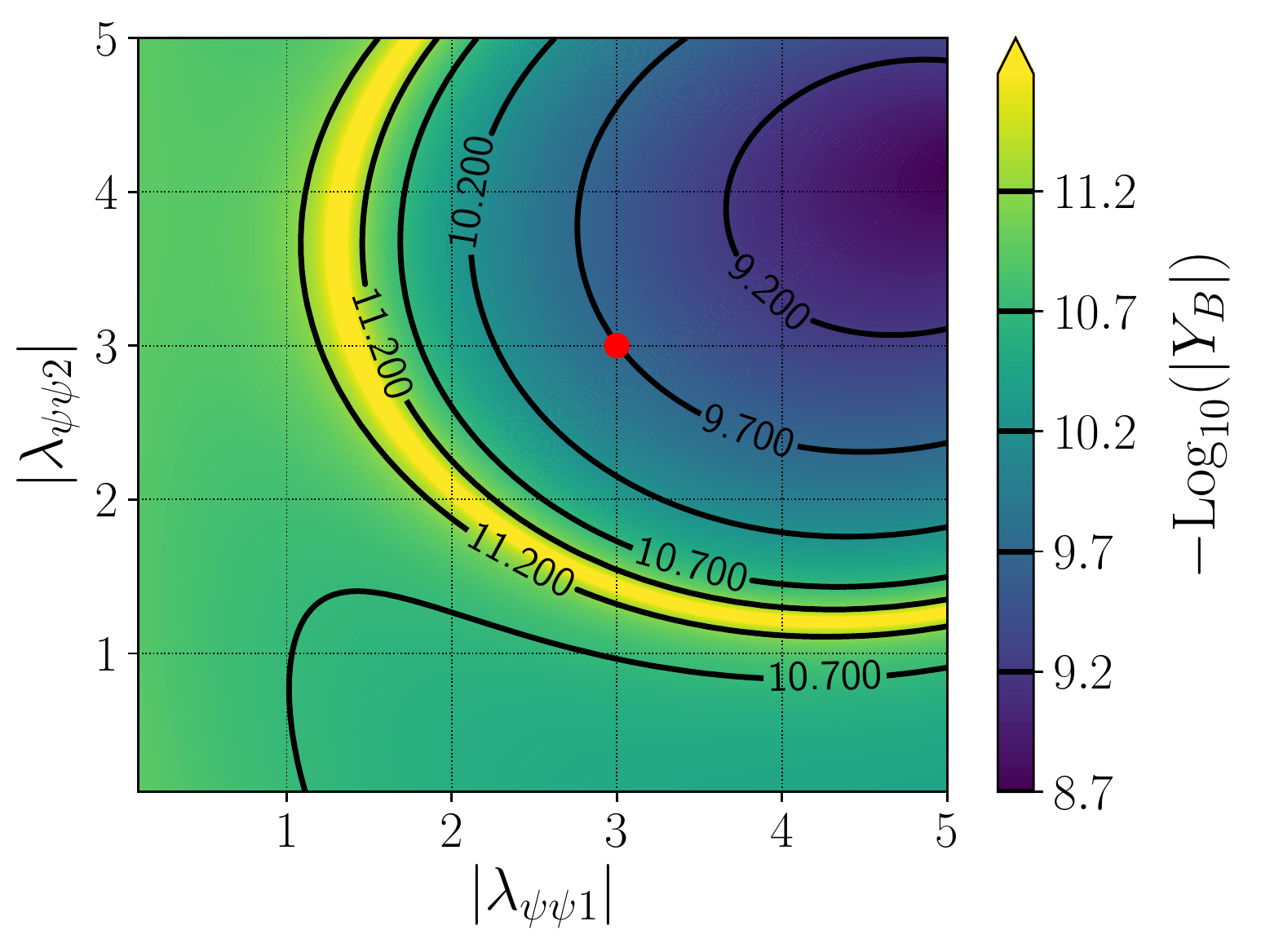}
    \caption{}
    \label{fig:YukawaContour2}
  \end{subfigure}
  \begin{subfigure}[b]{0.49\textwidth}
    \includegraphics[width=\textwidth]{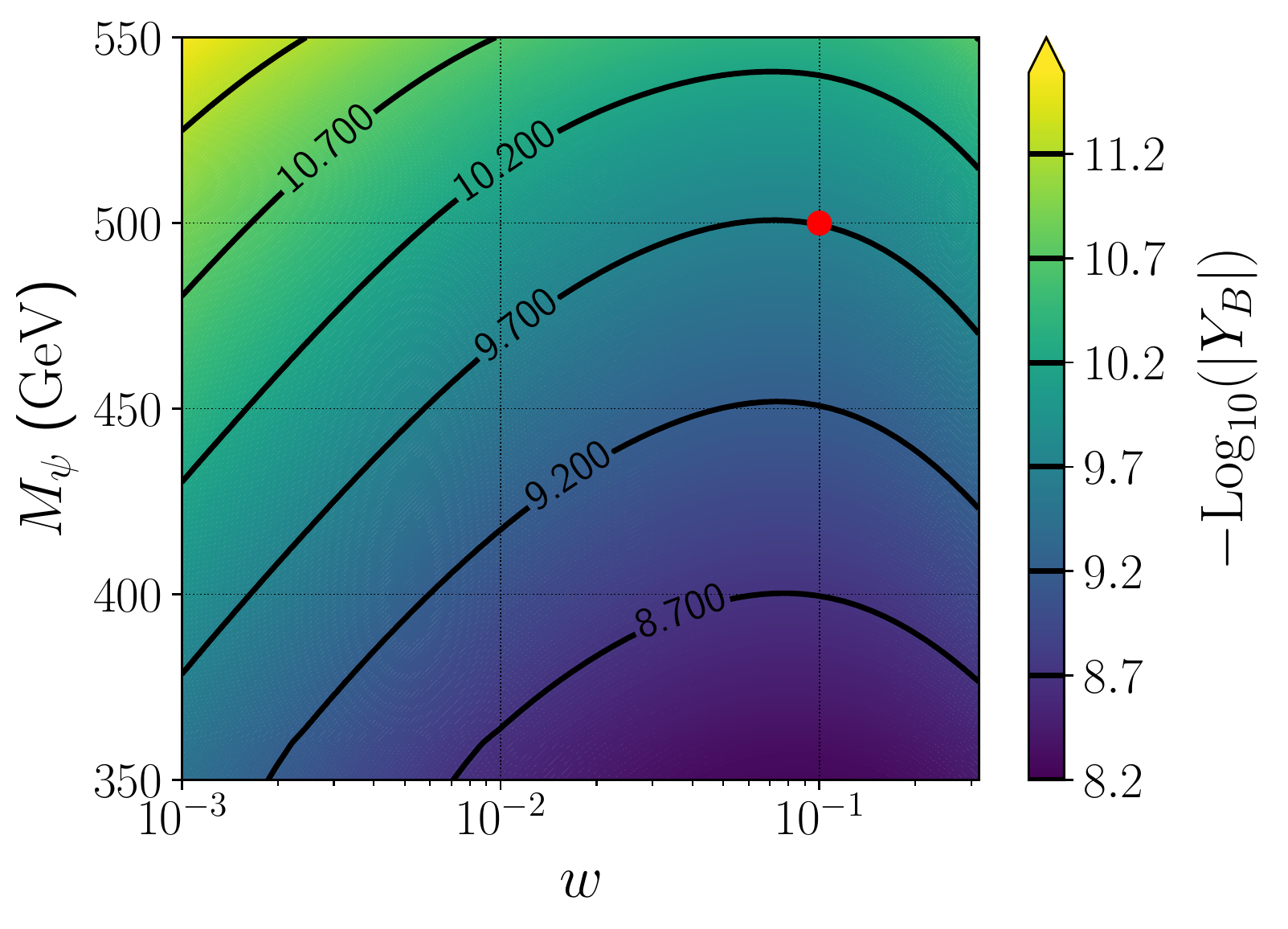}
    \caption{}
    \label{fig:MvContour}
  \end{subfigure}
  \begin{subfigure}[b]{0.49\textwidth}
    \includegraphics[width=\textwidth]{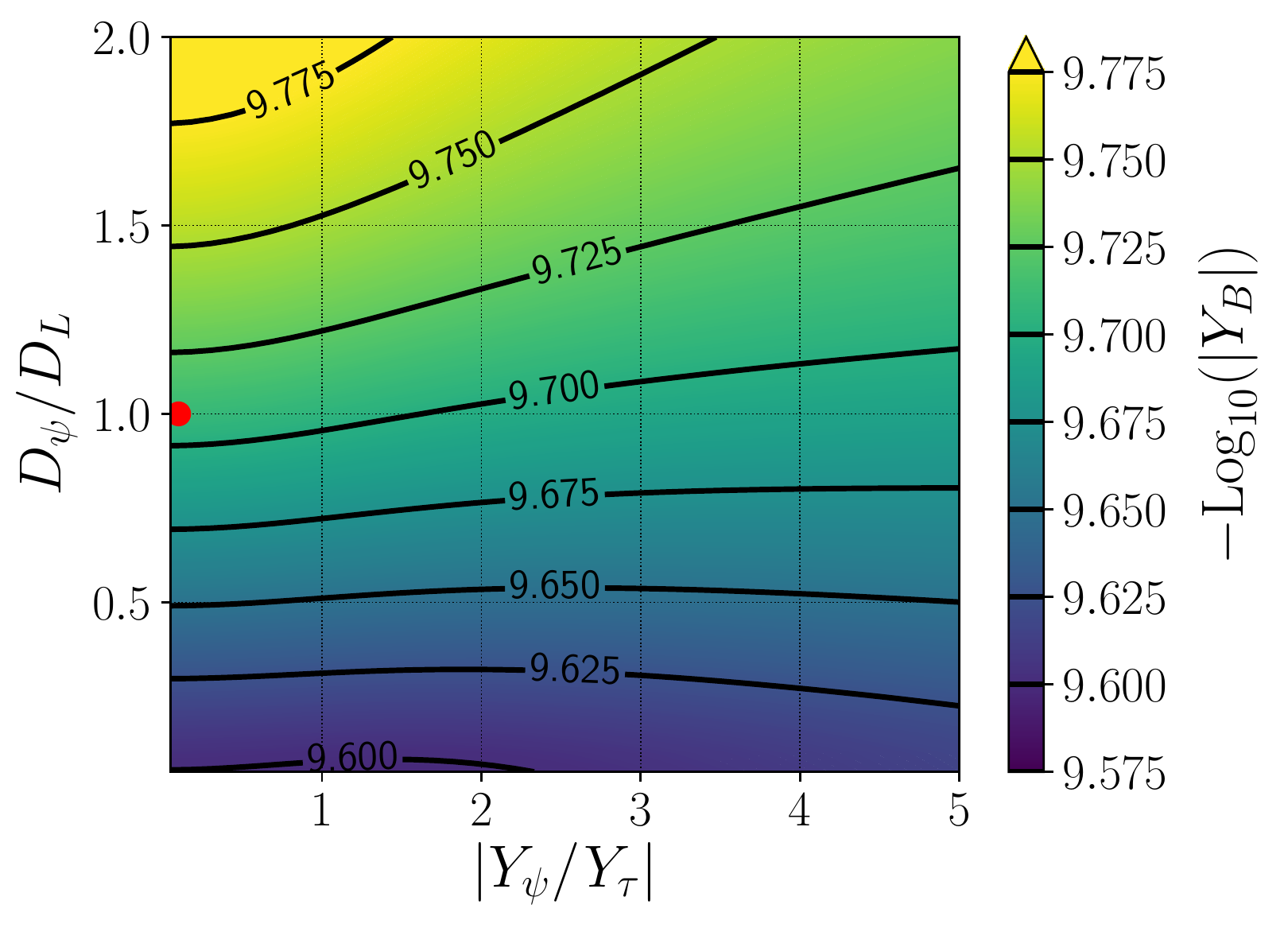}
    \caption{}
    \label{fig:YDContour}
  \end{subfigure}
  \caption{Contour plots showing the variation in the generated asymmetry as
    some of benchmark $A$'s parameters are varied, with the original benchmark values
    indicated by a red point. $Y_\tau$ and $D_L$ are held constant and used as
    reference values for the axes. The colour-mapping is different for each subplot. Darker colours indicate more asymmetry, brighter colours indicate less.}
  \label{fig:EWBGContours}
\end{figure}

The resulting baryon asymmetries generated by each of the benchmark points are
listed in Table~\ref{tab:EWBGBenchmarkResults}, for both the simplified and full
system of transport equations as given in Section~\ref{sec:QTE} and
Appendix~\ref{app:FullQTE}, respectively. The number densities obtained by
solving the simplified system of transport equations for benchmark $A$ are shown
in Fig.~\ref{fig:EWBGDensities}, where the bubble wall is located at $z=0$, with
$z>0$ corresponding to the interior of the bubble. The number density profiles
obtained for benchmark $A$ are illustrative of those obtained for the other
benchmarks.
  The effects of varying some of benchmark $A$'s parameters ($\lambda_{L \psi
    i}$, $\lambda_{\psi \psi i}$, $M_\psi$ and $Y_\psi$), the bubble velocity
  $w$ and the diffusion rate $D_\psi$ are illustrated in
  Fig.~\ref{fig:EWBGContours}. As the simplifying assumptions used to obtain the
  simplified transport equations no longer hold
  when $Y_\psi$ becomes large, we have used the full system of transport
  equations for the contour plots.

\begin{table}
  \centering
  \begin{tabular}{|c|c|c|c|c|} 
    \hline
    Benchmark  &   $A$   &    $B$ &    $C$ &    $D$  \\
    \hline
    $Y_B^{\mathrm{Approx}} \cdot 10^{10}$   & $1.97$ & $2.47$ & $1.15$ &  $1.15$    \\\hline
    $Y_B^{\mathrm{Full}} \cdot 10^{10}$ & $1.95$ & $2.45$ & $1.14$ &  $1.14$    \\\hline
  \end{tabular}
  \caption{Baryon asymmetry generated by each benchmark point using the full and approximate system of transport equations.}
  \label{tab:EWBGBenchmarkResults}
\end{table}

  Figures~\ref{fig:YukawaContour}~and~\ref{fig:YukawaContour2} show the
  dependence of the asymmetry on the scalar Yukawa couplings. If $\lvert
  \lambda_{L \psi i} \rvert \ll \lvert \lambda_{\psi \psi i}\rvert$, the role of
  the $\lambda_{L \psi i}$ is to generate mixing between the $\psi$ and $L_3$
  and produce large $\lambda'_{\L \psi i }$ which lead to a non zero CPV source
  term as given in eq.~\eqref{eq:sourceFunc}. The $\lambda_{\psi \psi i}$ play a
  dual role of both reducing the mass of the $\psi$ in the singlet phase, and
  contributing to $\lambda'_{L \psi i }$ and hence the CPV source term. Note
  that in Fig.~\ref{fig:YukawaContour2} there is a ring where the generated
  asymmetry crosses through zero. This occurs due to a cancellation of
  the imaginary parts of the $\Lambda_{L_3 \psi}^0$ function, which appears in
  the expression for the CPV source term eq.~\eqref{eq:sourceFunc} and is given in eq.~\eqref{eq:QTEIntegrals}.

  Figure~\ref{fig:MvContour} shows the dependence of the final asymmetry on the
  bubble wall velocity and $\psi$ mass. The decrease in the asymmetry when $w$
  becomes large is due to the inability of the particles to efficiently diffuse
  ahead of the wall, while the decrease for low $w$ is a result of the CPV
  source term being proportional to the velocity. Meanwhile decreasing the
  $\psi$ mass increases the resulting asymmetry, with a resonance occurring when
  $M_\psi \sim m_{L_3}$.

  
  Figure~\ref{fig:YDContour} shows that variations in $D_\psi$ or $Y_\psi$
  results in a relatively small change in the final asymmetry. The error
  introduced by utilising the massless fermion derivation of the diffusion
  constant ($D_\psi = D_L$) is thus likely negligible compared to other factors
  affecting the asymmetry generation. The asymmetry is much more strongly
  dependent on $D_L$ as the left-handed leptons diffusing ahead of the wall are
  what directly bias the sphalerons, and $D_\psi$ only acts to effectively
  modify the rate with which the $L_3$ equilibrates.
  


\section{Conclusion}
\label{sec:summary}

We have investigated the capability of vector-like leptons and scalar singlets
to generate the observed baryon asymmetry, with an emphasis on the phenomenology
of such a model. The singlets are necessary to induce a strongly first order
electroweak phase transition. The vector-like leptons introduce the CP-violating
interactions with the singlets and SM leptons needed to generate asymmetry
during the electroweak phase transition. Unlike some other electroweak
baryogenesis models featuring vector-like
fermions~\cite{VLFEWPTEWBGEDM,VLFDMEWPTEWBG,VLFDiphotonEWPT}, this model shares
a trait with multi-step EWPT models~\cite{TwoStep} in that the necessary CPV
does not come from interactions with the SM Higgs but from the new scalar
content.

From collider constraints we found that the presence of the new scalars will
generally lead to a larger mass bound than minimal vector-like lepton models
unless mixing is introduced in the scalar sector. The scalar mixing then
provides the primary contribution to the electron EDM. However, the generated
electron EDM is still a couple of orders of magnitude below current experimental
bounds. 
  An increase in the lower bound on the $\psi$ masses reduces the capability for
  asymmetry generation. This effect can be offset slightly by using large
  $\lambda_{\psi \psi i }$ Yukawa couplings and selecting their phases to make
  the vector-like fermions lighter in the early universe. However, the Yukawa
  couplings considered in our benchmarks are already uncomfortably large, a
  factor of three larger than the top quark Yukawa. Additional, it is these
  couplings and their phases that provide the dominant EDM contribution. Hence
  increased collider constraints alongside more precise electron EDM
  measurements will be sufficient to rule out this model.

The model examined in this paper, which is complementary to the one in
Ref.~\cite{ChaoSpontCP}, indicates that scalar singlet plus vector-like fermion
models can readily generate the observed baryon asymmetry during the EWPT. There
are a number of areas in which further work could be pursued, including more
thorough investigations of the moving wall dynamics, going beyond the
high-temperature approximation for a more rigorous treatment of the EWPT, and
resolution of uncertainties in the accuracy of the VEV-insertion approximation
formalism. A better understanding of these issues would lead to a more accurate
determination of the allowed parameter space of our model.

\acknowledgments We thank Phillip Basler, Christopher Lee, David Morrisey, Hiren
Patel and Gaham White for helpful discussions. Feynman diagrams were drawn using
the TikZ-Feynman package~\cite{TikZFeynman}. This work was supported by the
Australian Research Council. Michael J. Ramsey-Musolf was supported in part
under U.S. Department of Energy Contract DE-SC0011095.

\appendix

\section{Electron EDM Formulae}

  \label{app:EDM}
  \subsection{Mass Basis Weak Currents and Yukawa Interactions}
\label{app:mixing}
As mentioned in the main text, the notation we use for the weak currents
and Yukawa interactions are selected to match the notation of Ref.~\cite{mjrmEDMmssm}, such
that it is straightforward to apply their general electron EDM formulae if one
makes the following replacements,
\begin{equation}
  \begin{aligned}
    h^0, H^0 &\rightarrow {\phi_i},&
    \chi^-_a &\rightarrow {\mathcal{E}}_a,\\ 
    c_e^{{\phi_i}} &\rightarrow {\mathcal{P}_{i1}},&
    \chi^0_i &\rightarrow \mathcal{N}_i.
  \end{aligned}
\end{equation}
We define a matrix $M$ such that after moving to the mass basis the
charged weak currents become,
\begin{equation}
  \begin{aligned}
    j_{W^-}^\mu  \ & \supset\ 
    \frac{1}{ \sqrt{2}}  \bar{\tau} \gamma^\mu \mathbb{P}_L \nu_\tau \ + \  \frac{1}{\sqrt{2}}  \bar{E} \gamma^\mu  N  
    \\
    &
    = \ 
    -  \bar{{\mathcal{E}}'} \gamma^\mu \left( M^L \mathbb{P}_L + M^R \mathbb{P}_R \right)  \mathcal{N'},
  \end{aligned}
\end{equation}
where $\mathbb{P}_{L,R}$ are the standard left- or right-projection operators.
Similarly for the neutral currents we define a matrix $G$ such that,
\begin{equation}
  \begin{aligned}
    c_w j_Z^\mu  \  & \supset \ 
     \bar{\tau} \gamma^\mu \left( s_w^2 - \frac{1}{2} \mathbb{P}_L \right)  \tau  \ + \   \left(s_w^2 - \frac{1}{2}\right) \bar{E} \gamma^\mu  E
    \\
    &
    = \ 
     \bar{{\mathcal{E}}'} \gamma^\mu \left( G^R \mathbb{P}_R + G^L \mathbb{P}_L \right)   {\mathcal{E}}' .
  \end{aligned}
\end{equation}
For the Yukawa interactions we define matrices $D_i$ such that,  
\begin{equation}
  \begin{aligned}
    & \frac{h}{\sqrt{2}}
    \left( Y_\psi \bar{E} + Y_\tau \bar{\tau}  \right)
     \mathbb{P}_R \tau
    \ + \ 
    \sum_i s_i
    \left(  \lambda_{\psi \psi i} \bar{E}  + \lambda_{L \psi i} \bar{\tau}  \right)
    \mathbb{P}_R E
    \  + \mathrm{h.c.}
    \\
    & \  \ = 
    \frac{{\phi_i} e}{\sqrt{2} s_w} \bar{{\mathcal{E}}'} \left( D_{i}^R \mathbb{P}_R + D_{i}^L \mathbb{P}_L  \right) {\mathcal{E}}',
  \end{aligned}
\end{equation}
Using the mixing angles and phases defined in Section~\ref{sec:model}, these
matrices are given by,
\begin{subequations}
  \label{eq:currentAndYukawaMatrices}
  \begin{align}
M^L \ &\ = \frac{-1}{\sqrt{2}}
\begin{bmatrix}
 \cos{\theta_L} \ &\  \sin{\theta_L} \\
 -\sin{\theta_L} \ &\ \cos{\theta_L} \\
\end{bmatrix}
, & 
M^R \ &\ = \frac{-1}{\sqrt{2}}
\begin{bmatrix}
 \cos{\theta_R} \ &\  0 \\
 -\sin{\theta_R} \ &\  0 \\
\end{bmatrix}
,\\
G^L \ &\ = \left(s_w^2 - \frac{1}{2}\right)
\begin{bmatrix}
 1 \ &\  0 \\
 0 \ &\  1 \\
\end{bmatrix}
,  &
G^R \ &\ =    
\begin{bmatrix}
 {s_w}^2 - \frac{1}{2 } \cos^2 \theta_R \ &\  \frac{1}{4} \sin (2 \theta_R) \\
 \frac{1}{4} \sin (2 \theta_R) \ &\   {s_w}^2 -\frac{1}{2} \sin^2 \theta_R \\
\end{bmatrix},
\end{align}
\begin{align}
D^R_{\phi_1} &\approx \frac{s_w}{e} 
\begin{bmatrix}
 \sqrt{2} {\lvert\lambda_{\psi \psi 1}\rvert} e^{i {\delta_2}} {\mathcal{P}_{12}}+\sqrt{2} {\lvert\lambda_{\psi \psi 2}\rvert} e^{i {\delta_3}} {\mathcal{P}_{13}}+{\lvert Y_\psi \rvert} {\theta_R} \ & \  {\lvert Y_\psi \rvert}+{\theta_L} {Y_{\tau}} \\
 \sqrt{2} {\lvert \lambda_{L \psi 1}} \rvert e^{i ({\delta_1}+{\delta_4})} {\mathcal{P}_{12}}+\sqrt{2} e^{i {\delta_4}} { \lambda_{L \psi 2}} {\mathcal{P}_{13}}+{\theta_R} {Y_{\tau}} \ & \  {Y_{\tau}}-{\lvert Y_\psi \rvert} {\theta_L} \\
\end{bmatrix}
,\\
D^R_{\phi_2} &\approx \frac{s_w}{e} 
\begin{bmatrix}
 \sqrt{2} \left(e^{i {\delta_2}} {\lvert\lambda_{\psi \psi 1}\rvert}+{\lvert\lambda_{\psi \psi 2}\rvert} e^{i {\delta_3}} {\mathcal{P}_{23}}+{\lvert \lambda_{L \psi 1}} \rvert e^{i ({\delta_1}+{\delta_4})} {\theta_L}\right) \ & \  -{\lvert Y_\psi \rvert} {\mathcal{P}_{12}}-\sqrt{2} {\lvert\lambda_{\psi \psi 1}\rvert} e^{i {\delta_2}} {\theta_R} \\
 \sqrt{2} \left(e^{i ({\delta_1}+{\delta_4})} {\lvert \lambda_{L \psi 1}} \rvert+e^{i {\delta_4}} { \lambda_{L \psi 2}} {\mathcal{P}_{23}}-{\lvert\lambda_{\psi \psi 1}\rvert} e^{i {\delta_2}} {\theta_L}\right) \ & \  -\sqrt{2} {\lvert \lambda_{L \psi 1}} \rvert e^{i ({\delta_1}+{\delta_4})} {\theta_R}-{\mathcal{P}_{12}} {Y_{\tau}} \\
\end{bmatrix}
,\\
D^R_{\phi_3} &\approx \frac{s_w}{e} 
\begin{bmatrix}
 \sqrt{2} \left(e^{i {\delta_3}} {\lvert\lambda_{\psi \psi 2}\rvert}-{\lvert\lambda_{\psi \psi 1}\rvert} e^{i {\delta_2}} {\mathcal{P}_{23}}+e^{i {\delta_4}} { \lambda_{L \psi 2}} {\theta_L}\right) \ & \  -{\lvert Y_\psi \rvert} {\mathcal{P}_{13}}-\sqrt{2} {\lvert\lambda_{\psi \psi 2}\rvert} e^{i {\delta_3}} {\theta_R} \\
 \sqrt{2} \left(e^{i {\delta_4}} { \lambda_{L \psi 2}}-{\lvert \lambda_{L \psi 1}} \rvert e^{i ({\delta_1}+{\delta_4})} {\mathcal{P}_{23}}-{\lvert\lambda_{\psi \psi 2}\rvert} e^{i {\delta_3}} {\theta_L}\right) \ & \  -\sqrt{2} e^{i {\delta_4}} { \lambda_{L \psi 2}} {\theta_R}-{\mathcal{P}_{13}} {Y_{\tau}} \\
\end{bmatrix}, \\
    D_{\phi_i}^L &= (D_{\phi_i}^R)^\dagger .
\end{align}
\end{subequations}
In the expression for $D_{i}$ provided above we have dropped any terms second order in mixing angles
$\mathcal{P}_{i, j \neq i}$ and $\theta_{L/R}$, though the numerical calculations
included these terms.

\subsection{EDM Formulae}
\label{app:edmformulae}
Using the notation introduced in Section~\ref{app:mixing} and the results from Refs.~\cite{mjrmEDMmssm,VLQEDM,VLQEDMCalc}, the primary contributions to the electron EDM are given by
\begin{equation}
  d_e = \sum_{i=1}^{3} \left( d_e^{\gamma {\phi_i}} + d_e^{Z {\phi_i}} \right)  + d_e^{W^- W^-},
\end{equation}
\begin{equation}
  d_e^{\gamma {\phi_i}}
  = \frac{ e \alpha^2 {\mathcal{P}_{i1}} }{8 \sqrt{2}\pi^2 s_w^2 } \frac{m_e}{m_W m^2_{{\phi_i}}}
  \sum_{a=1}^{2} \mathrm{Im}(D_{{\phi_i}, a a}^R) M_{{\mathcal{E}}_a} \int_0^1 \dd x \frac{j\left( 0 , \frac{r_{{\mathcal{E}}_a h_1}}{x(1-x)} \right)}{x (1-x)},
\end{equation}
\begin{align}
  d_e^{Z {\phi_i}}
  &= \frac{e \alpha^2 {\mathcal{P}_{i1}}(s_w^2 - \frac{1}{2})}{8 \sqrt{2}\pi^2 c_w^2 s_w^4 } \frac{m_e}{m_W m^2_{{\phi_i}}}
    \sum_{a,b=1}^{2} \mathrm{Im}(G^R_{a b} D^R_{{\phi_i} b a} - G^L_{a b} D^L_{{\phi_i} b a}) m_{{\mathcal{E}}_b}
    \nonumber\\ & \qquad \times \ \int_0^1 \dd x \frac{1}{x}j\left( r_{Z {\phi_i}} , \frac{x r_{{\mathcal{E}}_a h_1} + (1-x) r_{{\mathcal{E}}_a {\phi_i}}}{x(1-x)} \right),
\end{align}
\begin{align}
  d_e^{W^- W^-}
  &=
    -\sum_{a,i=1}^2 \frac{e \alpha^2 m_e m_{{\mathcal{E}}_a} m_{\mathcal{N}_i}}{8 \pi^2 c_w^4 m_W^4} \mathrm{Im}\left( M^{L *}_{a i} M^R_{a i} \right) 
    \nonumber \\ & \qquad
                   \times \ \int_0^1 \dd z_1 \int_0^{1- z_1} \dd z_2  \frac{Z \mathrm{log}\left( \frac{K_{a i}}{Z} \right)}{ \left(K_{a i} - Z \right)^2 } - \frac{1}{K_{a i} - Z}.
\end{align}
Where we have used,
\begin{equation}
  \begin{aligned}
    j(x,y) &= \frac{\frac{x \mathrm{log}(x)}{x-1}  -  \frac{y \mathrm{log}(y)}{y-1}}{x-y}
    , &
    Z &= (z_1 + z_2)(1-z_1- z_2),
    \\
    K_{a i} &= r_{\mathcal{N}_i W} + z(r_{{\mathcal{E}}_a W} - r_{\mathcal{N}_i W})
    , \  & r_{x y} & =  \frac{m^2_x}{m^2_y}.
  \end{aligned}
\end{equation}
Substituting the matrices defined in
eq.~\eqref{eq:currentAndYukawaMatrices} one finds that in our case
$d_e^{W^- W^-}=0$. To first order in the mixing angles the only nonzero
contributions to the EDM come from terms proportional to
$(D^{L,R}_{\phi_i})_{11}$, which correspond to Barr-Zee style diagrams involving
an $E^+$-$E^-$ loop.

\section{Thermal Properties and Transport Equation Functions}
\label{app:EWBG}

\subsection{Thermal Properties}
\label{app:FTFT}

The thermal mass formulae and diffusion constants used are given in
Tables~\ref{tab:ThermalMasses}~and~\ref{tab:DiffusionConstants} respectively.
\begin{table}
  \centering
  \begin{tabular}{|c|c|c|c|c|}    
    \hline
    Particle & $Q, d, u$  & $L$  & $\tau$ & $H$
    \\\hline
    $D_i T$  &  $6$  & $100$ & $380$ & $110$ 
    \\\hline
  \end{tabular}
  \caption{Diffusion constants for the transport equations as derived in
    Ref.~\cite{DiffusionTerms}.}
  \label{tab:DiffusionConstants}
\end{table}
\begin{table}
  \centering
  \begin{tabular}{|c|c|}    
    \hline Particle  & $\delta m^2 / T^2$
    \\ \hline
    $H$ & $\frac{3}{16} g_2^2 + \frac{1}{16} g_1^2 + \frac{1}{4} \lvert Y_{u_3} \rvert^2 + \frac{1}{2} \lambda + \frac{1}{24} (b_{11}+ b_{22})$
    \\ \hline
    $S_1$ & $\frac{1}{4} a_{1111} + \frac{1}{6} b_{11} + \frac{1}{24} a_{1122}$
    \\ \hline
    $S_2$ & $\frac{1}{4} a_{2222} + \frac{1}{6} b_{22} + \frac{1}{24} a_{1122}$
    \\ \hline
    $Q_{i}$ & $\frac{1}{6} g_3^2  + \frac{3}{32} g_2^2  + \frac{1}{288} g_1^2 +    \frac{1}{16} \left(  \left\lvert Y_{u_i}\right\rvert^2  + \left\lvert Y_{d_i}\right\rvert^2\right)$ 
    \\ \hline 
    $u_i$ & $\frac{1}{6} g_3^2 + \frac{1}{18} g_1^2 + \frac{1}{8} \left\lvert Y_{u_i}\right\rvert^2$ 
    \\ \hline 
    $d_i$ & $\frac{1}{6} g_3^2 + \frac{1}{64} g_1^2  + \frac{1}{8} \left\lvert Y_{d_i}\right\rvert^2 $ 
    \\ \hline
    $L_{1,2}$ & $\frac{3}{32} g_2^2  + \frac{1}{32} g_1^2 $ 
    \\ \hline
    $L_3$ & $\frac{3}{32} g_2^2  + \frac{1}{32} g_1^2  + \frac{1}{16} \left( \left\lvert Y'_{\tau}\right\rvert^2   + \left\lvert\lambda'_{L \psi 1 }\right\rvert^2 + \left\lvert\lambda'_{L \psi 2 }\right\rvert^2 \right) $ 
    \\ \hline
    $\tau_R$ & $\frac{1}{8} g_1^2 + \frac{1}{8}(\left\lvert Y'_{\tau}\right\rvert^2 + \left\lvert Y'_{\psi}\right\rvert^2)$ 
    \\ \hline
    $\psi$ & $\frac{3}{32} g_2^2  + \frac{1}{32} g_1^2 + \frac{1}{16}\left( \left\lvert\lambda'_{\psi \psi 1}\right\rvert^2+ \left\lvert\lambda'_{\psi \psi 2}\right\rvert^2 + \left\lvert Y'_{\psi}\right\rvert^2 + \left\lvert\lambda'_{L \psi 1 }\right\rvert^2 + \left\lvert\lambda'_{L \psi 2}\right\rvert^2 \right) $
    \\ \hline
  \end{tabular}
  \caption{Thermal mass contributions for the fermions. }
  \label{tab:ThermalMasses}
\end{table}
For massless fermions the thermal widths at zero momentum are given by~\cite{widths1}
\begin{equation}
  \Gamma
  \approx \sum_i \frac{g_i^2 T C_{F,i} }{4 \pi},
\end{equation}
where $C_{F}$ denotes the quadratic Casimir invariant. In the limit of masses
heavy compared to the temperature one instead has,
\begin{equation}
  \Gamma \approx
  \sum_i \frac{g_i^2 T C_{F,i} }{8 \pi},
\end{equation}
which differs from the massless case by a factor of a half. The resulting widths for the relevant
particles are given in Table~\ref{tab:ThermalWidths}.

\begin{table}
  \centering
  \begin{tabular}{|c|c|}    
    \hline Particle & $\Gamma / T$
    \\ \hline 
    $Q_{i}$ & $\frac{4}{3} \frac{g_3^2}{4 \pi}  +  \frac{3}{4}\frac{g_2^2}{4 \pi}  + \frac{1}{36} \frac{g_1^2}{4 \pi}  $
    \\ \hline 
    $u_i$ & $\frac{4}{3} \frac{g_3^2}{4 \pi} +  \frac{4}{9} \frac{g_1^2}{4 \pi} $
    \\ \hline 
    $d_i$ & $\frac{4}{3} \frac{g_3^2}{4 \pi}  +  \frac{1}{9} \frac{g_1^2}{4 \pi} $
    \\ \hline
    $L_i$ & $ \frac{3}{4}\frac{g_2^2}{4 \pi} + \frac{1}{4}\frac{g_1^2}{4 \pi} $
    \\ \hline
    $\psi$ & $ \frac{3}{4}\frac{g_2^2}{8 \pi} + \frac{1}{4}\frac{g_1^2}{8 \pi} $
    \\ \hline
  \end{tabular}
  \caption{Thermal widths used for the VIA approximation calculations. }
  \label{tab:ThermalWidths}
\end{table}

\subsection{Full System of Transport Equations}
\label{app:FullQTE}

If the $\tau$ and $\psi$ Yukawa couplings are not negligible, the system of transport equations becomes significantly more complicated. We must then consider $n_H$, $n_{Q_3}$, $n_{u_3}$ in more detail. Neglecting the Yukawa interactions of the lighter fermion, instead of eqs.~\eqref{eq:QTEsimplification} we will get  
\begin{equation}
\mu_{L_1} = \mu_{L_2}, \quad  \mu_{Q_1}= \mu_{Q_2}, \quad \mu_{u_1} = \mu_{u_2} = \mu_{d_i} . \label{eq:quarkEquilibrium}
\end{equation}
The full system of quantum
transport equations to be solved is then,
\begin{subequations}
\label{eq:QTEs2}
\begin{align}
  \partial_\mu J_\psi^\mu =& 
   - \Gamma_{\psi {\tau_R} H} \left(   \frac{n_\psi}{k_\psi} -  \frac{n_{{\tau_R}}}{k_{{\tau_R}}} - \frac{n_H}{k_H} \right)
  - \left( \Gamma_{L_3 \psi S_i }^{-} + \Gamma_{{L_3} \psi  S_1}  + \Gamma_{{L_3} \psi  S_2}\right)
    \left( \frac{n_\psi}{k_\psi}  - \frac{n_{L_3}}{k_{L_3}} \right)
  \nonumber\\ & \
- \Gamma_{L_3 \psi S_i }^{+} \left( \frac{n_\psi}{k_\psi} + \frac{n_{L_3}}{k_{L_3}} \right)
        - \Gamma_{\psi {\tau_R} H}^{- } \left(   \frac{n_\psi}{k_\psi} -  \frac{n_{{\tau_R}}}{k_{{\tau_R}}}  \right)\
  - \Gamma_{\psi {\tau_R} H}^{+ } \left( \frac{n_\psi}{k_\psi} + \frac{n_{\tau_R}}{k_{\tau_R}} \right)
       + S_{L_3 \psi}^{CPV},\\
  \partial_\mu J_{L_3}^\mu =&
    \Gamma_{{L_3} \psi  S_i }^{+} \left( \frac{n_\psi}{k_\psi} + \frac{n_{L_3}}{k_{L_3}} \right)
  + \left( \Gamma_{\psi {L_3} S_i }^{ -} + \Gamma_{\psi {L_3} S_1}  + \Gamma_{\psi {L_3} S_2}\right)
    \left( \frac{n_\psi}{k_\psi}  - \frac{n_{L_3}}{k_{L_3}} \right)
  \nonumber\\ & \
       - \Gamma_{{L_3} {\tau_R} H} \left(   \frac{n_{L_3}}{k_{L_3}} -  \frac{n_{{\tau_R}}}{k_{{\tau_R}}} - \frac{n_H}{k_H}\right) - \Gamma_{L_3 {\tau_R} H}^{- } \left(   \frac{n_{L_3}}{k_{L_3}} -  \frac{n_{{\tau_R}}}{k_{{\tau_R}}} \right)\
  \\ &  \nonumber
  - \Gamma_{L_3 {\tau_R} H}^{+ } \left( \frac{n_L}{k_L} + \frac{n_{\tau_R}}{k_{\tau_R}} \right)
   - \Gamma_{\mathrm{WS}} \sum_j \left( \frac{3 n_{Q_j}}{k_{Q_j}} + \frac{n_{L_j}}{k_{L_j}} \right)
       - S^{CPV}_{L_3 \psi} ,\\         
  \partial_\mu J_{L_{1,2}}^\mu =& - \Gamma_{\mathrm{WS}} \sum_j \left( \frac{3 n_{Q_j}}{k_{Q_j}} + \frac{n_{L_j}}{k_{L_j}} \right), \\
  \partial_\mu J^\mu_{{\tau_R}} =& 
       \Gamma_{\psi {\tau_R} H} \left(   \frac{n_\psi}{k_\psi} -  \frac{n_{{\tau_R}}}{k_{{\tau_R}}} - \frac{n_H}{k_H}\right)+ \Gamma_{\psi {\tau_R} H }^{- } \left(   \frac{n_\psi}{k_\psi} -  \frac{n_{{\tau_R}}}{k_{{\tau_R}}} \right)
  + \Gamma_{\psi {\tau_R} H}^{+ } \left( \frac{n_\psi}{k_\psi} + \frac{n_{\tau_R}}{k_{\tau_R}} \right)
  \\ &  \nonumber
      +  \Gamma_{{L_3} {\tau_R} H} \left(   \frac{n_{L_3}}{k_{L_3}}        -  \frac{n_{{\tau_R}}}{k_{{\tau_R}}} - \frac{n_H}{k_H} \right) + \Gamma_{L_3 {\tau_R} H }^{- } \left(   \frac{n_{L_3}}{k_{L_3}}        -  \frac{n_{{\tau_R}}}{k_{{\tau_R}}} \right)
  + \Gamma_{L_3 {\tau_R} H }^{+ } \left( \frac{n_L}{k_L} + \frac{n_{\tau_R}}{k_{\tau_R}} \right)
                               ,\\
  \partial_\mu J^\mu_{Q_{1,2}}       =& 
 - 2 \Gamma_{\mathrm{SS}} \sum_j \left( \frac{2 n_{Q_j}}{k_{Q_j}} - \frac{n_{u_j}}{k_{u_j}} - \frac{n_{d_j}}{k_{d_j}} \right)
                - 3 \Gamma_{\mathrm{WS}} \sum_j \left( \frac{3 n_{Q_j}}{k_{Q_j}} + \frac{n_{L_j}}{k_{L_j}} \right)
                , \\
  \partial_\mu J^\mu_{Q_3}       =&   - 2 \Gamma_{\mathrm{SS}} \sum_j \left( \frac{2 n_{Q_j}}{k_{Q_j}} - \frac{n_{u_j}}{k_{u_j}} - \frac{n_{d_j}}{k_{d_j}} \right)
                - 3 \Gamma_{\mathrm{WS}} \sum_j \left( \frac{3 n_{Q_j}}{k_{Q_j}} + \frac{n_{L_j}}{k_{L_j}} \right) \\
   &  \nonumber
    - \Gamma_{{Q_3} {u_3} H} \left(  \frac{n_{Q_3}}{k_{Q_3}} - \frac{n_{u_3}}{k_{u_3}} + \frac{n_H}{k_H} \right) 
    - \Gamma_{Q_3 u_3 H }^{- }  \left(  \frac{n_{Q_3}}{k_{Q_3}} - \frac{n_{u_3}}{k_{u_3}} \right) 
  - \Gamma_{Q_3 u_3 H }^{+ } \left( \frac{n_{Q_3}}{k_{Q_3}} + \frac{n_{u_3}}{k_{u_3}} \right)
                , \\
   \partial_\mu J^\mu_{d_{1,2,3}}       =& \partial_\mu J^\mu_{u_{1,2}}  =
 \Gamma_{\mathrm{SS}} \sum_j \left( \frac{2 n_{Q_j}}{k_{Q_j}} - \frac{n_{u_j}}{k_{u_j}} - \frac{n_{d_j}}{k_{d_j}} \right)  , \\                                 
   \partial_\mu J^\mu_{u_3}       =& \partial_\mu J^\mu_{d_1}  
  - \partial_\mu J^\mu_{Q_3}       +  \partial_\mu J^\mu_{Q_{1}}, \\
  \partial_\mu J^\mu_{H}       =&   
    - \Gamma_{{Q_3} {u_3} H} \left(  \frac{n_{Q_3}}{k_{Q_3}} - \frac{n_{u_3}}{k_{u_3}} + \frac{n_H}{k_H} \right) 
       + \Gamma_{\psi {\tau_R} H} \left(   \frac{n_\psi}{k_\psi} -  \frac{n_{{\tau_R}}}{k_{{\tau_R}}} - \frac{n_H}{k_H} \right) 
  \\ &  \nonumber
       + \Gamma_{{L_3} {\tau_R} H} \left(   \frac{n_{L_3}}{k_{L_3}} -  \frac{n_{{\tau_R}}}{k_{{\tau_R}}} - \frac{n_H}{k_H}\right) 
.\end{align}
\end{subequations}
Here $n_H = n_{H^+}+n_{H^0}$ are the Higgs densities, and the scalar singlets
do not appear as they have zero chemical potential. Some of the relevant equilibration rates and CPV source terms are given in eqs.~\eqref{eq:rates} while the remainder are given here,
\begin{subequations}
\label{eq:rates2}
\begin{align}
  \Gamma_{\psi {\tau_R} H}^{\pm}(z) &= \frac{12}{T^2} \left\lvert Y'_\psi v_{H}(z) \right\rvert^2 \Lambda^{\pm}_{\psi {\tau_R}},
  \\
  \Gamma_{L_3 {\tau_R} H}^{\pm}(z) &= \frac{12}{T^2} \left\lvert Y'_\tau v_{H}(z) \right\rvert^2 \Lambda^{\pm}_{L_3 {\tau_R}},
  \\
  \Gamma_{Q_3 u_3 H}^{\pm}(z) &= \frac{12}{T^2} \left\lvert Y_{u_3} v_{H}(z) \right\rvert^2 \Lambda^{\pm}_{Q_3 u_3},
  \\
  \Gamma_{L_3 {{\tau_R}} H} & = \frac{12}{T^2} \left\lvert Y'_\tau \right\rvert^2 I_F(m_L, m_{{\tau_R}}, m_H),
  \\
  \Gamma_{\psi {{\tau_R}} H} & = \frac{12}{T^2} \left\lvert Y'_\psi \right\rvert^2 I_F(m_\psi, m_{{\tau_R}}, m_H),
  \\
  \Gamma_{Q_3 u_3 H} & = \frac{12}{T^2} \left\lvert Y_{u_3} \right\rvert^2 I_F(m_{Q_3}, m_{u_3}, m_H),
\end{align}
\end{subequations}

\subsection{Transport Equation Formulae}
\label{app:QTE}

The relaxation rates and source terms used in eqs.~\eqref{eq:rates} and
\eqref{eq:rates2} were derived using the VEV insertion approximation as outlined
in Ref.~\cite{ResRelax}. For the SM fermions, which are massless prior to EW
symmetry breaking, we make use of the interacting thermal propagator that has
particle (P) and hole (H) poles at energies $k^0 = \mathcal{E}^{X}_{P,H}(k)$
with residues $Z^{X}_{P,H}(k)$, where we use $X$ and $Y$ to denote some SM
fermion species. As the $\psi$ Dirac mass is significantly larger than the
thermal mass, we can safely utilise the free, non-interacting, thermal
propagator. Following Ref.~\cite{ResRelax}, we introduce the thermal widths
listed in Table~\ref{tab:ThermalWidths} by taking $\omega_k - i \epsilon
\rightarrow \mathcal{E}(k) = \omega_k - i \Gamma$. The rates and source terms
are then given by,
\begin{subequations}
\begin{equation}
\begin{aligned}
\label{eq:QTEIntegrals}
  \Lambda^0_{X \psi}  = \mathrm{Im}  \int_0^\infty \frac{ k^2  \dd k }{ \pi ^2 \omega_\psi }
  \bigg [ &
  Z^X_P \frac{\mathcal{E}_{\psi}+ k}{( \mathcal{E}_\psi + \mathcal{E}^{X}_{P} )^2} \left( n_f(\mathcal{E}_{\psi}) + n_f(\mathcal{E}^{X}_{P}) \right) 
  \\ + &
  Z^{X *}_P \frac{\mathcal{E}_{\psi}- k}{(\mathcal{E}_\psi - \mathcal{E}^{X *}_{P})^2} \left( n_f(\mathcal{E}_{\psi}) - n_f(\mathcal{E}^{X *}_{P}) \right) 
  \\ + &
  Z^X_H \frac{\mathcal{E}_{\psi}- k}{(\mathcal{E}_\psi + \mathcal{E}^{X}_{H})^2} \left( n_f(\mathcal{E}_{\psi}) + n_f(\mathcal{E}^{X}_{H})  \right) 
  \\ + &
  Z^{X *}_H \frac{\mathcal{E}_{\psi} + k}{(\mathcal{E}_\psi - \mathcal{E}^{X *}_{H})^2} \left( n_f(\mathcal{E}_{\psi}) - n_f(\mathcal{E}^{X *}_{H}) \right) \bigg ],
\end{aligned}
\end{equation}
\begin{equation}
\begin{aligned}
  \Lambda^\pm_{X \psi} =  -\mathrm{Im} \int_{0}^\infty \frac{ k^2 \dd k }{T \pi^2 \omega_\psi  }   \bigg[
  & Z^{X}_{P}  \frac{\mathcal{E}_{\psi} +k}{\mathcal{E}_{\psi} +\mathcal{E}^{X}_{P}}  \left(h_f(\mathcal{E}_{\psi} ) \mp  h_f(\mathcal{E}^{X}_{P}) \right)
  \\ + &
  Z^{X *}_{P}\frac{\mathcal{E}_{\psi}-k}{\mathcal{E}_{\psi}-\mathcal{E}^{X *}_{P}} \left( h_f(\mathcal{E}_{\psi}) \mp h_f(\mathcal{E}^{X *}_{P}) \right) 
  \\ + & Z^{X}_{H} \frac{\mathcal{E}_{\psi} -k}{\mathcal{E}_{\psi} +\mathcal{E}^{X}_{H}}    \left( h_f(\mathcal{E}_{\psi} ) \mp h_f(\mathcal{E}^{X}_{H}) \right) 
  \\ + &
  Z^{X *}_{H}\frac{\mathcal{E}_{\psi}+k}{\mathcal{E}_{\psi}-\mathcal{E}^{X *}_{H}}   \left( h_f(\mathcal{E}_{\psi}) \mp h_f(\mathcal{E}^{X *}_{H}) \right) \bigg],
\end{aligned}
\end{equation}
\begin{equation}
  \begin{aligned}
  \Lambda^\pm_{Y X} =  \mathrm{Im} \int_{0}^\infty \frac{ k^2 \dd k }{T \pi^2 }   \bigg[&
  \frac{Z^{X}_{P} Z^{Y}_{P}}{\mathcal{E}^{X}_{P} + \mathcal{E}^{Y}_{P}} \left( h_f(  \mathcal{E}^{X}_{P}) \mp h_f(  \mathcal{E}^{Y}_{P}) \right)
  \\ + &
  \frac{Z^{X}_{P} Z^{Y *}_{H}}{\mathcal{E}^{X}_{P} - \mathcal{E}^{Y *}_{H}} \left( h_f(  \mathcal{E}^{X}_{P}) \mp h_f(  \mathcal{E}^{Y *}_{H}) \right) \   + \    \left( P \leftrightarrow H \right) \bigg],
\end{aligned}
\end{equation}
\begin{align}
  \mathcal{E}_\psi(k) &= \omega_\psi(k) - i \Gamma_\psi = \sqrt{k^2 + {M'}^2_\psi } - i \Gamma_{\psi}
                        , &                                                                                                      
  n_f(x) &= \frac{1}{1 + e^{x/T}}
           , \\
               h_f(x) &= - T n_f'(x) = \frac{e^{x/T}}{(1 + e^{x/T})^2}  
                        .&& 
\end{align}
\end{subequations}
The formulae $\mathcal{E}^{X}_{P,H}(k)$ and $Z^{X}_{P,H}(k)$ are derived in Ref.~\cite{ThermalPropagators}. Due to the large vector-like mass the primary contributions to the integrals in $\Lambda^{0,\pm}_{X\psi}$ come from regions where the momenta are much larger than the thermal masses of the SM fermion. In this large momentum limit we have $Z^{X}_{P}\approx 1$, $Z^X_H \approx 0$ so that these integrals are in agreement with those used in Refs.~\cite{ChaoSpontCP, beautygenesis, MJRM-CHAO-VL-EWBG}.

The Yukawa-loop self energy function $I_F$ is taken from Ref.~\cite{ResRelaxFollowup},
\begin{subequations}
\begin{align}
  \nonumber	 &\mathcal{I}_F (m_1, m_2, m_\phi) 
               =
               \frac{1}{16 \pi^3 T} (m_1^2 + m_2^2 - m_\phi^2) \int_{m_1}^{\infty} \dd \omega_1 \int_{\omega_\phi^-}^{\omega_\phi^+} \dd \omega_\phi
  \\
             & \qquad \times \bigg\{
               n_B (\omega_\phi) [1 - n_F(\omega_1)]n_F(\omega_1 - \omega_\phi) [\theta(m_1 - m_2 - m_\phi) - \theta(m_\phi - m_1 - m_2)]
               \nonumber\\	
             &  \quad \qquad + n_B(\omega_\phi) n_F(\omega_1) [1 - n_F(\omega_1 + \omega_\phi)] \theta(m_2 - m_1 - m_\phi)
               \bigg\},
\end{align}
\begin{align}
  \nonumber \omega^{\pm}_\phi =& \frac{1}{2 m_1^2} \bigg( \omega_1 {\lvert m_\phi^2 + m_1^2 - m_2^2 \rvert} \\ & \quad \pm   \ \sqrt{                                         (\omega_1^2 - m_1^2) (m_1^2 - (m_2 + m_\phi)^2)(m_1^2 - (m_2 - m_\phi)^2)	} \bigg) .
\end{align}
\end{subequations}
In the event where $I_F = 0$ due to the mass-thresholds we utilise the
approximation for the four-body rates introduced in Refs.~\cite{ResRelaxFollowup, TwoStep}. For our
benchmarks this is only relevant for $Y_{Q_3 u_3 H}$. Due to the insignificance of
quark and Higgs densities this approximation is expected to
introduce negligible error. In units of $\mathrm{GeV}$, the rates evaluated for
benchmark point
$A$ are given in
Table~\ref{tab:NumericRates}. In this benchmark the CPV source across the bubble
wall is,
\begin{equation}
\int d z S_{L \psi}^{CPV}(z) = -8.60 \cdot 10^{-2} \ \mathrm{GeV}^3
\end{equation}
\begin{table}
  \centering
  \begin{tabular}{|c|c|c|}    
    \hline
    Rate & $\Gamma(z<0)\ (\mathrm{GeV})$ & $\Gamma(z>0)\ (\mathrm{GeV})$ \\ \hline
$\Gamma^{+}_{L \psi S_i }$ & $ -1.84\cdot 10^{-3} $ & $ 0 $ \\\hline    
$\Gamma^{-}_{L \psi S_i }$ & $ 1.63\cdot 10^{-2} $ & $ 0 $ \\\hline
$\Gamma^{+}_{\psi \tau_R H}$ & $ 0$ & $ 3.63\cdot 10^{-7}$ \\\hline
$\Gamma^{-}_{\psi \tau_R H}$ & $ 0$ & $ 2.52\cdot 10^{-6}$ \\\hline
$\Gamma^{+}_{L \tau_R H}$ & $ 0$ & $ 7.59\cdot 10^{-5}$ \\\hline
$\Gamma^{-}_{L \tau_R H}$ & $ 0$ & $ 3.49\cdot 10^{-4}$ \\\hline
$\Gamma^{+}_{Q_3 u_3 H}$ & $ 0$ & $ 9.43\cdot 10^{-1}$ \\\hline
$\Gamma^{-}_{Q_3 u_3 H}$ & $ 0$ & $ 4.88\cdot 10^{1}$ \\\hline
$\Gamma_{L \psi S_i}$ & $ 1.95\cdot 10^{-2}$ & $ 2.07\cdot 10^{-2}$ \\\hline
$\Gamma_{L \tau_R H}$ & $ 7.95\cdot 10^{-6}$ & $ 3.27\cdot 10^{-5}$ \\\hline
$\Gamma_{\psi \tau_R H}$ & $ 1.38\cdot 10^{-6}$ & $ 1.02\cdot 10^{-6}$ \\\hline
$\Gamma_{Q_3 u_3 H}$ & $ 1.42$ & $ 1.42$ \\\hline$\Gamma_{\mathrm{WS}}$ &$5.05 \cdot 10^{-4}$ &$0$ \\\hline 
$\Gamma_{\mathrm{SS}}$ & \multicolumn{2}{c|}{$1.55$}\\\hline
  \end{tabular}
  \caption{Numerical values for reaction rates used when solving the transport
    equations for benchmark point A.}
  \label{tab:NumericRates}
\end{table}
\bibliography{main}

\providecommand{\href}[2]{#2}\begingroup\raggedright\begin{thebibliography}{10}

\bibitem{planck2018}
{\scshape Planck} collaboration, N.~Aghanim et~al., \emph{{Planck 2018 results.
  VI. Cosmological parameters}},
  \href{https://arxiv.org/abs/1807.06209}{{\ttfamily 1807.06209}}.

\bibitem{Sakharov}
A.~D. Sakharov, \emph{{Violation of CP Invariance, c Asymmetry, and Baryon
  Asymmetry of the Universe}},
  \href{https://doi.org/10.1070/PU1991v034n05ABEH002497}{\emph{Pisma Zh. Eksp.
  Teor. Fiz.} {\bfseries 5} (1967) 32}.

\bibitem{crossover}
Y.~Aoki, F.~Csikor, Z.~Fodor and A.~Ukawa, \emph{{The Endpoint of the first
  order phase transition of the SU(2) gauge Higgs model on a four-dimensional
  isotropic lattice}},
  \href{https://doi.org/10.1103/PhysRevD.60.013001}{\emph{Phys. Rev.}
  {\bfseries D60} (1999) 013001}
  [\href{https://arxiv.org/abs/hep-lat/9901021}{{\ttfamily hep-lat/9901021}}].

\bibitem{notEnoughCPV}
M.~B. Gavela, P.~Hernandez, J.~Orloff, O.~Pene and C.~Quimbay, \emph{{Standard
  model CP violation and baryon asymmetry. Part 2: Finite temperature}},
  \href{https://doi.org/10.1016/0550-3213(94)00410-2}{\emph{Nucl. Phys.}
  {\bfseries B430} (1994) 382}
  [\href{https://arxiv.org/abs/hep-ph/9406289}{{\ttfamily hep-ph/9406289}}].

\bibitem{SingletColliderPheno1}
S.~Profumo, M.~J. Ramsey-Musolf and G.~Shaughnessy, \emph{{Singlet Higgs
  phenomenology and the electroweak phase transition}},
  \href{https://doi.org/10.1088/1126-6708/2007/08/010}{\emph{JHEP} {\bfseries
  08} (2007) 010} [\href{https://arxiv.org/abs/0705.2425}{{\ttfamily
  0705.2425}}].

\bibitem{SingletColliderPheno2}
S.~Profumo, M.~J. Ramsey-Musolf, C.~L. Wainwright and P.~Winslow,
  \emph{{Singlet-catalyzed electroweak phase transitions and precision Higgs
  boson studies}},
  \href{https://doi.org/10.1103/PhysRevD.91.035018}{\emph{Phys. Rev.}
  {\bfseries D91} (2015) 035018}
  [\href{https://arxiv.org/abs/1407.5342}{{\ttfamily 1407.5342}}].

\bibitem{CurtinColliderPheno}
D.~Curtin, P.~Meade and C.-T. Yu, \emph{{Testing Electroweak Baryogenesis with
  Future Colliders}},
  \href{https://doi.org/10.1007/JHEP11(2014)127}{\emph{JHEP} {\bfseries 11}
  (2014) 127} [\href{https://arxiv.org/abs/1409.0005}{{\ttfamily 1409.0005}}].

\bibitem{EDMSummary}
T.~Chupp, P.~Fierlinger, M.~Ramsey-Musolf and J.~Singh, \emph{{Electric Dipole
  Moments of the Atoms, Molecules, Nuclei and Particles}},
  \href{https://arxiv.org/abs/1710.02504}{{\ttfamily 1710.02504}}.

\bibitem{SingletEWPT1}
J.~R. Espinosa and M.~Quiros, \emph{{The Electroweak phase transition with a
  singlet}}, \href{https://doi.org/10.1016/0370-2693(93)91111-Y}{\emph{Phys.
  Lett.} {\bfseries B305} (1993) 98}
  [\href{https://arxiv.org/abs/hep-ph/9301285}{{\ttfamily hep-ph/9301285}}].

\bibitem{SingletEWPT2}
J.~Choi and R.~R. Volkas, \emph{{Real Higgs singlet and the electroweak phase
  transition in the Standard Model}},
  \href{https://doi.org/10.1016/0370-2693(93)91013-D}{\emph{Phys. Lett.}
  {\bfseries B317} (1993) 385}
  [\href{https://arxiv.org/abs/hep-ph/9308234}{{\ttfamily hep-ph/9308234}}].

\bibitem{SingletEWPT3}
J.~R. Espinosa, T.~Konstandin and F.~Riva, \emph{{Strong Electroweak Phase
  Transitions in the Standard Model with a Singlet}},
  \href{https://doi.org/10.1016/j.nuclphysb.2011.09.010}{\emph{Nucl. Phys.}
  {\bfseries B854} (2012) 592}
  [\href{https://arxiv.org/abs/1107.5441}{{\ttfamily 1107.5441}}].

\bibitem{SingletEWPT4}
C.-W. Chiang, Y.-T. Li and E.~Senaha, \emph{{Revisiting electroweak phase
  transition in the standard model with a real singlet scalar}},
  \href{https://doi.org/10.1016/j.physletb.2018.12.017}{\emph{Phys. Lett.}
  {\bfseries B789} (2019) 154}
  [\href{https://arxiv.org/abs/1808.01098}{{\ttfamily 1808.01098}}].

\bibitem{StepInto}
H.~H. Patel and M.~J. Ramsey-Musolf, \emph{{Stepping Into Electroweak Symmetry
  Breaking: Phase Transitions and Higgs Phenomenology}},
  \href{https://doi.org/10.1103/PhysRevD.88.035013}{\emph{Phys. Rev.}
  {\bfseries D88} (2013) 035013}
  [\href{https://arxiv.org/abs/1212.5652}{{\ttfamily 1212.5652}}].

\bibitem{MorriseyTwoStep}
N.~Blinov, J.~Kozaczuk, D.~E. Morrissey and C.~Tamarit, \emph{{Electroweak
  Baryogenesis from Exotic Electroweak Symmetry Breaking}},
  \href{https://doi.org/10.1103/PhysRevD.92.035012}{\emph{Phys. Rev.}
  {\bfseries D92} (2015) 035012}
  [\href{https://arxiv.org/abs/1504.05195}{{\ttfamily 1504.05195}}].

\bibitem{TwoStep}
S.~Inoue, G.~Ovanesyan and M.~J. Ramsey-Musolf, \emph{{Two-Step Electroweak
  Baryogenesis}}, \href{https://doi.org/10.1103/PhysRevD.93.015013}{\emph{Phys.
  Rev.} {\bfseries D93} (2016) 015013}
  [\href{https://arxiv.org/abs/1508.05404}{{\ttfamily 1508.05404}}].

\bibitem{ColorTwoStep}
M.~J. Ramsey-Musolf, P.~Winslow and G.~White, \emph{{Color Breaking
  Baryogenesis}}, \href{https://doi.org/10.1103/PhysRevD.97.123509}{\emph{Phys.
  Rev.} {\bfseries D97} (2018) 123509}
  [\href{https://arxiv.org/abs/1708.07511}{{\ttfamily 1708.07511}}].

\bibitem{VLQEDM}
C.-Y. Chen, S.~Dawson and Y.~Zhang, \emph{{Higgs CP Violation from Vectorlike
  Quarks}}, \href{https://doi.org/10.1103/PhysRevD.92.075026}{\emph{Phys. Rev.}
  {\bfseries D92} (2015) 075026}
  [\href{https://arxiv.org/abs/1507.07020}{{\ttfamily 1507.07020}}].

\bibitem{ChaoSpontCP}
W.~Chao, \emph{{CP Violation at the Finite Temperature}},
  \href{https://arxiv.org/abs/1706.01041}{{\ttfamily 1706.01041}}.

\bibitem{VLFDiphotonEWPT}
H.~Davoudiasl, I.~Lewis and E.~Ponton, \emph{{Electroweak Phase Transition,
  Higgs Diphoton Rate, and New Heavy Fermions}},
  \href{https://doi.org/10.1103/PhysRevD.87.093001}{\emph{Phys. Rev.}
  {\bfseries D87} (2013) 093001}
  [\href{https://arxiv.org/abs/1211.3449}{{\ttfamily 1211.3449}}].

\bibitem{VLFDMEWPTEWBG}
M.~Fairbairn and P.~Grothaus, \emph{{Baryogenesis and Dark Matter with
  Vector-like Fermions}},
  \href{https://doi.org/10.1007/JHEP10(2013)176}{\emph{JHEP} {\bfseries 10}
  (2013) 176} [\href{https://arxiv.org/abs/1307.8011}{{\ttfamily 1307.8011}}].

\bibitem{MJRM-CHAO-VL-EWBG}
W.~Chao and M.~J. Ramsey-Musolf, \emph{{Electroweak Baryogenesis, Electric
  Dipole Moments, and Higgs Diphoton Decays}},
  \href{https://doi.org/10.1007/JHEP10(2014)180}{\emph{JHEP} {\bfseries 10}
  (2014) 180} [\href{https://arxiv.org/abs/1406.0517}{{\ttfamily 1406.0517}}].

\bibitem{VLFEWPTEWBGEDM}
D.~Egana-Ugrinovic, \emph{{The minimal fermionic model of electroweak
  baryogenesis}}, \href{https://doi.org/10.1007/JHEP12(2017)064}{\emph{JHEP}
  {\bfseries 12} (2017) 064}
  [\href{https://arxiv.org/abs/1707.02306}{{\ttfamily 1707.02306}}].

\bibitem{VLFEWPT}
A.~Angelescu and P.~Huang, \emph{{Multistep Strongly First Order Phase
  Transitions from New Fermions at the TeV Scale}},
  \href{https://doi.org/10.1103/PhysRevD.99.055023}{\emph{Phys. Rev.}
  {\bfseries D99} (2019) 055023}
  [\href{https://arxiv.org/abs/1812.08293}{{\ttfamily 1812.08293}}].

\bibitem{VLLPheno}
N.~Kumar and S.~P. Martin, \emph{{Vectorlike leptons at the Large Hadron
  Collider}}, \href{https://doi.org/10.1103/PhysRevD.92.115018}{\emph{Phys.
  Rev.} {\bfseries D92} (2015) 115018}
  [\href{https://arxiv.org/abs/1510.03456}{{\ttfamily 1510.03456}}].

\bibitem{higgstautau}
{\scshape CMS} collaboration, A.~M. Sirunyan et~al., \emph{{Observation of the
  Higgs boson decay to a pair of $\tau$ leptons with the CMS detector}},
  \href{https://doi.org/10.1016/j.physletb.2018.02.004}{\emph{Phys. Lett.}
  {\bfseries B779} (2018) 283}
  [\href{https://arxiv.org/abs/1708.00373}{{\ttfamily 1708.00373}}].

\bibitem{mjrmEDMmssm}
Y.~Li, S.~Profumo and M.~Ramsey-Musolf, \emph{{Higgs-Higgsino-Gaugino Induced
  Two Loop Electric Dipole Moments}},
  \href{https://doi.org/10.1103/PhysRevD.78.075009}{\emph{Phys. Rev.}
  {\bfseries D78} (2008) 075009}
  [\href{https://arxiv.org/abs/0806.2693}{{\ttfamily 0806.2693}}].

\bibitem{asperge}
A.~Alloul, J.~D'Hondt, K.~De~Causmaecker, B.~Fuks and M.~Rausch~de Traubenberg,
  \emph{{Automated mass spectrum generation for new physics}},
  \href{https://doi.org/10.1140/epjc/s10052-013-2325-x}{\emph{Eur. Phys. J.}
  {\bfseries C73} (2013) 2325}
  [\href{https://arxiv.org/abs/1301.5932}{{\ttfamily 1301.5932}}].

\bibitem{feynrules}
A.~Alloul, N.~D. Christensen, C.~Degrande, C.~Duhr and B.~Fuks,
  \emph{{FeynRules 2.0 - A complete toolbox for tree-level phenomenology}},
  \href{https://doi.org/10.1016/j.cpc.2014.04.012}{\emph{Comput. Phys. Commun.}
  {\bfseries 185} (2014) 2250}
  [\href{https://arxiv.org/abs/1310.1921}{{\ttfamily 1310.1921}}].

\bibitem{UFO}
C.~Degrande, C.~Duhr, B.~Fuks, D.~Grellscheid, O.~Mattelaer and T.~Reiter,
  \emph{{UFO - The Universal FeynRules Output}},
  \href{https://doi.org/10.1016/j.cpc.2012.01.022}{\emph{Comput. Phys. Commun.}
  {\bfseries 183} (2012) 1201}
  [\href{https://arxiv.org/abs/1108.2040}{{\ttfamily 1108.2040}}].

\bibitem{madgraph}
J.~Alwall, R.~Frederix, S.~Frixione, V.~Hirschi, F.~Maltoni, O.~Mattelaer
  et~al., \emph{{The automated computation of tree-level and next-to-leading
  order differential cross sections, and their matching to parton shower
  simulations}}, \href{https://doi.org/10.1007/JHEP07(2014)079}{\emph{JHEP}
  {\bfseries 07} (2014) 079} [\href{https://arxiv.org/abs/1405.0301}{{\ttfamily
  1405.0301}}].

\bibitem{madWidth}
J.~Alwall, C.~Duhr, B.~Fuks, O.~Mattelaer, D.~G. Öztürk and C.-H. Shen,
  \emph{{Computing decay rates for new physics theories with FeynRules and
  MadGraph 5\_aMC@NLO}},
  \href{https://doi.org/10.1016/j.cpc.2015.08.031}{\emph{Comput. Phys. Commun.}
  {\bfseries 197} (2015) 312}
  [\href{https://arxiv.org/abs/1402.1178}{{\ttfamily 1402.1178}}].

\bibitem{pythia1}
T.~Sjostrand, S.~Mrenna and P.~Z. Skands, \emph{{PYTHIA 6.4 Physics and
  Manual}}, \href{https://doi.org/10.1088/1126-6708/2006/05/026}{\emph{JHEP}
  {\bfseries 05} (2006) 026}
  [\href{https://arxiv.org/abs/hep-ph/0603175}{{\ttfamily hep-ph/0603175}}].

\bibitem{pythia2}
T.~Sjostrand, S.~Mrenna and P.~Z. Skands, \emph{{A Brief Introduction to PYTHIA
  8.1}}, \href{https://doi.org/10.1016/j.cpc.2008.01.036}{\emph{Comput. Phys.
  Commun.} {\bfseries 178} (2008) 852}
  [\href{https://arxiv.org/abs/0710.3820}{{\ttfamily 0710.3820}}].

\bibitem{checkmate}
D.~Dercks, N.~Desai, J.~S. Kim, K.~Rolbiecki, J.~Tattersall and T.~Weber,
  \emph{{CheckMATE 2: From the model to the limit}},
  \href{https://doi.org/10.1016/j.cpc.2017.08.021}{\emph{Comput. Phys. Commun.}
  {\bfseries 221} (2017) 383}
  [\href{https://arxiv.org/abs/1611.09856}{{\ttfamily 1611.09856}}].

\bibitem{CLS}
A.~L. Read, \emph{{Presentation of search results: The CL(s) technique}},
  \href{https://doi.org/10.1088/0954-3899/28/10/313}{\emph{J. Phys.} {\bfseries
  G28} (2002) 2693}.

\bibitem{fastjet2}
M.~Cacciari and G.~P. Salam, \emph{{Dispelling the $N^{3}$ myth for the $k_t$
  jet-finder}},
  \href{https://doi.org/10.1016/j.physletb.2006.08.037}{\emph{Phys. Lett.}
  {\bfseries B641} (2006) 57}
  [\href{https://arxiv.org/abs/hep-ph/0512210}{{\ttfamily hep-ph/0512210}}].

\bibitem{cmjet}
M.~Cacciari, G.~P. Salam and G.~Soyez, \emph{{The Anti-k(t) jet clustering
  algorithm}}, \href{https://doi.org/10.1088/1126-6708/2008/04/063}{\emph{JHEP}
  {\bfseries 04} (2008) 063} [\href{https://arxiv.org/abs/0802.1189}{{\ttfamily
  0802.1189}}].

\bibitem{fastjet1}
M.~Cacciari, G.~P. Salam and G.~Soyez, \emph{{FastJet User Manual}},
  \href{https://doi.org/10.1140/epjc/s10052-012-1896-2}{\emph{Eur. Phys. J.}
  {\bfseries C72} (2012) 1896}
  [\href{https://arxiv.org/abs/1111.6097}{{\ttfamily 1111.6097}}].

\bibitem{delphes}
{\scshape DELPHES 3} collaboration, J.~de~Favereau, C.~Delaere, P.~Demin,
  A.~Giammanco, V.~Lemaître, A.~Mertens et~al., \emph{{DELPHES 3, A modular
  framework for fast simulation of a generic collider experiment}},
  \href{https://doi.org/10.1007/JHEP02(2014)057}{\emph{JHEP} {\bfseries 02}
  (2014) 057} [\href{https://arxiv.org/abs/1307.6346}{{\ttfamily 1307.6346}}].

\bibitem{LeptonKFactor}
R.~Ruiz, \emph{{QCD Corrections to Pair Production of Type III Seesaw Leptons
  at Hadron Colliders}},
  \href{https://doi.org/10.1007/JHEP12(2015)165}{\emph{JHEP} {\bfseries 12}
  (2015) 165} [\href{https://arxiv.org/abs/1509.05416}{{\ttfamily
  1509.05416}}].

\bibitem{atlas170603731}
{\scshape ATLAS} collaboration, M.~Aaboud et~al., \emph{{Search for
  supersymmetry in final states with two same-sign or three leptons and jets
  using 36 fb$^{-1}$ of $\sqrt{s}=13$ TeV $pp$ collision data with the ATLAS
  detector}}, \href{https://doi.org/10.1007/JHEP09(2017)084}{\emph{JHEP}
  {\bfseries 09} (2017) 084}
  [\href{https://arxiv.org/abs/1706.03731}{{\ttfamily 1706.03731}}].

\bibitem{atlas170807875}
{\scshape ATLAS} collaboration, M.~Aaboud et~al., \emph{{Search for the direct
  production of charginos and neutralinos in final states with tau leptons in
  $\sqrt{s}=$ 13 TeV $pp$ collisions with the ATLAS detector}},
  \href{https://doi.org/10.1140/epjc/s10052-018-5583-9}{\emph{Eur. Phys. J.}
  {\bfseries C78} (2018) 154}
  [\href{https://arxiv.org/abs/1708.07875}{{\ttfamily 1708.07875}}].

\bibitem{CMSSUS16039}
{\scshape CMS} collaboration, A.~M. Sirunyan et~al., \emph{{Search for
  electroweak production of charginos and neutralinos in multilepton final
  states in proton-proton collisions at $\sqrt{s}=$ 13 TeV}},
  \href{https://doi.org/10.1007/JHEP03(2018)166}{\emph{JHEP} {\bfseries 03}
  (2018) 166} [\href{https://arxiv.org/abs/1709.05406}{{\ttfamily
  1709.05406}}].

\bibitem{No:2013wsa}
J.~M. No and M.~Ramsey-Musolf, \emph{{Probing the Higgs Portal at the LHC
  Through Resonant di-Higgs Production}},
  \href{https://doi.org/10.1103/PhysRevD.89.095031}{\emph{Phys. Rev.}
  {\bfseries D89} (2014) 095031}
  [\href{https://arxiv.org/abs/1310.6035}{{\ttfamily 1310.6035}}].

\bibitem{xSM}
T.~Huang, J.~M. No, L.~Pernié, M.~Ramsey-Musolf, A.~Safonov, M.~Spannowsky
  et~al., \emph{{Resonant di-Higgs boson production in the $b{\bar b}WW$
  channel: Probing the electroweak phase transition at the LHC}},
  \href{https://doi.org/10.1103/PhysRevD.96.035007}{\emph{Phys. Rev.}
  {\bfseries D96} (2017) 035007}
  [\href{https://arxiv.org/abs/1701.04442}{{\ttfamily 1701.04442}}].

\bibitem{ATLASHiggsCubic}
{\scshape ATLAS Collaboration} collaboration, \emph{{Combination of searches
  for Higgs boson pairs in $pp$ collisions at 13 TeV with the ATLAS
  experiment.}},  Tech. Rep. ATLAS-CONF-2018-043, CERN, Geneva, Sep, 2018.

\bibitem{Kotwal:2016tex}
A.~V. Kotwal, M.~J. Ramsey-Musolf, J.~M. No and P.~Winslow,
  \emph{{Singlet-catalyzed electroweak phase transitions in the 100 TeV
  frontier}}, \href{https://doi.org/10.1103/PhysRevD.94.035022}{\emph{Phys.
  Rev.} {\bfseries D94} (2016) 035022}
  [\href{https://arxiv.org/abs/1605.06123}{{\ttfamily 1605.06123}}].

\bibitem{SingletPairAtHLLHC}
C.-Y. Chen, J.~Kozaczuk and I.~M. Lewis, \emph{{Non-resonant Collider
  Signatures of a Singlet-Driven Electroweak Phase Transition}},
  \href{https://doi.org/10.1007/JHEP08(2017)096}{\emph{JHEP} {\bfseries 08}
  (2017) 096} [\href{https://arxiv.org/abs/1704.05844}{{\ttfamily
  1704.05844}}].

\bibitem{SingletPairAtHELC}
D.~Buttazzo, D.~Redigolo, F.~Sala and A.~Tesi, \emph{{Fusing Vectors into
  Scalars at High Energy Lepton Colliders}},
  \href{https://doi.org/10.1007/JHEP11(2018)144}{\emph{JHEP} {\bfseries 11}
  (2018) 144} [\href{https://arxiv.org/abs/1807.04743}{{\ttfamily
  1807.04743}}].

\bibitem{ACMEEDM}
{\scshape ACME} collaboration, V.~Andreev et~al., \emph{{Improved limit on the
  electric dipole moment of the electron}},
  \href{https://doi.org/10.1038/s41586-018-0599-8}{\emph{Nature} {\bfseries
  562} (2018) 355}.

\bibitem{packageX}
H.~H. Patel, \emph{{Package-X: A Mathematica package for the analytic
  calculation of one-loop integrals}},
  \href{https://doi.org/10.1016/j.cpc.2015.08.017}{\emph{Comput. Phys. Commun.}
  {\bfseries 197} (2015) 276}
  [\href{https://arxiv.org/abs/1503.01469}{{\ttfamily 1503.01469}}].

\bibitem{VLQZVert}
E.~C. Leskow, T.~A. Martin and A.~de~la Puente, \emph{Vector-like quarks with a
  scalar triplet},
  \href{https://doi.org/10.1016/j.physletb.2015.02.071}{\emph{Physics Letters
  B} {\bfseries 743} (2015) 366 }.

\bibitem{PDG2016}
{\scshape Particle Data Group} collaboration, C.~Patrignani et~al.,
  \emph{{Review of Particle Physics}},
  \href{https://doi.org/10.1088/1674-1137/40/10/100001}{\emph{Chin. Phys.}
  {\bfseries C40} (2016) 100001}.

\bibitem{LeptonPheno1}
Z.~Poh and S.~Raby, \emph{{Vectorlike leptons: Muon g-2 anomaly, lepton flavor
  violation, Higgs boson decays, and lepton nonuniversality}},
  \href{https://doi.org/10.1103/PhysRevD.96.015032}{\emph{Phys. Rev.}
  {\bfseries D96} (2017) 015032}
  [\href{https://arxiv.org/abs/1705.07007}{{\ttfamily 1705.07007}}].

\bibitem{ResRelax}
C.~Lee, V.~Cirigliano and M.~J. Ramsey-Musolf, \emph{{Resonant relaxation in
  electroweak baryogenesis}},
  \href{https://doi.org/10.1103/PhysRevD.71.075010}{\emph{Phys. Rev.}
  {\bfseries D71} (2005) 075010}
  [\href{https://arxiv.org/abs/hep-ph/0412354}{{\ttfamily hep-ph/0412354}}].

\bibitem{MJRMgaugeDep}
H.~H. Patel and M.~J. Ramsey-Musolf, \emph{{Baryon Washout, Electroweak Phase
  Transition, and Perturbation Theory}},
  \href{https://doi.org/10.1007/JHEP07(2011)029}{\emph{JHEP} {\bfseries 07}
  (2011) 029} [\href{https://arxiv.org/abs/1101.4665}{{\ttfamily 1101.4665}}].

\bibitem{Origsphaleron}
F.~R. Klinkhamer and N.~S. Manton, \emph{A saddle-point solution in the
  weinberg-salam theory},
  \href{https://doi.org/10.1103/PhysRevD.30.2212}{\emph{Phys. Rev. D}
  {\bfseries 30} (1984) 2212}.

\bibitem{singletSphaleron}
A.~Ahriche, \emph{{What is the criterion for a strong first order electroweak
  phase transition in singlet models?}},
  \href{https://doi.org/10.1103/PhysRevD.75.083522}{\emph{Phys. Rev.}
  {\bfseries D75} (2007) 083522}
  [\href{https://arxiv.org/abs/hep-ph/0701192}{{\ttfamily hep-ph/0701192}}].

\bibitem{sphaleronEnergyForRepresentations}
A.~Ahriche, T.~A. Chowdhury and S.~Nasri, \emph{{Sphalerons and the Electroweak
  Phase Transition in Models with Higher Scalar Representations}},
  \href{https://doi.org/10.1007/JHEP11(2014)096}{\emph{JHEP} {\bfseries 11}
  (2014) 096} [\href{https://arxiv.org/abs/1409.4086}{{\ttfamily 1409.4086}}].

\bibitem{MJRMgaugeDep2}
C.-W. Chiang, M.~J. Ramsey-Musolf and E.~Senaha, \emph{{Standard Model with a
  Complex Scalar Singlet: Cosmological Implications and Theoretical
  Considerations}},
  \href{https://doi.org/10.1103/PhysRevD.97.015005}{\emph{Phys. Rev.}
  {\bfseries D97} (2018) 015005}
  [\href{https://arxiv.org/abs/1707.09960}{{\ttfamily 1707.09960}}].

\bibitem{wallVelocityKonstandin}
T.~Konstandin, G.~Nardini and I.~Rues, \emph{{From Boltzmann equations to
  steady wall velocities}},
  \href{https://doi.org/10.1088/1475-7516/2014/09/028}{\emph{JCAP} {\bfseries
  1409} (2014) 028} [\href{https://arxiv.org/abs/1407.3132}{{\ttfamily
  1407.3132}}].

\bibitem{wallVelocity}
J.~Kozaczuk, \emph{{Bubble Expansion and the Viability of Singlet-Driven
  Electroweak Baryogenesis}},
  \href{https://doi.org/10.1007/JHEP10(2015)135}{\emph{JHEP} {\bfseries 10}
  (2015) 135} [\href{https://arxiv.org/abs/1506.04741}{{\ttfamily
  1506.04741}}].

\bibitem{newKonstandinWall}
G.~C. Dorsch, S.~J. Huber and T.~Konstandin, \emph{{Bubble wall velocities in
  the Standard Model and beyond}},
  \href{https://arxiv.org/abs/1809.04907}{{\ttfamily 1809.04907}}.

\bibitem{TroddenReview}
M.~Trodden, \emph{{Electroweak baryogenesis: A Brief review}},  in
  \emph{{Proceedings, 33rd Rencontres de Moriond 98 electrowek interactions and
  unified theories: Les Arcs, France, Mar 14-21, 1998}}, pp.~471--480, 1998,
  \href{https://arxiv.org/abs/hep-ph/9805252}{{\ttfamily hep-ph/9805252}}.

\bibitem{MorrisseyMJRMReview}
D.~E. Morrissey and M.~J. Ramsey-Musolf, \emph{{Electroweak baryogenesis}},
  \href{https://doi.org/10.1088/1367-2630/14/12/125003}{\emph{New J. Phys.}
  {\bfseries 14} (2012) 125003}
  [\href{https://arxiv.org/abs/1206.2942}{{\ttfamily 1206.2942}}].

\bibitem{GrahamReview}
A.~Mazumdar and G.~White, \emph{{Cosmic phase transitions: their applications
  and experimental signatures}},
  \href{https://arxiv.org/abs/1811.01948}{{\ttfamily 1811.01948}}.

\bibitem{CosmoTransitions}
C.~L. Wainwright, \emph{{CosmoTransitions: Computing Cosmological Phase
  Transition Temperatures and Bubble Profiles with Multiple Fields}},
  \href{https://doi.org/10.1016/j.cpc.2012.04.004}{\emph{Comput. Phys. Commun.}
  {\bfseries 183} (2012) 2006}
  [\href{https://arxiv.org/abs/1109.4189}{{\ttfamily 1109.4189}}].

\bibitem{LindeFTFTVacuumDecay}
A.~D. Linde, \emph{{Decay of the False Vacuum at Finite Temperature}},
  \href{https://doi.org/10.1016/0550-3213(83)90293-6,
  10.1016/0550-3213(83)90072-X}{\emph{Nucl. Phys.} {\bfseries B216} (1983)
  421}.

\bibitem{grahamGravWaves}
D.~Croon and G.~White, \emph{{Exotic Gravitational Wave Signatures from
  Simultaneous Phase Transitions}},
  \href{https://arxiv.org/abs/1803.05438}{{\ttfamily 1803.05438}}.

\bibitem{GW1}
E.~Witten, \emph{Cosmic separation of phases},
  \href{https://doi.org/10.1103/PhysRevD.30.272}{\emph{Phys. Rev. D} {\bfseries
  30} (1984) 272}.

\bibitem{GW2}
C.~J. Hogan, \emph{{Gravitational radiation from cosmological phase
  transitions}}, \href{https://doi.org/10.1093/mnras/218.4.629}{\emph{Monthly
  Notices of the Royal Astronomical Society} {\bfseries 218} (1986) 629}
  [\href{https://arxiv.org/abs/http://oup.prod.sis.lan/mnras/article-pdf/218/4/629/3299141/mnras218-0629.pdf}{{\ttfamily
  http://oup.prod.sis.lan/mnras/article-pdf/218/4/629/3299141/mnras218-0629.pdf}}].

\bibitem{GW3}
M.~S. Turner and F.~Wilczek, \emph{Relic gravitational waves and extended
  inflation}, \href{https://doi.org/10.1103/PhysRevLett.65.3080}{\emph{Phys.
  Rev. Lett.} {\bfseries 65} (1990) 3080}.

\bibitem{GW4}
A.~Kosowsky and M.~S. Turner, \emph{{Gravitational radiation from colliding
  vacuum bubbles: envelope approximation to many bubble collisions}},
  \href{https://doi.org/10.1103/PhysRevD.47.4372}{\emph{Phys. Rev.} {\bfseries
  D47} (1993) 4372} [\href{https://arxiv.org/abs/astro-ph/9211004}{{\ttfamily
  astro-ph/9211004}}].

\bibitem{GW5}
M.~Kamionkowski, A.~Kosowsky and M.~S. Turner, \emph{{Gravitational radiation
  from first order phase transitions}},
  \href{https://doi.org/10.1103/PhysRevD.49.2837}{\emph{Phys. Rev.} {\bfseries
  D49} (1994) 2837} [\href{https://arxiv.org/abs/astro-ph/9310044}{{\ttfamily
  astro-ph/9310044}}].

\bibitem{multistepGravWave}
T.~Vieu, A.~P. Morais and R.~Pasechnik, \emph{{Multi-peaked signatures of
  primordial gravitational waves from multi-step electroweak phase
  transition}},  \href{https://arxiv.org/abs/1802.10109}{{\ttfamily
  1802.10109}}.

\bibitem{semiclassicalForce}
L.~Fromme and S.~J. Huber, \emph{{Top transport in electroweak baryogenesis}},
  \href{https://doi.org/10.1088/1126-6708/2007/03/049}{\emph{JHEP} {\bfseries
  03} (2007) 049} [\href{https://arxiv.org/abs/hep-ph/0604159}{{\ttfamily
  hep-ph/0604159}}].

\bibitem{semiclassicalForce2}
J.~M. Cline, K.~Kainulainen and M.~Trott, \emph{{Electroweak Baryogenesis in
  Two Higgs Doublet Models and B meson anomalies}},
  \href{https://doi.org/10.1007/JHEP11(2011)089}{\emph{JHEP} {\bfseries 11}
  (2011) 089} [\href{https://arxiv.org/abs/1107.3559}{{\ttfamily 1107.3559}}].

\bibitem{DiffusionTerms}
M.~Joyce, T.~Prokopec and N.~Turok, \emph{{Nonlocal electroweak baryogenesis.
  Part 1: Thin wall regime}},
  \href{https://doi.org/10.1103/PhysRevD.53.2930}{\emph{Phys. Rev.} {\bfseries
  D53} (1996) 2930} [\href{https://arxiv.org/abs/hep-ph/9410281}{{\ttfamily
  hep-ph/9410281}}].

\bibitem{ResEWBG}
A.~Riotto, \emph{{The More relaxed supersymmetric electroweak baryogenesis}},
  \href{https://doi.org/10.1103/PhysRevD.58.095009}{\emph{Phys. Rev.}
  {\bfseries D58} (1998) 095009}
  [\href{https://arxiv.org/abs/hep-ph/9803357}{{\ttfamily hep-ph/9803357}}].

\bibitem{ResRelaxFollowup}
V.~Cirigliano, M.~J. Ramsey-Musolf, S.~Tulin and C.~Lee, \emph{{Yukawa and
  tri-scalar processes in electroweak baryogenesis}},
  \href{https://doi.org/10.1103/PhysRevD.73.115009}{\emph{Phys. Rev.}
  {\bfseries D73} (2006) 115009}
  [\href{https://arxiv.org/abs/hep-ph/0603058}{{\ttfamily hep-ph/0603058}}].

\bibitem{KPSW}
K.~Kainulainen, T.~Prokopec, M.~G. Schmidt and S.~Weinstock, \emph{{First
  principle derivation of semiclassical force for electroweak baryogenesis}},
  \href{https://doi.org/10.1088/1126-6708/2001/06/031}{\emph{JHEP} {\bfseries
  06} (2001) 031} [\href{https://arxiv.org/abs/hep-ph/0105295}{{\ttfamily
  hep-ph/0105295}}].

\bibitem{FlavQTE}
V.~Cirigliano, C.~Lee, M.~J. Ramsey-Musolf and S.~Tulin, \emph{{Flavored
  Quantum Boltzmann Equations}},
  \href{https://doi.org/10.1103/PhysRevD.81.103503}{\emph{Phys. Rev.}
  {\bfseries D81} (2010) 103503}
  [\href{https://arxiv.org/abs/0912.3523}{{\ttfamily 0912.3523}}].

\bibitem{ResFlavQTE}
V.~Cirigliano, C.~Lee and S.~Tulin, \emph{{Resonant Flavor Oscillations in
  Electroweak Baryogenesis}},
  \href{https://doi.org/10.1103/PhysRevD.84.056006}{\emph{Phys. Rev.}
  {\bfseries D84} (2011) 056006}
  [\href{https://arxiv.org/abs/1106.0747}{{\ttfamily 1106.0747}}].

\bibitem{GammaWS2}
G.~D. Moore, \emph{{Do we understand the sphaleron rate?}},  in \emph{{Strong
  and electroweak matter. Proceedings, Meeting, SEWM 2000, Marseille, France,
  June 13-17, 2000}}, pp.~82--94, 2000,
  \href{https://arxiv.org/abs/hep-ph/0009161}{{\ttfamily hep-ph/0009161}},
  \href{https://doi.org/10.1142/9789812799913_0007}{DOI}.

\bibitem{GammaWS1}
G.~D. Moore, \emph{{Sphaleron rate in the symmetric electroweak phase}},
  \href{https://doi.org/10.1103/PhysRevD.62.085011}{\emph{Phys. Rev.}
  {\bfseries D62} (2000) 085011}
  [\href{https://arxiv.org/abs/hep-ph/0001216}{{\ttfamily hep-ph/0001216}}].

\bibitem{GammaSS}
G.~D. Moore and M.~Tassler, \emph{{The Sphaleron Rate in SU(N) Gauge Theory}},
  \href{https://doi.org/10.1007/JHEP02(2011)105}{\emph{JHEP} {\bfseries 02}
  (2011) 105} [\href{https://arxiv.org/abs/1011.1167}{{\ttfamily 1011.1167}}].

\bibitem{beautygenesis}
T.~Liu, M.~J. Ramsey-Musolf and J.~Shu, \emph{{Electroweak Beautygenesis: From
  $b \to s$ CP-violation to the Cosmic Baryon Asymmetry}},
  \href{https://doi.org/10.1103/PhysRevLett.108.221301}{\emph{Phys. Rev. Lett.}
  {\bfseries 108} (2012) 221301}
  [\href{https://arxiv.org/abs/1109.4145}{{\ttfamily 1109.4145}}].

\bibitem{TikZFeynman}
J.~Ellis, \emph{{TikZ-Feynman: Feynman diagrams with TikZ}},
  \href{https://doi.org/10.1016/j.cpc.2016.08.019}{\emph{Comput. Phys. Commun.}
  {\bfseries 210} (2017) 103}
  [\href{https://arxiv.org/abs/1601.05437}{{\ttfamily 1601.05437}}].

\bibitem{VLQEDMCalc}
A.~Avilez-Lopez, H.~Novales-Sanchez, G.~Tavares-Velasco and J.~J. Toscano,
  \emph{{Bound on the anomalous tbW coupling from two-loop contribution to
  neutron electric dipole moment}},
  \href{https://doi.org/10.1016/j.physletb.2007.08.012}{\emph{Phys. Lett.}
  {\bfseries B653} (2007) 241}
  [\href{https://arxiv.org/abs/0708.1786}{{\ttfamily 0708.1786}}].

\bibitem{widths1}
P.~Elmfors, K.~Enqvist, A.~Riotto and I.~Vilja, \emph{{Damping rates in the
  MSSM and electroweak baryogenesis}},
  \href{https://doi.org/10.1016/S0370-2693(99)00169-0}{\emph{Phys. Lett.}
  {\bfseries B452} (1999) 279}
  [\href{https://arxiv.org/abs/hep-ph/9809529}{{\ttfamily hep-ph/9809529}}].

\bibitem{ThermalPropagators}
H.~A. Weldon, \emph{Effective fermion masses of order $\mathrm{gT}$ in
  high-temperature gauge theories with exact chiral invariance},
  \href{https://doi.org/10.1103/PhysRevD.26.2789}{\emph{Phys. Rev. D}
  {\bfseries 26} (1982) 2789}.

\end{thebibliography}\endgroup

\end{document}